\documentclass[10pt,twocolumn,aps,prx,superscriptaddress,longbibliography,eprint]
{revtex4-2}

\usepackage{graphicx}
\usepackage{amsmath}
\usepackage{amssymb}
\usepackage{float}
\usepackage{braket}   
\usepackage{enumitem}
\usepackage[dvipsnames]{xcolor}
\usepackage[colorlinks=true, urlcolor=blue, linkcolor=blue,citecolor=blue, hypertexnames=false]{hyperref}
\usepackage[detect-weight=true,separate-uncertainty=true]{siunitx}
\usepackage{comment}
\usepackage[normalem]{ulem}

\usepackage{cleveref}
\crefname{equation}{Eq.}{Eqs.}
\Crefname{equation}{Equation}{Equations}
\crefname{figure}{Fig.}{Figs.}
\Crefname{figure}{Figure}{Figures}
\crefname{section}{Sec.}{Sects.}
\Crefname{section}{Section}{Sections}
\crefname{table}{Table}{Tables}
\crefname{appendix}{Appendix}{Apps.}
\Crefname{appendix}{Appendix}{Apps.}

\newcommand{\ha}{\hat{a}}

\newcommand{\hn}{\hat{n}}
\newcommand{\hvarphi}{\hat{\varphi}}

\newcommand{\hH}{\hat{H}}

\begin{document}

\title{How the Kerr-Cat Qubit Dies---And How to Rescue It}

\author{Othmane Benhayoune-Khadraoui}
\email{othmane.benhayoune.khadraoui@usherbrooke.ca}
\affiliation{Institut Quantique and D\'epartement de Physique, Universit\'e de Sherbrooke, Sherbrooke J1K 2R1 Quebec, Canada}

\author{Crist\'{o}bal Lled\'{o}}
\affiliation{Institut Quantique and D\'epartement de Physique, Universit\'e de Sherbrooke, Sherbrooke J1K 2R1 Quebec, Canada}
\affiliation{Departamento de Física, Facultad de Ciencias Físicas y Matemáticas, Universidad de Chile, Santiago 837.0415, Chile}

\author{Alexandre Blais}
\affiliation{Institut Quantique and D\'epartement de Physique, Universit\'e de Sherbrooke, Sherbrooke J1K 2R1 Quebec, Canada}
\affiliation{CIFAR, Toronto, ON M5G 1M1, Canada}

\date{\today}

\begin{abstract}
Kerr-cat qubits have been experimentally shown to exhibit a large noise bias, with one decay channel suppressed by several orders of magnitude. In superconducting implementations, increasing the microwave drive on the nonlinear oscillator that hosts the Kerr-cat qubit should, in principle, further enhance this bias. Instead, experiments reveal that above a critical drive amplitude the tunneling time, which is the less dominant decay channel, ceases to increase and even decreases. Here, we show that this breakdown arises from the multimode nature of the circuit implementation. 
Specifically, additional modes, including the buffer mode used to deliver the stabilizing drive and higher modes of the Josephson junction array, can induce multiphoton resonances that sharply degrade Kerr-cat coherence. 
We uncover this mechanism by retaining the full circuit nonlinearities and treating the strong drive exactly within a Floquet–Markov framework that accounts for quasidegeneracies of the Kerr-cat spectrum. Our results not only provide an explanation for the sudden reduction of the tunneling time but also demonstrate that the Kerr-cat qubit can be very robust in the presence of a carefully engineered electromagnetic environment. Beyond the Kerr-cat qubit, the tools developed here apply broadly to strongly driven dissipative systems with quasidegenerate spectra, including superconducting devices under subharmonic driving (e.g., parametric amplifiers and couplers) and protected qubits where quasidegeneracies similarly govern coherence.
\end{abstract}

\maketitle

\section{Introduction} \label{sec:intro}

Bosonic codes are promising candidates for achieving fault tolerance with minimal hardware overhead, a feature often referred to as hardware efficiency \cite{Chuang1997, Cochran1999, Gottesman2001, Niset2008, Leghtas2013, Michael2016,Lemonde2024}. These codes leverage the large Hilbert space of an oscillator to encode logical information in a carefully chosen $d$-dimensional
subspace. A particularly compelling example is the cat qubit, where the logical states are encoded in the manifold spanned by superposition of coherent states $\ket\alpha$ referred to as Schrödinger cat states, $\ket{C_{\alpha}^{\pm}} = (\ket{\alpha} \pm \ket{-\alpha})/N_\alpha^{\pm}$ with $N_\alpha^{\pm}$ the normalization \cite{lecturesnotes_dissipativecats}. This choice is motivated by the fact that local noise---acting locally in the oscillator's phase space---induces transitions between $\ket{+\alpha}$ and $\ket{-\alpha}$ with rates that are exponentially suppressed in $|\alpha|^2$~\cite{Mirrahimi:2014, Leghtas:2015, Puri:2017}.
This suppression of phase-flip errors is only modestly counterbalanced by a linear in $|\alpha|^2$ increase in the rate of bit-flip errors---that is, transitions between the cat states $\ket{C_\alpha^+}$ and $\ket{C_\alpha^-}$---induced by local noise. The resulting large noise bias can be leveraged to implement error correction protocols with minimal hardware resources~\cite{Panos2008,Aliferis2009, Tucket2018,Tuckett2019,Guillaud2019,Darmawan2021,Bonilla2021,Guillaud2021,Chamberland2022}.

In superconducting quantum circuits, the two main types of cat qubits are the dissipative cat qubit~\cite{Leghtas:2015, Touzard2018, Lescanne2020, Berdou2023, Reglade2024, Marquet2024, Putterman2025, Putterman2025Hardware} and the Kerr-cat qubit \cite{Goto2016,Puri:2017,Grimm:2020, Frattini2024, Venkatraman_deltacat,Hajr2024,ding2025, dealbornoz2024}.
In the dissipative approach, engineered dissipation stabilizes a steady-state manifold that spans the desired cat-states subspace. By contrast, the Kerr-cat approach relies on Hamiltonian engineering where the cat states emerge as degenerate eigenstates of the squeezed Kerr Hamiltonian
\begin{align} \label{eq:H squeezed Kerr}
\begin{split}
\hat{H}_{\textrm{SK}} &= -K\ha^{\dag^2} \ha^2 + \varepsilon_2 \ha^{\dag^2} + \varepsilon_2^* \ha^2, \\
&= -K (\hat{a}^{\dagger 2} - \alpha^{*2})(\hat{a}^2 - \alpha^2) + K|\alpha|^4,
\end{split}
\end{align}
where $\hat{a}$ is the annihilation operator of the oscillator, $\varepsilon_2$ is the strength of the two-photon drive, and $K$ is the self-Kerr nonlinearity. The factorized form of $\hat H_\textrm{SK}$ makes it explicit that the coherent states $\ket{\pm \alpha}$ with $\alpha = \sqrt{\varepsilon_2/K}$ are degenerate eigenstates of the Hamiltonian~\cite{Puri:2017}. Their symmetric and antisymmetric superpositions are the cat states $\ket{C_{\alpha}^{\pm}}$, whose average photon number is given by $|\alpha|^2 = |\varepsilon_2|/K$.

\begin{figure}
    \centering
    \includegraphics{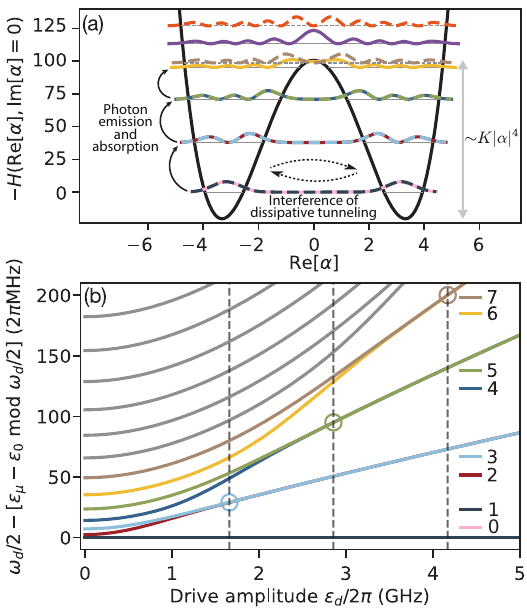}
    \caption{
    (a) Cut along $\text{Re}[\alpha]$ of the metapotential associated with the shifted squeezed Kerr Hamiltonian $\hat H_\text{SK} - \varepsilon_2^2/K$ for $\varepsilon_2 = 10$ and $K = 1$. Horizontal gray lines indicate the eigenenergies, and the colored curves the marginals of the Wigner distributions of the corresponding eigenstates. For the quasidegenerate level pairs 0--1, 2--3, and 4--5, dissipative tunneling across the double well is suppressed due to destructive interference between two relaxation pathways. 
    Population transfer between wells can instead occur by climbing the energy ladder—via photon emission and absorption—until nondegenerate levels (here 6 and 7) are populated. The number of quasidegenerate level pairs within the well is controlled by the photon number $|\alpha|^2 = \varepsilon_2/K = 10$. 
     (b)  Floquet quasienergy spectrum of the  Kerr-cat qubit showing spectral kissings (circles) where the quasienergy differences between pairs of excited states asymptotically approach $\omega_d/2$. 
     Dashed lines indicate drive amplitudes at which these differences become comparable to the decay rate. 
    The circuit parameters are chosen close to those of Ref.~\cite{Frattini2024}, with $E_J/2\pi = 272.436$~GHz, $E_C/2\pi = 107.8$~MHz, $\alpha = 0.046$, and $\varphi_\text{ext} = 0.33 \times 2\pi$, yielding a 0--1 transition frequency $\omega_{01}/2\pi = 6.094$ GHz and a self-Kerr nonlinearity $K/2\pi = 1.18$~MHz.
    }
    \label{fig:metapotential}
\end{figure}

To gain intuition into the structure of this Hamiltonian, it is helpful to analyze its phase-space representation $H(\text{Re}[\alpha], \text{Im}[\alpha])$, obtained via the Weyl correspondence~\cite{Polkovnikov2010}. This metapotential, shown in \cref{fig:metapotential}(a), exhibits a double-well structure with a barrier height $\sim K |\alpha|^4$ separating two minima centered at $\pm \alpha$, corresponding to the locations of the Kerr-cat coherent states.
As the two-photon drive and hence $|\alpha|^2$ is increased, pairs of states successively enter the wells. As they drop deeper into the potential landscape, tunneling between them is rapidly suppressed because of the barrier. This leads to spectral kissing, where pairs of energy levels become nearly degenerate; see \cref{fig:metapotential}(b), with circles indicating the kissing of different pairs \cite{Frattini2024}.

In practice, however, the exponential suppression of tunneling between the coherent states at the bottom of the well does not persist indefinitely with increasing $|\alpha|^2$. Thermal excitation from these coherent states to higher-lying states that have not yet kissed can restore tunneling between the wells~\cite{Putterman2022,Ruiz2023, su2024}. This phenomenon repeats itself as each new pair of states enters the wells, resulting in a staircase-like dependence of the tunneling time with $|\alpha|^2$. This behavior has been measured experimentally~\cite{Frattini2024, Hajr2024} and studied theoretically using perturbative methods~\cite{Venkatraman:2024}. More strikingly, at still larger photon numbers, the staircase behavior breaks down: the tunneling time is observed to decrease with increasing photon number, accompanied by sharp features at specific values of the drive amplitude~\cite{dealbornoz2024, Brock_private_communication}. While existing theoretical models capture the qualitative features of the staircase regime, they do not quantitatively match the experimental data. Moreover, the breakdown of the staircase pattern is not reproduced by current theoretical approaches, raising the question of whether this discrepancy reflects limitations of the perturbative treatment or the presence of additional mechanisms not yet accounted for in the theory.

In this work, we answer this question by showing that both the breakdown of the staircase behavior of the tunneling time and the appearance of sharp resonances originate from the multimode nature of the circuit implementation of the Kerr-cat qubit in Refs.~\cite{Frattini2024,dealbornoz2024} and can be captured by a model foregoing any perturbative expansion but including additional modes of the circuit. In the absence of additional modes, we find that the Kerr-cat qubit is remarkably robust. This observation suggests that engineering the Kerr-cat's electromagnetic environment such as to avoid the presence of these modes can lead to significantly longer tunneling times than currently observed experimentally.  

These results are obtained using three key ingredients. First, we retain the full nonlinear potential of the circuit, as any finite-order expansion of the potential risks missing the resonances between the nonlinear oscillator and the other degrees of freedom in the circuit. Second, the drive is treated exactly using the Floquet formalism, ensuring that multiphoton processes and nonlinear effects are fully accounted for. This type of multiphoton process has already been shown to be at the origin of ionization of highly-excited atoms driven by a microwave field~\cite{Breuer1989_Floquet_Ionization} and of drive-induced resonances of the transmon qubit~\cite{MIST_1, MIST_2,Shillito2022Dynamics,Cohen2023,Dumas2024,Fechant2025,Wang2025, Mingkang2025, connolly:2025, dai:2025}. Third, we incorporate the interplay between the drive and dissipation by employing the partial secular Floquet–Markov master equation~\cite{Oelschlagel1993,Kohler:1997,Grifoni:1998}. This approach captures the coherent interplay 
between different relaxation channels---an essential feature to accurately model dissipation in the Kerr-cat qubit.  

This paper is organized as follows. In \cref{sec:single mode} we review the standard single-mode approximation of the Kerr-cat qubit. We then show how Floquet theory reproduces essential features of this system, such as spectral kissing, and introduce a form of the Floquet–Markov Lindblad master equation that accounts for these near-degeneracies. Using this framework, we compute the tunneling time of the Kerr-cat qubit, demonstrating its intrinsic robustness. In \cref{sec: boxmode}, we incorporate the buffer mode used to drive the Kerr-cat qubit in experiments, and show that it leads to an abrupt reduction of the tunneling time above a critical drive amplitude. In \cref{sec: array modes,appendix: inductance} we examine the effects of SNAIL array modes and a stray geometric inductance on the qubit's coherence. Further details of the derivations can be found in the Appendices. Our findings are summarized in \cref{sec: conclusion}.

\section{Single-mode description} \label{sec:single mode}

\begin{figure}
    \centering
    \includegraphics[width=\linewidth]{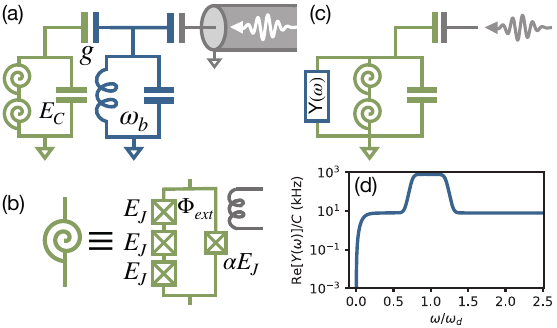}
    \caption{(a) Circuit implementation of the Kerr-cat qubit, consisting of an array of SNAILs shunted by a large capacitor (green) and driven through a buffer mode (blue) whose resonance frequency is close to the drive frequency. 
    (b) Circuit diagram of a single SNAIL composed of an array of large Josephson junctions with energy $E_J$ in parallel with a smaller junction of energy $\alpha E_J$, forming a loop threaded by an external magnetic flux (gray).    
    (c) Semiclassical representation of the circuit in (a), the driven buffer mode is replaced by a classical drive (gray) and a dissipative environment modeled by an admittance $\mathrm{Y}(\omega)$ (blue). 
    (d) Typical spectral density $J(\omega) \propto \textrm{Re}[\mathrm{Y}(\omega)]/C$ featuring a broad peak around the drive frequency $\omega_d$ reflecting the low quality factor of the buffer mode. The values around $J(m\omega_d/2)$ for $m\in \mathbb{Z}^+$ are taken from Ref.~\cite{Venkatraman:2024}. 
    }
    \label{fig:circuits}
\end{figure}

Experimental realizations of the Kerr-cat qubit in circuit QED rely on superconducting nonlinear asymmetric inductive elements (SNAILs) \cite{Frattini2017,Sivak:2019} shunted by a large capacitance and controlled via a buffer mode, see \cref{fig:circuits}(a) \cite{Grimm:2020,Frattini2024,Venkatraman_deltacat,Hajr2024,ding2025,dealbornoz2024}. In this section, we analyze the behavior of the Kerr-cat qubit in the single-mode approximation of this circuit, identifying the key assumptions underlying this simplification. While sufficient to reproduce the physics of the spectral kissing and the observed plateaus in the tunneling time \cite{Frattini2024,Hajr2024, Mata2024}, we find that this reduced model does not capture the observed abrupt reduction of the tunneling time above a drive amplitude threshold. Building on these observations, in the next section, we show how this breakdown is captured when accounting for the multimode nature of the circuit of \cref{fig:circuits}(a). The present section introduces the key ideas and methods necessary to understand this more general multimode case. 

Accounting for the buffer mode, the Hamiltonian describing the circuit of \cref{fig:circuits}(a) reads ($\hbar =1$)
 \begin{align}\label{eq:full_H}
     \begin{split}
     \hH'' &= \omega_b \ha_b^\dagger\ha_b + \hH_s + ig\hn \left(\ha_b^\dagger -\ha_b\right) +i\Omega (t)\left(\ha_b^\dagger -\ha_b\right).
     \end{split}
 \end{align} 
In this expression, $\omega_b$ is the frequency of the buffer mode with annihilation operator $\ha_b$. The implementation of the buffer mode varies depending on the experiment: it may correspond to a 3D box mode defined by the cavity geometry \cite{Grimm:2020, Frattini2024,Venkatraman_deltacat,dealbornoz2024}, or a stripline resonator that facilitates parametric driving of the SNAILs \cite{ding2025}. 
In all cases, the presence of this mode is intentional, with its resonance frequency engineered to be close to that of the drive frequency to allow driving the SNAIL. The second term of \cref{eq:full_H}, $\hH_s$, is the SNAIL array Hamiltonian defined below. The third term corresponds to the coupling of the buffer mode to the SNAIL array, with $g$ the strength of that coupling and $\hn$ the SNAIL array charge operator. The last term corresponds to a charge drive $\Omega(t) = \Omega_0 \sin(\omega_d t)$ applied to the buffer mode at frequency $\omega_d$. 
To see more clearly how the drive on the buffer effectively acts on the SNAIL array, we displace the buffer mode by its classical response using a displacement transformation \cite{Mollow1975}. This results in the displaced Hamiltonian
\begin{align} \label{Hamiltonian_buffer}
    \hH' = \omega_b \ha_b^\dagger\ha_b + \hH_s  + ig\hn \left(\ha_b^\dagger -\ha_b\right) +\varepsilon_d \cos(\omega_d t) \hn,
\end{align}
where $\varepsilon_d$ is the amplitude of the effective drive proportional to the square root of the buffer mode photon population. So far, no approximations have been made; the displacement transformation is exact, as seen in Ref.~\cite{Lledo_cloaking}, for instance. In the following, we use $\varepsilon_d$ rather than $\Omega_0$ as a measure of the drive strength.
As a first approximation, following Refs.~\cite{Cohen2023,Dumas2024}, we ignore quantum fluctuations in the buffer mode resulting in the simplified Hamiltonian 
\begin{align} \label{singlemode_Hamiltonian}
    \hH \simeq \hH_s + \varepsilon_d \cos(\omega_d t) \hn,
\end{align}
which describes a voltage-driven SNAIL, as illustrated in \cref{fig:circuits}(c).
We go beyond this approximation in \cref{sec: boxmode} by accounting for the buffer mode exactly.

As illustrated in \cref{fig:circuits}(a,b), the circuit is built from an array of two SNAILs, each comprised of an array of three identical large Josephson junctions with energy $E_J$ in parallel with a smaller junction of energy $\alpha E_J$ (with $\alpha < 1$). Assuming that the total phase drop across the array is evenly distributed among the six large junctions allows the SNAIL array to be described by a single degree of freedom. This single-phase approximation relies on the assumption that this forms a weakly nonlinear oscillator whose resonance frequency $\omega_0$ is much smaller than the plasma frequency of the individual large junctions.
 This condition is satisfied provided the stray capacitance $C_{J,g}$ to ground of each island in the array is small ($C_{J,g} \ll C_J/6^2$)~\cite{Masluk:2012, Ferguson:2013, Viola2015}. We go beyond this approximation in \cref{sec: array modes} where we analyze the influence of higher-frequency collective modes of the array on both the spectrum and lifetime of the Kerr-cat qubit.

Under the single-phase approximation, the Hamiltonian of the capacitively shunted double SNAIL is given by
\begin{align} \label{eq:doublesnail_singlemode}
\begin{split}
    \hH_s =& 4 E_C\hn^2-6E_J\cos\left(\frac{\hvarphi}{6}\right)\\
    &-2\alpha E_J\cos\left(\frac{\hvarphi}{2}+\varphi_{\textrm{ext}}\right),\\  
    \end{split}
\end{align}
where $E_C$ denotes the charging energy dominated by the shunting capacitance, and $\varphi_{\textrm{ext}} = 2\pi \Phi_{\textrm{ext}}/\Phi_0$ the external reduced flux bias of each SNAIL. The operators represent the phase $\hat{\varphi}$ across the SNAIL array and its canonically conjugate charge $\hat{n}$.

Introducing the annihilation operator $\ha$ of the shunted SNAIL such that $\hvarphi = \varphi_\text{min} + \varphi_\mathrm{zpf} (\ha+\ha^\dag)$, where $\varphi_\text{min}$ denotes the phase at the minimum of the potential, and $\varphi_\mathrm{zpf}$ represents the amplitude of quantum fluctuations of the phase operator, we expand the potential energy terms of \cref{eq:doublesnail_singlemode} to fourth order in $\varphi_{\mathrm{zpf}}$. Choosing the drive frequency $\omega_d \approx 2\omega_0$, and moving to the appropriate frame and applying the rotating-wave approximation (RWA), the single-mode Hamiltonian of \cref{singlemode_Hamiltonian} takes the standard form of the Kerr-cat Hamiltonian \cref{eq:H squeezed Kerr} \cite{Grimm:2020}. 
This approximate model is sufficient to capture the basic properties of the system \cite{Puri:2017}. However, in the context of measurement-induced state transitions in the transmon \cite{MIST_1,Lescanne:2019,Verney2019,Shillito2022Dynamics,Dumas2024}, making these approximations (cosine potential expansion and RWA) substantially modifies the threshold drive amplitude before transitions are observed. We expect these approximations to similarly modify the threshold of the breakdown of the staircase behavior of the Kerr-cat. For this reason, in this work we use the full cosine potential of \cref{eq:doublesnail_singlemode}.  

To accurately account for the effect of the strong drive, we use Floquet theory. At any time, the state of the system can be decomposed in terms of periodic Floquet modes $\ket{\phi_\mu(t)}=\ket{\phi_\mu(t+T)}$ and their corresponding quasienergies $\epsilon_\mu$, which satisfy the eigenvalue equation
\begin{equation}
\hat{U}(t+T,t)\ket{\phi_\mu(t)} = e^{-i\epsilon_\mu T}\ket{\phi_\mu(t)},
\end{equation}
where $\hat U(t+T,t)$ is the unitary propagator, generated by the Hamiltonian in \cref{singlemode_Hamiltonian}, over one period $T=2\pi/\omega_d$ of the drive~\cite{Grifoni:1998}. The Floquet modes and quasienergies are obtained by numerically diagonalizing this operator. Since they are obtained as the argument of a phase, the quasienergies are defined modulo the drive frequency, $\epsilon_\mu \in [-\omega_d/2, \omega_d/2)$.
 Hereinafter, we denote $\ket{\phi_\mu} \equiv \ket{\phi_\mu(0)}$ the Floquet modes at time $t=0$. 
At $\varepsilon_d=0$, the quasienergies coincide with the eigenenergies (modulo $\omega_d$) of the undriven double-SNAIL Hamiltonian $\hat H_s$, and the Floquet modes reduce to the corresponding undriven eigenstates.
Starting from this point, we increment $\varepsilon_d$ in small steps and assign labels to the Floquet spectrum at each step by maximizing their modes' overlap with the modes at the previous drive amplitude. This procedure enables a consistent tracking of each mode based on its origin in the undriven spectrum~\cite{Breuer1989QuantumPhases,Breuer1989_Floquet_Ionization,Dumas2024}.

\begin{figure}
    \centering
    \includegraphics[width=\linewidth]{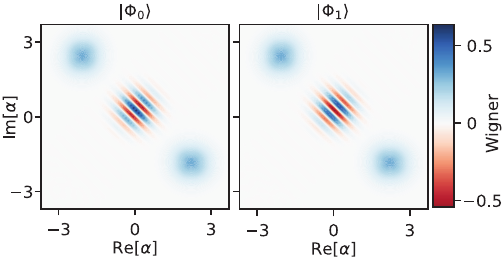}
    \caption{ Wigner distributions of the first two Floquet modes, $\ket{\phi_0}$ and $\ket{\phi_1}$, at a drive amplitude $\varepsilon_d/2\pi = 2.5~\text{GHz}$ corresponding to approximately nine photons in the cat manifold. These two modes are the logical cat states.
     }
    \label{fig:cats_spectralkissing}
\end{figure}

Increasing the drive amplitude, the two lowest Floquet modes, $\mu=0,1$, continuously evolve from the ground and first excited states of the undriven Hamiltonian $\hH_s$ into (displaced) cat states, see \cref{fig:cats_spectralkissing} showing the Wigner distributions of these modes for $\varepsilon_d/2\pi=\SI{2.5}{GHz}$. Unlike in the approximate squeezed-Kerr Hamiltonian \cref{eq:H squeezed Kerr}, where the cat eigenstates are centered at the origin, here they are slightly displaced because they are displayed in the laboratory frame and not the displaced frame.
Importantly, in these calculations, to compensate for the ac-Stark shift the drive frequency $\omega_d$ is adjusted at each value of the drive amplitude to match twice the Floquet quasienergy difference between the two lowest modes, $\omega_d = 2 |\epsilon_1 - \epsilon_0|$. This mirrors the experimental procedures used, for example, in Refs.~\cite{Frattini2024,dealbornoz2024,ding2025}. Away from this choice of frequency, the Floquet modes $\mu=0$ and 1 associated with the Hamiltonian in \cref{singlemode_Hamiltonian} no longer correspond to symmetric superpositions of opposite coherent states in phase space~\cite{Venkatraman_deltacat, dealbornoz2024}.

\subsection{Spectral kissing in the presence of strong drive and full nonlinearities}

Having obtained the Floquet spectrum as a function of the drive amplitude, it is useful to present it in a way that facilitates comparison with the spectroscopy measurements reported in Refs.~\cite{Frattini2024, Venkatraman_deltacat, Hajr2024,ding2025}, see \cref{appendix: single mode details} for more details. To do so, we first unfold the quasienergies from the Floquet Brillouin zone $[-\omega_d/2, \omega_d/2)$ onto the real line, and then re-fold them modulo $\omega_d/2$ relative to the ground state energy. This results in the spectrum shown in \cref{fig:metapotential}(b) with a series of pairwise degeneracies, or spectral kissings (circles). This spectral kissing is a crucial feature of the ideal squeezed Kerr oscillator Hamiltonian of \cref{eq:H squeezed Kerr} which we also observe---consistent with experiments---when using the full nonlinear model of \cref{eq:doublesnail_singlemode}. 

Remarkably, the quasienergy spectrum of the Kerr-cat remains free of avoided crossings indicative of unwanted resonances up to $\varepsilon_d/2\pi = \SI{10}{GHz}$, corresponding to 36 photons—the maximum amplitude considered in our simulations (not shown).
This is in stark contrast to other qubits under strong drives, such as the transmon, where the quasienergy spectra show a multitude of avoided crossings resulting from unwanted multiphoton resonances~\cite{MIST_1, MIST_2,Shillito2022Dynamics,Cohen2023,Dumas2024,Fechant2025,Wang2025, Mingkang2025}.
Crucially, in the transmon the quasienergies corresponding to the computational states experience an ac-Stark shift of opposite sign to that of states near the top of the transmon's cosine potential (where the anharmonicity changes sign).  
This can lead to a collision of quasienergies at a `critical' drive amplitude and to unwanted transitions from the computational states to highly-excited states. Because the transmon's cosine potential is relatively shallow and only supports of the order of $\sim 10$ states, these collisions typically occur at moderate drive amplitudes.
The situation is very different in the Kerr-cat qubit, whose robustness to multiphoton resonances results from its very deep potential well. Indeed, for the parameters used in this work, which are close to those of the experiment of Ref.~\cite{Frattini2024} (see the caption of \cref{fig:metapotential}(b)), the number of bound states in the SNAIL well is approximately 530. As a result, all the relevant levels---including the cat-states and higher-excited levels---experience an ac-Stark shift of the same sign. Collisions of quasienergies are thus avoided up to very large drive amplitudes. This observation suggests that the single-mode approximation of the Kerr-cat qubit does not contain the necessary ingredients to capture the resonances associated with the sudden decrease of the tunneling time with drive amplitude. The same conclusion can be reached by considering this system from the point of classical chaos~\cite{Cohen2023,Chavez-Carlos2025}. 


\subsection{Floquet-Markov-Lindblad with quasidegerate transitions}

To fully account for the effect of the drive and SNAIL nonlinearity, we employ the Floquet-Markov master equation~\cite{Blumel:1991,Kohler:1997,Grifoni:1998}. In contrast to the standard quantum optics master equation---where dissipation is treated as incoherent transitions between eigenstates of a time-independent Hamiltonian~\cite{BRE:2002}---the Floquet-Markov approach describes dissipation in terms of transitions between Floquet modes. Crucially, the corresponding rates account for processes involving simultaneous absorption or emission of drive photons and environment photons~\cite{Grifoni:1998,Mori2023}.

In the context of the Kerr cat, it is essential for the master equation to capture the spectral kissing of the quasienergies. As a result, in the derivation of the master equation we cannot employ the commonly used secular approximation which assumes that all transition frequencies are well separated, leading to a sum of independent dissipators, one for each transition~\cite{Grifoni:1998}. Instead, we use a partial secular approximation accounting for the near degeneracy of certain transitions while discarding fast-oscillating off-diagonal terms between widely separated transition frequencies~\cite{Oelschlagel1993,Kohler:1997}. This approach allows for an accurate description of the dynamics near spectral degeneracies, which are central to the protection of the Kerr-cat coherence.

More precisely, the Floquet-Markov master equation we use is given in the interaction picture by
\begin{equation} \label{Lindbladian}
\partial_t \hat \rho = \mathcal D_\text{inc}\hat \rho + \mathcal D_\text{coh}\hat \rho \equiv \mathcal L\hat \rho,
\end{equation}
with two different collections of dissipators. The first collection
\begin{equation}
\mathcal D_\text{inc} = \sum_{\mu, \nu, k} \kappa(\Delta_{\mu \nu k}) \mathcal D\left[X_{\mu \nu k} \ket{\phi_\mu}\bra{\phi_\nu}\right],
\end{equation}
with $\mathcal D[\hat L]\hat \rho \equiv \hat L \hat \rho \hat L^\dag - \{\hat L^\dag \hat L, \hat \rho\}/2$, takes the form of an incoherent sum of all processes involving transitions that are not nearly degenerate. In this expression, 
\begin{equation}
\Delta_{\mu \nu k} = \epsilon_\mu - \epsilon_\nu + k\omega_d
\end{equation}
is the detuning between Floquet modes $\mu$
 and $\nu$ up to $k\in \mathbb{Z}$ drive photons, and 
\begin{equation} \label{eq:matrix_elements}
X_{\mu\nu k} = \frac{1}{T}\int\limits_0^T e^{-ik \omega_dt} \bra{\phi_\mu(t)}i(\hat a^\dag - \hat a)\ket{\phi_\nu(t)} dt
\end{equation}
is a Fourier coefficient of the $(\mu,\nu)$ matrix element of the SNAIL's charge operator $\hat n/n_{\text{zpf}} = i(\hat a^\dag - \hat a)$ in the Floquet mode basis at time $t$. In \cref{eq:matrix_elements}, the zero-point fluctuations of the charge operator, $n_{\text{zpf}}$, are implicitly included in the definition of $\kappa(\omega)$. The index $k$ accounts for the fact that transitions can involve the emission or absorption of drive photons. The second collection takes the form
\begin{equation}\label{eq:D_coh}
\begin{split}
\mathcal D_\text{coh} =
\sum_{\mu,\nu,k} \kappa(\Delta_{\mu \nu k}) \, 
\mathcal{D}\biggl[
\sum_{\substack{\mu',\nu',k'\\ \sim (\mu,\nu,k)}} 
X_{\mu'\nu'k'} 
\ket{\phi_{\mu'}}\bra{\phi_{\nu'}}
\biggr]
\end{split}
\end{equation}
and is the sum of all processes involving transitions that are quasidegenerate, which are coherently added together inside the dissipators. The prime indices $(\mu',\nu',k')$ run over triplets for which the associated transition frequency $\Delta_{\mu' \nu' k'}$ is quasidegenerate with $\Delta_{\mu \nu k}$. For the parameters used in this work, we define two transitions $\Delta_{\mu \nu k}$ and $\Delta_{\mu' \nu' k'}$ as quasidegenerate when $|\Delta_{\mu \nu k}-\Delta_{\mu' \nu' k'}|<2\pi \times 100$~kHz, which is about ten times the qubit decay rate adopted below.

\begin{figure}
    \centering
\includegraphics[width=\linewidth]{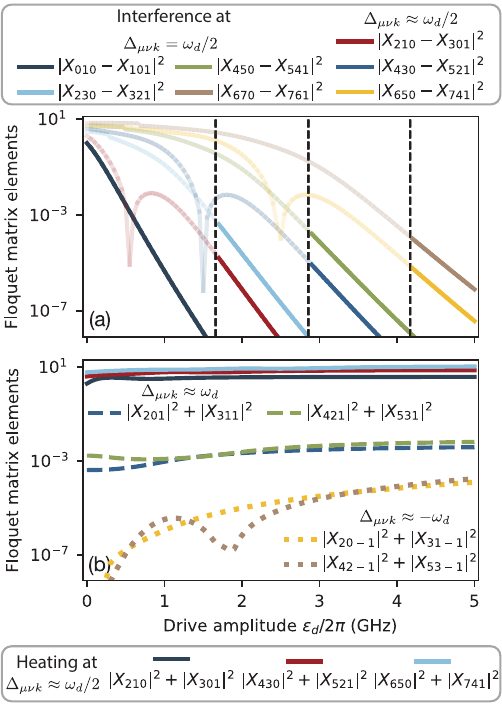}
    \caption{Matrix elements of the SNAIL charge operator in the Floquet mode basis as a function of drive amplitude, responsible for (a) direct tunneling and (b) intrawell leakage. In (b), we highlight three types of heating processes: solid lines indicate photon absorption at $\omega_d/2$, dashed lines at $\omega_d$, and dotted lines denote drive-induced processes that persist at zero temperature. 
    Before the kissing between $\epsilon_2$ and $\epsilon_3$ (left-most vertical dashed line), tunneling is dominated by the intermediate transitions to these modes outside of the metapotential double well, with a rate proportional to $|X_{210}|^2 + |X_{301}|^2$, with additional contributions from other transitions whose matrix elements are shown in (b). After the kissing, these dominant processes interfere, and their contribution $|X_{210} - X_{301}|^2$ to the tunneling rate becomes suppressed. Furthermore, leakage to these modes does not lead to direct tunneling because the rate $|X_{230} - X_{321}|^2$ also becomes suppressed.
    This pattern repeats for higher modes at each of the other two spectral kissings indicated by the vertical dashed lines.
    }
\label{fig:Floquet_matrix_elements}
\end{figure}

In both collections, the rates are given by
\begin{equation} \label{kappa}
\kappa(\omega) = n_\text{th}(\omega) J(\omega) + \left[1+n_\text{th}(-\omega)\right] J(-\omega),
\end{equation}
with $n_\text{th}(\omega)$ the thermal population of the environment at frequency $\omega$, and $J(\omega) = \theta(\omega)(\omega_0/\omega)\times \operatorname{Re}[\textrm{Y}(\omega)]/C$ the bath spectral density expressed in terms of the external admittance $\textrm{Y}(\omega)$ \cite{Esteve:1986, Vool_Devoret, Labarca2024}, see \cref{fig:circuits}(c) and (d). With $\theta(\omega)$ the Heaviside function, $\kappa(\omega>0)$ corresponds to absorption from the bath while $\kappa(\omega<0)$ to emission. The derivation of this master equation can be found in \cref{appendix: master equation}.

For the Kerr-cat qubit, transitions between the relevant Floquet modes cluster around harmonics of $\omega_d/2$ or, in other words, $\Delta_{\mu\nu k} \sim m \omega_d/2$ with $m \in \mathbb{Z}$; see \cref{fig:metapotential}(b) and \cref{fig:fullspectrum_photonnumber}(b) in \cref{appendix: single mode details}.  We assume that the spectral density $J(\omega)$ is flat in the vicinity of each of these harmonics, consistent with the Markovian approximation \cite{Rivas:2010}. Accordingly, we parametrize the spectral density by its values at the relevant harmonics. Since the buffer mode acts as a bandpass filter at $\omega_d$, we take $J(\omega_d) \gg J(\omega_d/2) = J(3\omega_d/2)$, as illustrated in \cref{fig:circuits}(c). 
Moreover, following Ref.~\cite{Venkatraman:2024} we account for imperfect thermalization of the buffer mode resulting in a higher frequency at the drive frequency than at the other harmonics, $T(\omega_d) > T(\omega_d/2)=T(3\omega_d/2)$.
In \cref{singlemode: Temperature and spectral densities}, we explore in more detail how the spectral density and the thermal population influence the tunneling rate in the Kerr-cat qubit.

\subsection{Examples of processes entering $\mathcal D_\text{coh}$ and $\mathcal D_\text{inc}$}
\label{sec: examples of processes entering Dcoh and Dinc}

Before presenting results obtained using the above Floquet-Markov master equation, it is instructive to first consider a few examples of the kinds of processes that enter the dissipators $\mathcal D_\text{coh}$ and $\mathcal D_\text{inc}$. 
First, given the choice of drive frequency $\omega_d/2 = |\epsilon_1-\epsilon_0|$, there are exact degeneracies $\Delta_{010}=\Delta_{101}=\omega_d/2$ and $\Delta_{100}=\Delta_{01,-1}=-\omega_d/2$ for all drive amplitudes. The corresponding transitions join the same dissipators in $\mathcal D_\text{coh}$. For example, focusing on the transitions of frequency $\omega_d/2$, the sum of operators entering $\mathcal{D}$ in \cref{eq:D_coh} is $X_{010}\ket{\phi_0}\bra{\phi_1}+ X_{101}\ket{\phi_1}\bra{\phi_0}$. Because the cat states of amplitude $\pm\alpha$ are approximately given by $\ket{\phi_{0,1}} \approx (\ket{\beta+\alpha}\pm e^{i\theta}\ket{\beta-\alpha})/\sqrt{2}$ where $\beta$ is the displacement from the origin, this expression can be written as
\begin{equation} \label{phi0-phi1}
\begin{split}  &X_{010}\ket{\phi_0}\bra{\phi_1}+ X_{101}\ket{\phi_1}\bra{\phi_0} = \\&\frac{X_{010}+X_{101}}{2}\left(\ket{\beta+\alpha}\bra{\beta+\alpha}-\ket{\beta-\alpha}\bra{\beta-\alpha}\right)\\
&+\frac{X_{010}-X_{101}}{2}\ket{\beta+\alpha}\bra{\beta-\alpha} \\&-\frac{X_{010}-X_{101}}{2}\ket{\beta-\alpha}\bra{\beta+\alpha} . 
\end{split}
\end{equation}
The first term on the right-hand side induces bit flips (logical Pauli $\hat{X}_L$) at a rate proportional to $|X_{010}+X_{101}|^2$ within the cat qubit manifold. The second and third terms induce transitions between the coherent states corresponding to phase flips (logical Pauli $\hat{Z}_L$) at a rate proportional to $|X_{010}-X_{101}|^2$. In other words, these last two terms lead to tunneling between the two wells of the metapotential. As expected from the discussion in \cref{sec:intro} and shown in \cref{fig:Floquet_matrix_elements}~(a), this rate decreases exponentially with increasing drive amplitude (dark blue line) \cite{Puri:2017,Puri2019}.

\begin{figure}[t!]
    \centering
\includegraphics[width=\linewidth]{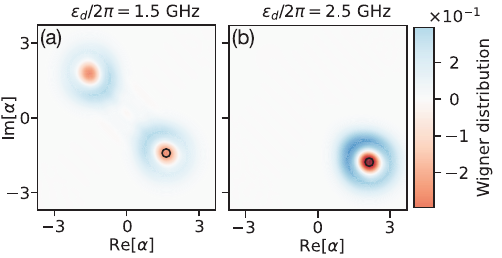}
    \caption{
    Wigner distributions illustrating how the action of the dissipator affects the initially localized coherent state $\ket{\beta+\alpha}$ (indicated by the black circle).
    (a) At $\varepsilon_d/2\pi = \SI{1.5}{GHz}$, before the spectral kissing, the jump operators $\hat L_1 = X_{210}\ket{\phi_2}\bra{\phi_1}$ and $\hat L_2 = X_{301}\ket{\phi_3}\bra{\phi_0}$ act incoherently. The resulting state is $p_1 \hat \rho_1 + p_2 \hat \rho_2$, where $\hat \rho_{1,2} = \hat L_{1,2} \ket{\beta+\alpha}\bra{\beta+\alpha} \hat L_{1,2}^\dag /p_{1,2}$ and $p_{1,2} = \bra{\beta+\alpha} \hat L_{1,2}^\dag \hat L_{1,2} \ket{\beta+\alpha}$. The state becomes delocalized across the double-well.
    (b) At $\varepsilon_d/2\pi = \SI{2.5}{GHz}$, after the kissing, the jump operators interfere coherently. The resulting state is $\hat L \ket{\beta+\alpha} / \sqrt{\bra{\beta+\alpha} \hat L^\dag \hat L \ket{\beta+\alpha}}$ with $\hat L = \hat L_1 + \hat L_2$, preserving localization in the original well.}
\label{fig:action_cops}
\end{figure}

This noise bias is significantly reduced when higher levels are considered~\cite{Putterman2022, Ruiz2023, su2024}, something that is captured in the Floquet-Markov master equation of \cref{Lindbladian} by the incoherent or coherent addition of transitions before and after level kissings. 
For example, before kissing, the two heating processes from the states at the bottom of the well, $\{\ket{\phi_0},\ket{\phi_1}\}$, to the next pair of states, $\{\ket{\phi_2},\ket{\phi_3}\}$,  are accounted for in $\mathcal D_\text{inc}$ by 
an incoherent sum of dissipators, $\kappa(\Delta_{210})\mathcal{D}[X_{210}\ket{\phi_2}\bra{\phi_1}] + \kappa(\Delta_{301})\mathcal{D}[X_{301}\ket{\phi_3}\bra{\phi_0}]$. As discussed above, upward transition to states that have not yet kissed leads to tunneling across the double well potential happening here at a rate $\propto|X_{210}|^2+|X_{301}|^2$ given by the dark blue line in \cref{fig:Floquet_matrix_elements}(b). This effect is illustrated in \cref{fig:action_cops}(a) which shows 
how the coherent state $\ket{\beta +\alpha}$, initially localized in one well of the metapotential, transforms under the action of this incoherent sum of dissipators, resulting in a mixture of states localized in the two wells.

After spectral kissing of $\epsilon_2$ and $\epsilon_3$, the above two transitions coherently add contributing to $\mathcal D_\text{coh}$ as $\kappa(\Delta_{210})\mathcal{D}[X_{210}\ket{\phi_2}\bra{\phi_1}+ X_{301}\ket{\phi_3}\bra{\phi_0}]$. As illustrated in \cref{fig:action_cops}(b), the action of this joint jump operator 
on $\ket{\beta+\alpha}$ no longer leads to delocalization across the metapotential. To understand this we express the Floquet modes $\ket{\phi_{2,3}}$ as superpositions of states localized in each well
\begin{equation}
\ket{\phi_{2,3}} \approx (\hat{D}(\beta+\alpha)\ket{1}\mp e^{i\theta'}\hat{D}(\beta-\alpha)\ket{1})/\sqrt{2},    
\end{equation}
where $\hat{D}$ is the displacement operator and $\ket{1}$ is the one-photon Fock state~\cite{Chamberland2022}. Using this expression to perform a similar decomposition as in \cref{phi0-phi1}, we find that the above dissipator generates two types of processes. First, it contributes to intrawell leakage $\ket{\beta\pm\alpha} \to \hat{D}(\beta\pm\alpha)\ket{1}$ at a rate  $\propto |X_{210} + X_{301}|^2 \approx 2(|X_{210}|^2 + |X_{301}|^2) $. Second, it contributes to interwell tunneling $\ket{\beta\pm\alpha} \to \hat{D}(\beta\mp\alpha)\ket{1}$ at a rate $\propto |X_{210} - X_{301}|^2$. Crucially, this interwell transition rate is exponentially suppressed as the drive amplitude increases beyond the kissing point, as shown by the red line in \cref{fig:Floquet_matrix_elements}(a), and further supported by analytical results presented in \cref{appendix: SW on Shirley}.
The intrawell leakage rate, though, is not suppressed; see the dark blue line in \cref{fig:Floquet_matrix_elements}~(b).

If the high-energy Floquet modes were limited to $\ket{\phi_2}$ and $\ket{\phi_3}$, tunneling across the double-well barrier would remain exponentially suppressed after the first kissing event, as shown by the light blue line in \cref{fig:Floquet_matrix_elements}(a). However, the master equation~\cref{Lindbladian} accounts for dissipators connecting $\ket{\phi_{2,3}}$ to higher modes, causing the above scenario of tunneling through higher levels to repeat. Before each new spectral kissing, an incoherent sum of dissipative pathways enables tunneling. After the kissing, interference between two competing transitions suppresses tunneling, leading to only intrawell leakage; see in \cref{fig:Floquet_matrix_elements}(a) how further matrix-element differences become suppressed with $\varepsilon_d$.

\subsection{Tunneling and coherence times of the Kerr-cat qubit}
\label{subsec: Results on tunneling single mode}

\begin{figure}[t]
    \centering
\includegraphics[width=\linewidth]{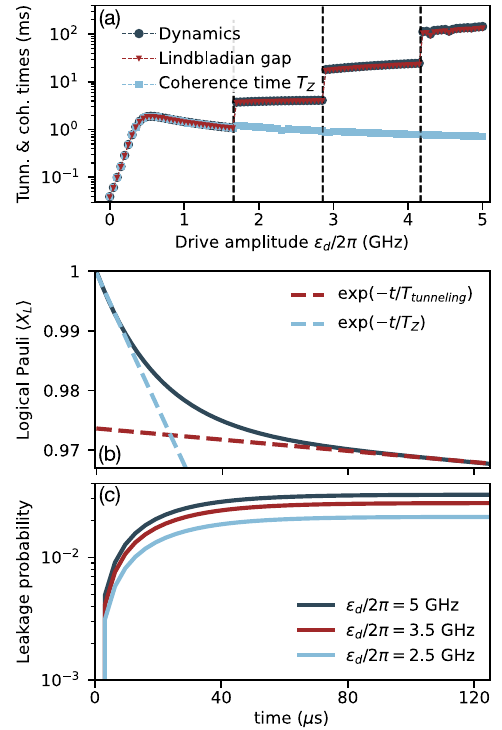}
    \caption{(a) Tunneling time of the Kerr-cat qubit extracted from time dynamics (dark blue) and the inverse of the Lindbladian gap (red). The coherence time, defined by the initial decay of the logical Pauli operator $\hat{X}_L = \ket{\beta+\alpha}\bra{\beta+\alpha} - \ket{\beta-\alpha}\bra{\beta-\alpha}$, is shown in light blue. Dashed vertical lines mark the drive amplitudes corresponding to spectral kissing events.
    (b) Time evolution of the expectation value of $\hat{X}_L$ for an initial state $\ket{\beta+\alpha}$ at drive amplitude $\varepsilon_d/2\pi = 3.5$ GHz. Red and light blue dashed lines indicate exponential fits at short and long times, respectively.
    (c) Intrawell leakage probability as a function of time for three values of drive amplitudes. System parameters are identical to those used in \cref{fig:cats_spectralkissing}. The bath parameters are taken from Ref.~\cite{Venkatraman:2024}, with $J(\omega_d) = 796$~kHz and $T(\omega_d) = 350$~mK; at other harmonics $J(\omega_d/2) = 7.96$~kHz and $T(\omega_d/2) = 50$~mK.
    }
\label{fig:Tunneling_and_leakage}
\end{figure}

We now turn to the main results of this section: (1) Under the single-mode approximation of \cref{singlemode_Hamiltonian}, the tunneling time increase in a staircase-like fashion and continues to grow with drive amplitude without saturation; and (2) the tunneling time does not coincide with the Kerr-cat coherence time $T_Z$, which initially increases but then decreases as the drive amplitude is further increased. 
\Cref{fig:Tunneling_and_leakage}(a) shows the Kerr-cat tunneling time as a function of drive amplitude, obtained from two methods: i) the discrimination of the time-evolved distribution of the coherent states $\ket{\beta \pm \alpha}$ spanning the cat states (blue circles) and ii) the smallest gap of the Lindbladian $\mathcal L$ in \cref{Lindbladian} (red triangles). The agreement between these approaches is excellent. 

In the first method, for each value of the drive amplitude $\varepsilon_d$, the system is initialized in one of two coherent states, $\ket{\beta+\alpha}$ or $\ket{\beta-\alpha}$, constructed from superpositions of the lowest Floquet modes $\ket{\phi_{0}}$ and $\ket{\phi_{1}}$. These coherent states are then evolved under the Floquet-Markov master equation \cref{Lindbladian}. At each time step, the Husimi Q distributions of these evolved states are computed and used to construct the log-likelihood function from which the assignment error of a hypothetical heterodyne measurement is evaluated. By analyzing the time dependence of the assignment error and fitting it to an exponential decay, we obtain the tunneling time. This procedure closely mirrors the experimental approach used to measure tunneling via heterodyne detection~\cite{Frattini2024,Hajr2024,ding2025,dealbornoz2024}; see \cref{appendix: time dynamics} for details. In the second method, the tunneling time is obtained as the inverse of the Lindbladian gap, which defines the slowest relaxation rate in the system. It is obtained from the diagonalization of the Lindbladian on the right-hand side of \cref{Lindbladian} in the Floquet-mode basis.

In agreement with experimental observations \cite{Frattini2024,Hajr2024}, we find in \cref{fig:Tunneling_and_leakage}(a) that the tunneling time increases in a staircase-like manner with the drive amplitude, with each plateau ending precisely at a spectral kissing point (vertical dashed lines). 
Importantly, these results do not rely on a perturbative expansion in the nonlinearity or the drive strength. Rather, it follows from minimal assumptions: the single-mode approximation of the system Hamiltonian of \cref{singlemode_Hamiltonian} and the standard approximations of weak, short-time correlated dissipation. Interestingly, the staircase behavior can also be reproduced within a perturbative framework, based on an expansion of the cosine nonlinearity and an effective Lindbladian derived by retaining terms beyond the standard rotating-wave approximation~\cite{Venkatraman:2024}. However, both approaches---the exact single-mode treatment presented here and the perturbative analysis---fail to capture the sharp decrease in tunneling time at large drive amplitudes observed in experiments.

We also find that the tunneling time differs dramatically from the coherence time $T_Z$ of the Kerr-cat, as defined by the decay time of the expectation value of the logical Pauli operator $\hat{X}_L = \ket{\beta+\alpha}\bra{\beta+\alpha}-\ket{\beta-\alpha}\bra{\beta-\alpha}$ at short times; see the light blue squares in \cref{fig:Tunneling_and_leakage}(a). Indeed, the latter quantity reaches a maximum value at a low drive amplitude and does not show the characteristic staircase behavior of the tunneling time. As shown in \cref{fig:Tunneling_and_leakage}(b), the expectation value of $\hat{X}_L$ exhibits a two-stage decay: an initial drop over a timescale $T_Z$ (light blue dashed line), followed by a much slower decay governed by the tunneling time (red dashed line). These two timescales coincide at low drive amplitude—specifically before the first spectral kissing, see \cref{fig:Tunneling_and_leakage}(a) where $T_Z$ matches the tunneling time before the first vertical dashed line---but begin to differ as the drive amplitude increases beyond that point. This difference arises because, although interwell tunneling is strongly suppressed as $\varepsilon_d$ increases, the {\it intrawell leakage} rate continues to grow, reducing the occupation probability of the coherent states. This is illustrated in \cref{fig:Tunneling_and_leakage}(c), which shows the intrawell leakage probability as a function of time for three values of drive amplitude. Here we define the leakage probability as $p_\pm - p_{\beta\pm\alpha}$, where $p_{\beta\pm\alpha}$ is the probability of occupying the coherent state $\ket{\beta\pm\alpha}$, and $p_\pm$ denotes the probability of being on the side of phase space where the coherent state $\ket{\beta\pm\alpha}$ is localized. Intrawell leakage accumulates over time and saturates at a value that increases with drive amplitude. Further details about this leakage probability can be found in \cref{appendix: time dynamics}.

Finally, we note that the absence of any breakdown in the resulting tunneling time at large drive amplitudes, despite the minimal assumptions used here, suggests a limitation in one of our underlying approximations. In \cref{sec: boxmode,sec: array modes}, we show---using the same general formalism presented in this section---how the multimode nature of the circuit leads to a breakdown of the tunneling time at moderate photon number.


\subsection{Heating and drive-induced dissipation} \label{singlemode: Temperature and spectral densities}

\begin{figure}[t!]
    \centering
\includegraphics[width=\linewidth]{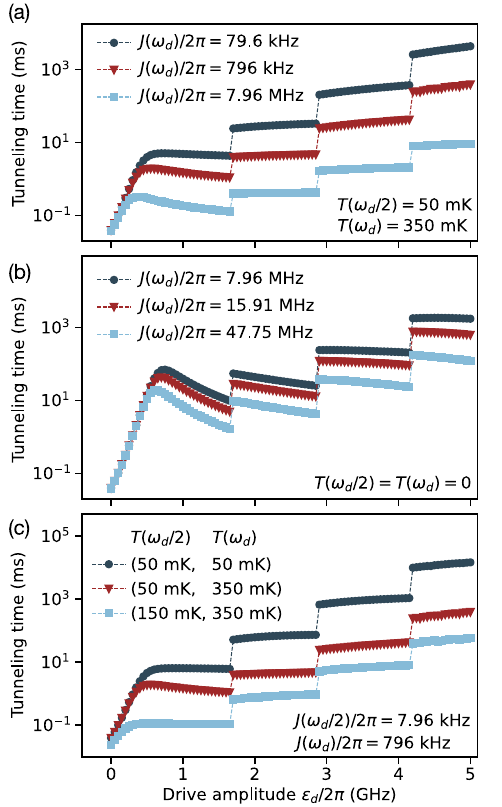}
\caption{Tunneling time as a function of the drive amplitude for different bath spectral densities and temperatures.
(a) Fixed temperature with three different spectral densities $J(\omega)$.
(b) Zero temperature with three different spectral densities $J(\omega)$.
(c) Fixed $J(\omega_d)$ with three combinations of bath temperatures $T(\omega_d/2)$ and $T(\omega_d)$. 
The specific values of $J(\omega)$ and $T(\omega)$ are indicated in the plots.
}
\label{fig:spectraldensity_temperature}
\end{figure}

In \cref{sec: examples of processes entering Dcoh and Dinc}, we have discussed processes associated with the absorption of a photon at $\omega_d/2$ from the bath and contributing to interwell tunneling and intrawell leakage, such as $|X_{210}|^2$ and $|X_{301}|^2$. Here, we describe other heating processes that limit the coherence and tunneling times of the Kerr-cat qubit and explain the roles of the spectral density and the effective temperature.

In addition to the processes discussed above, the environment can induce leakage near $\omega_d$, where the spectral density $J(\omega)$ peaks---see \cref{fig:circuits}(d)---despite smaller associated matrix elements.
In particular, the matrix elements $X_{20\pm1}$ and $X_{31\pm1}$ correspond to transitions that connect states of the same parity, such as $\ket{\phi_0} \to \ket{\phi_2}$ and $\ket{\phi_1} \to \ket{\phi_3}$. Before the first spectral kissing, these transitions contribute to tunneling with a total rate given by
\begin{equation}\label{eq:rate_at_omega_d}
\frac{1}{2} \kappa(\omega_d)(|X_{201}|^2 + |X_{311}|^2) + \frac{1}{2} \kappa(-\omega_d)(|X_{20{-}1}|^2 + |X_{31{-}1}|^2).    
\end{equation} 
The first term corresponds to photon absorption from the bath at frequency $\omega_d$ and is present even at zero drive amplitude. Indeed, in the limit $\varepsilon_d=0$, the associated Floquet matrix elements reduce to the charge matrix elements $\langle 0|\hn|2\rangle/n_{\textrm{zpf}}$ and $\langle 1|\hn|3\rangle/n_{\textrm{zpf}}$ between bare SNAIL states $\ket{i}$. These matrix elements are small but nonzero---of order $10^{-2}$ for the parameters used here---reflecting the fact that the double-SNAIL Hamiltonian $\hH_s$ does not have parity symmetry. Increasing the drive amplitude, the matrix elements $|X_{201}|^2 + |X_{311}|^2$ are further enhanced; see the blue dashed line in \cref{fig:Floquet_matrix_elements}(b). This is supported by analytical results presented in \cref{appendix: SW on Shirley}. Importantly, this leakage mechanism is suppressed in circuit implementations that preserve parity symmetry---i.e., those for which the Hamiltonian is invariant under $\hn \to -\hn$ and $\hat{\varphi} \to -\hat{\varphi}$---such as symmetric SQUID-based designs~\cite{Puri:2017,bhandari2024}, see \cref{appendix: SW on Shirley} for more details.

On the other hand, the second term of \cref{eq:rate_at_omega_d} corresponds to a more subtle drive-induced emission to the bath at frequency $\omega_d$ which persists at zero temperature. As shown by the dotted yellow and brown lines in \cref{fig:Floquet_matrix_elements}(b), the corresponding matrix elements vanish as $\varepsilon_d \to 0$ but grow by several orders of magnitude as $\varepsilon_d$ increases. Perturbative expressions for these rates are given in \cref{appendix: SW on Shirley}. This type of drive-induced emission to the bath occurring concurrently with qubit leakage has been shown to be a limiting factor for dispersive readout in the high-frequency regime~\cite{connolly:2025, dai:2025}.

For the bath parameters used in \cref{fig:Tunneling_and_leakage}---taken from Ref.~\cite{Venkatraman:2024}---we find $\kappa(\omega_d)/\kappa(\omega_d/2) \sim 10^4$. 
As a result, we find that
\begin{equation}\nonumber
    \kappa(\omega_d)(|X_{201}|^2 + |X_{311}|^2) \sim \kappa(\omega_d/2)(|X_{210}|^2 + |X_{301}|^2),
\end{equation}
even though $|X_{201}|^2 + |X_{311}|^2 \gg |X_{210}|^2 + |X_{301}|^2$; see \Cref{fig:Floquet_matrix_elements}.
Moreover, these two rates are much larger than $\kappa(-\omega_d)(|X_{20{-}1}|^2 + |X_{31{-}1}|^2)$, indicating that drive-induced processes are negligible for this choice of parameters. Importantly, these different rates depend sensitively on the specific form of the spectral density and the bath temperature at the relevant frequencies. For instance, in the case of a flat spectral density and a constant temperature over the whole frequency range, leakage would be dominated by the heating process at $\omega_d/2$ if $T \gtrsim \SI{20}{mK}$. In contrast, at lower temperatures, the dominant contribution would come from drive-assisted processes. 

As an illustration, \cref{fig:spectraldensity_temperature} shows the tunneling time as a function of drive amplitude, using the same system parameters as in \cref{fig:Tunneling_and_leakage}, but with different values of the bath spectral density $J(\omega_d)$ and temperatures $T(\omega_d)$ and  $T(\omega_d/2)$. First, we find that both the overall scale of the tunneling time and the heights of its staircase-like features are highly sensitive to these parameters: decreasing either $J(\omega_d)$ or the temperature increases the tunneling time, in agreement with the above discussion. 

In panel (a), we vary $J(\omega_d)$ while keeping $J(\omega_d/2)$ and both temperatures fixed. As a result, we observe a transition from a regime dominated by heating at $\omega_d$ (light blue) to one dominated by heating at $\omega_d/2$ (dark blue). The red curve represents the intermediate regime, previously discussed, where both channels contribute comparably. A key signature of $\omega_d$-induced heating is the gradual decrease of the first plateau, rather than a flat one. This behavior arises because the corresponding matrix elements $|X_{201}|^2+|X_{311}|^2$ grow with drive amplitude, as seen in the blue dashed line of \cref{fig:Floquet_matrix_elements}(b).

In panel (b), we consider the zero-temperature limit with $T(\omega)=0 $ for all frequencies such that that drive-induced processes constitute the sole heating mechanism. In this regime, the plateaus between kissing events have a steeper decline with increasing drive amplitude due to the strong enhancement of the relevant matrix elements ($|X_{20-1}|^2$+$|X_{31-1}|^2$ and $|X_{42-1}|^2$+$|X_{53-1}|^2$) with drive amplitude, as shown by the orange and brown dotted lines in \cref{fig:Floquet_matrix_elements}(b).

Panel (c) shows a similar effect as in (a), but now by varying the temperatures at $\omega_d$ and $\omega_d/2$ while keeping the spectral density fixed. The dark and light blue curves again highlight regimes where heating at $\omega_d/2$ dominates, while the red curve is reproduced from (a) for comparison.


\section{Buffer mode}
\label{sec: boxmode}

\begin{figure*}[t]
    \centering
\includegraphics[]{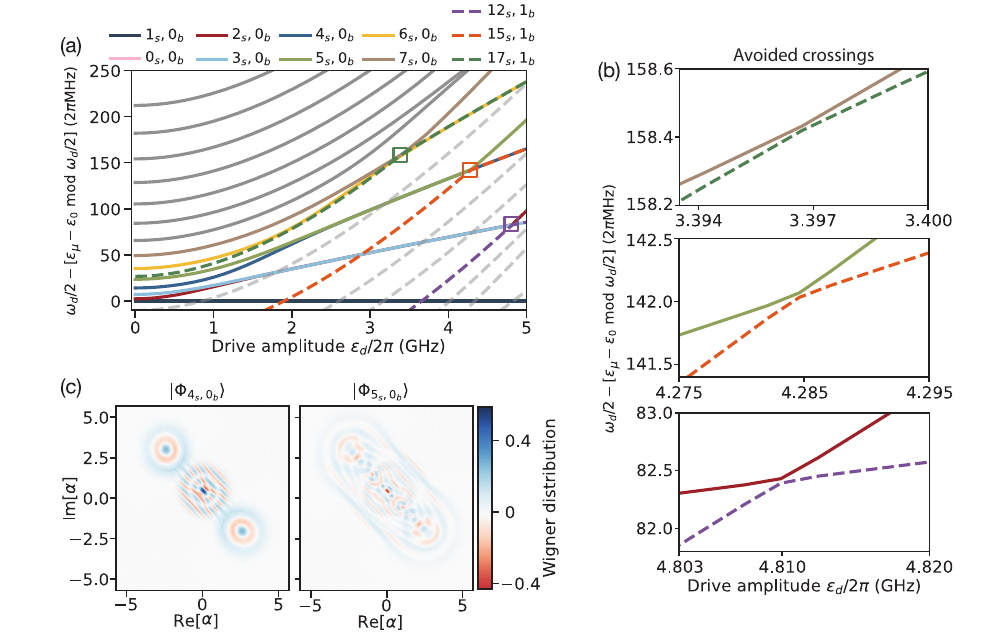}
    \caption{(a) Spectral kissing of the quasienergies in the presence of the buffer mode, whose frequency lies above the drive frequency. Solid lines indicate quasienergies without buffer excitation, while dashed lines correspond to quasienergies with one buffer excitation. Grey dashed lines correspond to quasienergies $\epsilon_{i_s,1_b}$ with $i_s=16,14,13,11,10,9$ from left to right. 
    (b) Detailed view of three avoided crossings highlighted by squares in panel (a) and which interrupt the spectral kissing of low-lying quasienergies.
    (c) Wigner functions of the modes $\ket{\phi_{4_s,0_b}}$ (left) and $\ket{\phi_{5_s,0_b}}$ (right) near the avoided crossing at $\varepsilon_d/2\pi = 4.81$ GHz.
    }
\label{fig:buffer_spectralkissing_wigner}
\end{figure*}

The single-mode approximation of \cref{singlemode_Hamiltonian}, when combined with a Floquet-Markov treatment and the full nonlinear SNAIL potential, captures many of the key features of the Kerr-cat qubit. In particular, it reproduces spectral kissing and the staircase-like growth of the tunneling time with increasing drive amplitude. However, it fails to capture the breakdown of this staircase structure observed at large drive amplitudes, illustrating the intrinsic robustness of this qubit.

A possible cause of the breakdown is mechanisms leading to unwanted resonances in the Floquet spectrum that can mediate leakage to higher energy states, and therefore to a transfer of the occupation probability between the wells of the metapotential. Such spectral resonances have been identified in previous work as the main mechanism behind ionization in driven highly-excited hydrogen atoms \cite{Breuer1989_Floquet_Ionization} and superconducting circuits~\cite{Dumas2024,MIST_1,MIST_2,Shillito2022Dynamics,Cohen2023,Xiao:2024, singh:2025, Mingkang2025, Fechant2025, Wang2025}. Given the presence of  additional degrees of freedom in the circuit implementation of the Kerr cat qubit, it is natural to ask whether their presence can induce such resonances.

To address this possibility, we now reintegrate the buffer mode into our description of the system and investigate its impact on the Kerr-cat qubit. Our starting point is the exact displaced-frame Hamiltonian of \cref{Hamiltonian_buffer}, which includes the interaction between the cat mode and the buffer. For experimentally relevant parameters---specifically, a buffer-drive detuning of $|\Delta_{bd}|/2\pi = |\omega_b - \omega_d|/2\pi \approx \SI{285}{MHz}$ and a coupling strength of $g/2\pi \approx \SI{100}{MHz}$---this interaction leads to a dispersive shift of order $\sim \SI{10}{kHz}$
between the buffer and the double-SNAIL oscillator~\cite{Frattini2024,Venkatraman_deltacat,ding2025}. As we show below, this additional degree of freedom introduces the resonant processes absent in the single-mode model and results in the breakdown of the staircase behavior. In what follows, we focus first on positive detuning, where the buffer lies above the drive frequency by $\Delta_{bd}/2\pi \approx \SI{285}{MHz}$, as realized in Refs.~\cite{Frattini2024,dealbornoz2024} (see Subsecs.~\ref{subsec: buffer mode spectral kissing} and \ref{subsec: buffer tunneling time}). For completeness, we also analyze the case of negative detuning, $\Delta_{bd}/2\pi \approx -\SI{285}{MHz}$, in Subsec.~\ref{subsec: buffer negative detuning}.

\subsection{Impact of the buffer mode on the Floquet spectrum}
\label{subsec: buffer mode spectral kissing}

The analysis follows the same steps as above, with the particularity that special care must be taken in identifying the dressed states of the joint SNAIL–buffer system. To do so, we begin by diagonalizing the undriven Hamiltonian,
\begin{align}
\hH_{sb} = \hH_s + \omega_b \ha_b^\dagger \ha_b + i g \hn (\ha_b^\dagger - \ha_b),
\end{align}
yielding a set of dressed eigenstates $\{\ket \lambda\}$ with corresponding dressed energies $\{E_\lambda\}$. Following Ref.~\cite{Shillito2022Dynamics}, we label the spectrum by first identifying the dressed states $\{\ket{i_s, 0_b}\}$ with $i_s$ excitations of the SNAIL and zero excitation of the buffer mode, by comparing them to the bare states with no buffer excitation. We then apply $\hat{a}_b^\dagger$ to each of these dressed states to generate candidates for the $\{\ket{i_s, 1_b}\}$ states, selecting from the remaining ${\ket{\lambda}}$ the ones with the largest overlaps. This iterative process is repeated to construct and label the higher-excitation states $\{\ket{i_s, j_b}\}$. As in the previous section, we truncate the Hilbert space to include 160 levels in the double-SNAIL mode and up to five excitations in the buffer mode. 

Working in the displaced frame, we then include the drive on the double-SNAIL mode as in \cref{Hamiltonian_buffer} and obtain the Floquet spectrum as a function of the drive amplitude following the procedure discussed in \cref{sec:single mode} using a step-size increment of $\delta \varepsilon_d/2\pi = \SI{2.5}{MHz}$ in the tracking of the spectrum; see Ref.~\cite{Dumas2024} for a discussion of the choice of increment $\delta \varepsilon_d$. In this extended system, the cat states correspond to the Floquet modes $\ket{\phi_{0_s,0_b}}$ and $\ket{\phi_{1_s,0_b}}$.
To facilitate comparison with the single-mode approximation, we show in \cref{fig:buffer_spectralkissing_wigner}(a) the Floquet spectrum (folded at $\omega_d/2$) for the lowest modes with zero buffer excitation (full lines), alongside nine relevant modes with one buffer excitation (dashed lines). 

As in the single-mode approximation, for drive amplitudes up to $\varepsilon_d/2\pi \approx \SI{3.4}{GHz}$ we recover the spectral kissings involving, for example, the quasienergy pairs $(\epsilon_{2_s,0_b}, \epsilon_{3_s,0_b})$ and $(\epsilon_{4_s,0_b}, \epsilon_{5_s,0_b})$.
However, avoided crossings that interrupt the spectral kissing emerge between low-energy quasienergies without buffer-mode excitation and higher-energy states involving one buffer-mode excitation. These features are marked by squares in panel (a) and shown in detail in the zoomed-in view of panel (b).
The first anticrossing is between $\epsilon_{7_s,0_b}$ and $\epsilon_{17_s,1_b}$ at $\varepsilon_d/2\pi \approx \SI{3.4}{GHz}$ ---corresponding to $\bra{\phi_{0(1)}}\hat a^\dag \hat a \ket{\phi_{0(1)}} \approx 12$ photons in the cat manifold. Similar avoided crossings occur between $\epsilon_{5_s,0_b}$ and $\epsilon_{15_s,1_b}$ at $\varepsilon_d/2\pi \approx \SI{4.29}{GHz}$, and between $\epsilon_{2_s,0_b}$ and $\epsilon_{12_s,1_b}$ at $\varepsilon_d/2\pi \approx \SI{4.81}{GHz}$.

Crucially, the three avoided crossings highlighted above are not the only ones present, but merely those resolved with the finite step size $\delta\varepsilon_d$ of drive amplitude used in our Floquet spectrum tracking. Indeed, in the absence of any selection rule, as is the case here, there is finite coupling between all pairs of quasienergies, such that crossings are generically replaced by avoided crossings. Thus, when the quasienergies indicated by the dashed lines in \cref{fig:buffer_spectralkissing_wigner}(a) intersect with low-lying quasienergies, they in fact form avoided crossings. 
These involve hybridization between higher-lying states---such as $\ket{13_s,1_b}$, $\ket{14_s,1_b}$, and $\ket{16_s,1_b}$---and lower-lying states involved in spectral kissing, and they occur densely within the drive range $\varepsilon_d/2\pi \in [2.8, 5]$~GHz. While their splitting are small, on the order of $\sim \SI{10}{kHz}$ as in those shown in \cref{fig:buffer_spectralkissing_wigner}(b), their presence nonetheless has significant consequences on the tunneling time, as further discussed in the next subsection. 

These avoided crossings can be resolved despite their small gap size because they are wide: the involved quasienergies remain nearly parallel over a broad range of drive amplitudes, so their relative separation evolves slowly compared to the size of the step-size increment $\delta \varepsilon_d/2\pi = \SI{2.5}{MHz}$. 
As a result, the modes hybridize appreciably over an extended range of drive amplitude,
even if the eigenstate labels do not swap at the avoided crossing. Further reducing the step-size increment in the tracking, although possible, is not necessary for resolving these features, and we avoid it due to its computational cost.

The importance of these avoided crossings---where spectral kissing is interrupted---is that, at any one of them, no superposition of the Floquet modes involved in kissings can be localized within a single well of the Kerr-cat metapotential. This is illustrated in \cref{fig:buffer_spectralkissing_wigner}(c) which shows the Wigner distributions of the Floquet modes $\ket{\phi_{4_s,0_b}}$ and $\ket{\phi_{5_s,0_b}}$ at $\varepsilon_d/2\pi = \SI{4.81}{GHz}$ corresponding to the avoided crossing between $\epsilon_{5_s,0_b}$ and $\epsilon_{15_s,1_b}$. Prior to the avoided crossing, $\ket{\phi_{5_s,0_b}}$ is approximately related to $\ket{\phi_{4_s,0_b}}$ by a displaced parity transformation; as a result, their superpositions yield states of the form $\hat D(\beta \pm \alpha)\ket{n=2}$, which are localized in a single well \cite{Chamberland2022}. At the avoided crossing, however, $\ket{\phi_{5_s,0_b}}$ becomes significantly hybridized with
a higher-energy mode---as manifested by its Wigner distribution---and can no longer be combined with $\ket{\phi_{4_s,0_b}}$ to produce a well-localized state.

The avoided crossings, along with the resulting successive interruption of spectral kissing, are the result of the ac-Stark shifts of the energy levels as well as the moderate detuning between the buffer mode and the drive frequency. To illustrate this, we focus on the avoided crossing between $\epsilon_{7_s,0_b}$ and  $\epsilon_{17_s,1_b}$. In the absence of  a drive, the energy difference between the corresponding dressed states is approximately
\begin{equation} \label{stark_shift_collisions}
E_{17_s,1_b} - E_{7_s,0_b}\approx 6\omega_d + \Delta_{bd} - (17\cdot 16 - 7\cdot 6)K,    
\end{equation}
where the last term arises from the self-Kerr nonlinearity which is here $K/2\pi = \SI{1.18}{MHz}$. This expression shows that, in the presence of a drive, these two states can become resonant (modulo $\omega_d$) if the ac-Stark shift is strong enough to compensate for the detuning $\Delta_{bd} - (17\cdot 16 - 7\cdot 6)K$.
This situation is likely to occur for this or other transitions when $\Delta_{bd} > 0$, as considered here and as is the case in recent experiments~\cite{Frattini2024,dealbornoz2024}. This is because the drive induces a negative ac-Stark shift to the transition from $E_{i_s,0_b}$ to $E_{j_s,1_b}$ for $j_s>i_s$, leading to a multiphoton resonance condition whenever $\Delta_{bd} \gtrsim [j_s(j_s-1) - i_s(i_s-1)]K$. For the range of drive amplitudes considered here, detunings $\Delta_{bd} - [j_s(j_s-1) - i_s(i_s-1)]K$ of $\sim 2\pi\times \SI{100}{MHz}$ or less are problematic, as the ac-Stark shift can bridge that gap at moderate drive amplitudes. In the single-mode case, these types of resonances cannot occur as energy differences like $E_{17_s} - E_{7_s}\approx 5 \omega_d - (17\cdot 16 - 7\cdot 6)K$ cannot be compensated by the negative shift induced by the drive.

Moreover, because a small Kerr implies a slowly increasing ac-Stark shift with growing $\varepsilon_d$, the near-resonance condition can persist over a wide range of drive amplitudes. As a result, even a weak dispersive interaction can lead to strong hybridization between these Floquet modes. A similar argument explains the other two avoided crossings highlighted in \cref{fig:buffer_spectralkissing_wigner}. Importantly, capturing this hybridization accurately would be difficult---if not impossible---using a perturbative expansion of the nonlinearity; see also Refs.~\cite{Breuer1989_Floquet_Ionization,Dumas2024}. In Appendix~\ref{appendix: buffer mode}, we show that the leading-order process responsible for the coupling between $\ket{\phi_{7_s, 0_b}}$ and $\ket{\phi_{17_s, 1_b}}$ arises only at eleventh order in the expansion of the nonlinear potential around its minimum.

These unwanted resonances between low-energy and high-energy Floquet modes underscore the importance of considering the joint dynamics of the double-SNAIL and the buffer mode, highlighting the limitations of the single-mode approximation. As we show below, the hybridization of the double-SNAIL mode with the buffer mode impacts the tunneling time of the Kerr-cat qubit.

\begin{figure*}[t!]
    \centering
\includegraphics{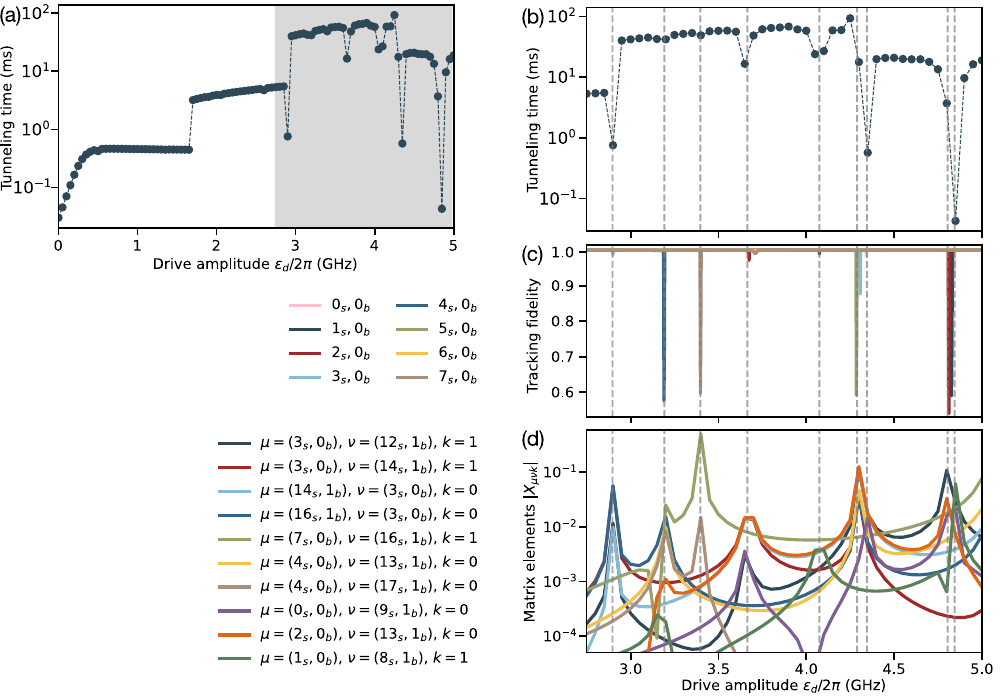}
\caption{
(a) Tunneling time as a function of drive amplitude in the presence of the buffer mode whose frequency lies above the drive frequency.
(b) Same as panel (a) in the reduced range $\varepsilon_d/2\pi \in (2.75, 5)$ GHz corresponding to cat states approximately 10 to 17 photons. See the legend on the left of the panel.
(c) Fidelity of Floquet tracking, $|\langle \phi_\mu[\varepsilon_d]|\phi_\mu[\varepsilon_d + \delta \varepsilon_d] \rangle |^2$. Drops in fidelity (highlighted by dashed lines) indicate the presence of resonances.
(d) Representative Floquet matrix elements $|X_{\mu \nu k}^s|$ as a function of the drive amplitude. The drops in the tracking fidelity coincide with resonant enhancements in matrix elements connecting low-lying modes to highly excited states, leading to the observed dips in tunneling time. See the legend on the left of the panel.
}
\label{fig:buffer_tunnelingtime}
\end{figure*}

\subsection{Impact of the buffer mode on tunneling time}
\label{subsec: buffer tunneling time}

We extend the Floquet-Markov formalism introduced in \cref{sec:single mode}  to account for the fact that the SNAILs and the buffer mode are coupled to distinct environments. The master equation now includes two types of dissipative channels, associated with the matrix elements of the SNAIL charge operator
\begin{equation}
X^{s}_{\mu\nu k} = \frac{1}{T}\int\limits_0^T e^{-ik \omega_dt} \bra{\phi_\mu(t)}i(\hat a^\dag - \hat a)\ket{\phi_\nu(t)} dt,
\end{equation}
and the buffer charge operator
\begin{equation}
X^{b}_{\mu\nu k} = \frac{1}{T}\int\limits_0^T e^{-ik \omega_dt} \bra{\phi_\mu(t)}i(\hat a_b^\dag - \hat a_b)\ket{\phi_\nu(t)} dt.
\end{equation}
In these expressions,  each index $\mu$ and $\nu$ labels a pair $(i_s, j_b)$ of SNAIL and buffer excitation. The first type of matrix elements, $X^s_{\mu\nu k}$, describes dissipation due to the intrinsic bath of the double-SNAIL accounting, e.g., for dielectric losses.
Importantly, in contrast to the similar-looking expression \cref{eq:matrix_elements} in the single-mode approximation, this bath is not filtered by the buffer mode. We therefore consider a flat spectral density $J_s(\omega) = \theta(\omega) \times \SI{7.98}{kHz}$ and a single effective temperature of $T_s = \SI{50}{mK}$. The second type of matrix elements, $X^b_{\mu\nu k}$, describes the coupling of the buffer mode to its environment, dominated by the coupling to the drive line.
This bath is likewise assumed to have a flat spectral density $J_b(\omega) = \theta(\omega) \times \SI{798}{kHz}$ and a constant effective temperature of $T_b = 350$~mK. The specific numerical values for the spectral densities and temperatures are chosen so that the dissipative dynamics considered in \cref{sec:single mode} are recovered in the single-mode approximation.

With these definitions in hand, we now follow the approach discussed in \cref{subsec: Results on tunneling single mode} and compute the tunneling time as a function of drive amplitude from the assignment error, as shown in \cref{fig:buffer_tunnelingtime}(a).
For drive amplitudes below $ \varepsilon_d/2\pi \approx 2.89$ GHz, the tunneling time exhibits a staircase-like increase thanks to the spectral kissing of the lowest-lying quasienergies and the induced interference of dissipative processes, just like in the case of the semiclassical treatment of the buffer mode in \cref{sec:single mode}. For larger drive amplitudes, where the spectral kissing is interrupted by avoided crossings, the tunneling time is considerably affected. \Cref{fig:buffer_tunnelingtime}(b) shows the tunneling time in the range $\varepsilon_d/2\pi \in (2.75,5)$~GHz. This is accompanied in panel (c) by the fidelity of the Floquet tracking ($|\langle \phi_\mu[\varepsilon_d]|\phi_\mu[\varepsilon_d + \delta \varepsilon_d] \rangle |^2$), which measures the squared overlap between each Floquet mode before and after each drive increment, and in panel (d) by representative Floquet matrix elements $|X_{\mu \nu k}|=|X_{\mu \nu k }^s|$. While we focus here on matrix elements of the first type, we note that the second type, $X_{\mu \nu k}^b$, exhibits similar features.

At $\varepsilon_d/2\pi \approx 2.9$~GHz, a sharp dip is observed in the tunneling time. As shown by the light blue line in \cref{fig:buffer_tunnelingtime}(c), this coincides with a sudden drop in the tracking overlap of $\ket{\phi_{3_s,0_b}}$, which hybridizes with $\ket{\phi_{15_s, 1_b}}$. 
Although hybridization between these states is relatively weak at this particular value of $\varepsilon_d$---on the order of $2\%$---
it is sufficient to disrupt the staircase-like increase of the tunneling time because transition rates to high‑energy modes grow substantially. Indeed, as shown in \cref{fig:buffer_tunnelingtime}(d), the Floquet matrix elements $|X^s_{\mu \nu k}|$ between $(3_s, 0_b)$ and $(12_s, 1_b)$ (dark blue line), $(14_s, 1_b)$ (red and light blue lines), and $(16_s, 1_b)$ (blue line) increase by several orders of magnitude. This hybridization not only opens a new leakage channel from $(3_s, 0_b)$ to states outside the double-well barrier, but also enables direct transitions from the other low‑lying states $(0_s, 0_b)$, $(1_s, 0_b)$ and $(2_s, 0_b)$ to those high-energy states.

Above $\varepsilon_d/2\pi \approx \SI{2.9}{GHz}$, the tunneling time exhibits a sharp increase followed by a new plateau, due to spectral kissing between $\epsilon_{4_s,0_b}$ and $\epsilon_{5_s,0_b}$---a behavior reminiscent of the single-mode case. However, this plateau is interrupted several times whenever one of the low-lying modes—namely, $\ket{0_s,0_b}$ through $\ket{5_s,0_b}$—hybridizes with higher-energy states.
For instance, at $\varepsilon_d/2\pi \approx \SI{3.65}{GHz}$ the mode $\ket{\phi_{2_s,0_b}}$ hybridizes with $\ket{\phi_{14_s,1_b}}$ leading to the dip in the tracking fidelity of the former, as shown by the red line in \cref{fig:buffer_tunnelingtime}(c). This hybridization disrupts the coherent interference of jump operators responsible for suppression of tunneling between $\ket{\phi_{2_s,0_b}}$ and $\ket{\phi_{3_s,0_b}}$ (see, e.g., \cref{fig:action_cops}). At the same time, it also enhances matrix elements connecting $\ket{\phi_{2_s,0_b}}$ and other low-lying modes to high-energy modes, as shown, e.g., by the orange and red curves in \cref{fig:buffer_tunnelingtime}(d), ultimately resulting in a pronounced drop in tunneling time.

Interestingly, the hybridization between $\ket{\phi_{7_s,0_b}}$ and $\ket{\phi_{17_s,1_b}}$ discussed in \cref{subsec: buffer mode spectral kissing} and highlighted here by the large dip in tracking fidelity (brown curve) in \cref{fig:buffer_tunnelingtime}(c) does not lead to a sharp drop of the tunneling time
since the quasienergies $\epsilon_{7_s,0_b}$ and $\epsilon_{8_s,0_b}$ have not yet kissed at $\varepsilon_d/2\pi \approx \SI{3.4}{GHz}$. As a result, leakage to these states limits the tunneling time independently of their hybridization with other higher-energy modes.

Subsequent dips in tunneling time are caused by further resonances, including hybridizations between $\ket{\phi_{1_s,0_b}}$ and $\ket{\phi_{9_s,1_b}}$, and between $\ket{\phi_{3_s,0_b}}$ and $\ket{\phi_{13_s,1_b}}$.
Beyond $\varepsilon_d/2\pi \approx \SI{4.3}{GHz}$, these hybridizations become sufficiently strong for the cat and first excited states $\ket{\phi_{i_s,0_b}}$ with $i_s=0,1,2$ to successively erode the staircase-like growth in tunneling time, see the pink, dark-blue  and red curves in \cref{fig:buffer_tunnelingtime}(c), and the corresponding matrix element enhancements in \cref{fig:buffer_tunnelingtime}(d).

The overall decrease in tunneling time---punctuated by sharp dips---as a function of drive amplitude beyond a certain threshold was experimentally observed in Ref.~\cite{dealbornoz2024}, but its origin has remained an open question. Our results suggest that the cause of this observation is multiphoton resonances involving an additional degree of freedom, which interrupt the destructive interference of dissipative channels responsible for suppressing tunneling and open new leakage channels to states above the double-well barrier of the Kerr-cat's metapotential.

\subsection{Impact of a buffer mode below the drive frequency}
\label{subsec: buffer negative detuning}

The preceding discussion would suggest that when the buffer lies below the drive frequency, i.e., $\Delta_{bd} < 0$, no multiphoton resonances should arise within the range of drive amplitudes explored here since the negative ac-Stark shift of the transitions $\epsilon_{j_s,1_b} - \epsilon_{i_s,0_b}$ with $j_s > i_s$ has the same sign as the detuning at zero drive, $E_{j_s,1_b} - E_{i_s,0_b}$, which is set by $\Delta_{bd}$. Yet, as we show below, in the negative detuning regime a different detrimental mechanism comes into play, which likewise leads to the breakdown of the Kerr-cat qubit's tunneling time. This negative detuning regime has been realized experimentally in Ref.~\cite{ding2025}.  Instead of the sign of $\Delta_{bd}$, this regime can also be mimicked by flipping the sign of the self-Kerr nonlinearity $K$ using an external flux control ~\cite{Rodrigo_private_communication, dealbornoz2024}.

\begin{figure}[t!]
    \centering
\includegraphics{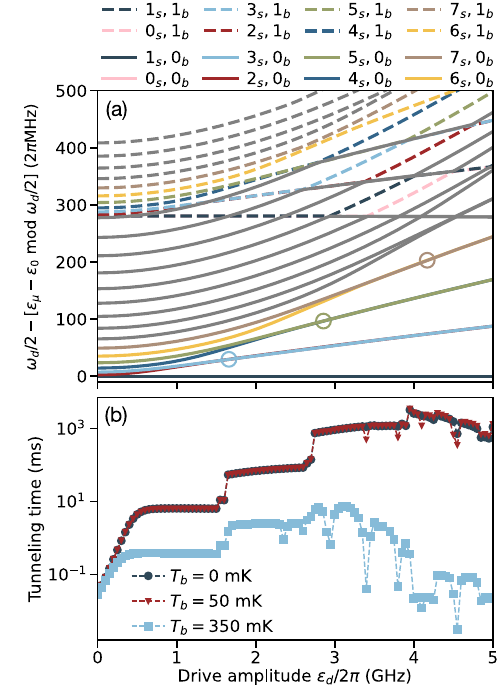}
\caption{(a) Floquet spectrum folded at $\omega_d/2$ with the buffer mode frequency below the drive frequency. Solid lines correspond to low-lying quasienergies with zero buffer excitations, with the even and odd cat states highlighted in pink and dark blue, respectively. Dashed lines indicate states with one buffer excitation.
(b) Tunneling time as a function of drive amplitude for a buffer mode below the drive frequency, shown at different buffer bath temperatures $T_b$.
}
\label{fig:buffer_negative_detuning}
\end{figure}

This alternative breakdown mechanism can be understood from \cref{fig:buffer_negative_detuning}(a) where we show the Floquet spectrum including not only the relevant levels of the cat with zero buffer excitation $\epsilon_{i_s,j_b=0}$ (solid lines), but also the corresponding levels with one buffer excitation $\epsilon_{i_s,j_b=1}$ (dashed lines). As before, the cat states are defined by the lowest two levels, $\epsilon_{i_s=0,j_b=0}$ and $\epsilon_{i_s=1,j_b=0}$. Consistent with the discussion above, the levels $\epsilon_{i_s,j_b=0}$ with $i_s \leq 7$ do not undergo collisions with higher-lying states, and the spectral kissing of these levels remains uninterrupted in the range of drive amplitudes displayed here---unlike in the case of positive detuning. However, for the same reason as in the positive-detuning case discussed in \cref{subsec: buffer mode spectral kissing}, the levels $\epsilon_{i_s,j_b=1}$ (colored dashed lines) collide with excited cat states with zero buffer excitation, $\epsilon_{j_s,j_b=0}$ with $j_s > i_s$ (solid gray lines), as seen in \cref{fig:buffer_negative_detuning}(a), where the gray solid lines cross the colored dashed lines multiple times.

If the buffer is constrained to remain close to its vacuum state (in the displaced frame), as would be the case at zero temperature $T_b=0$, these resonances are inconsequential: the cat states and the relevant states below the metapotential do not hybridize with high-energy modes, and the tunneling time increases with $\varepsilon_d$ without breakdown. This is confirmed in \cref{fig:buffer_negative_detuning}(b), which shows the tunneling time as a function of drive amplitude for different buffer-bath temperatures $T_b$, with all other bath parameters ($J_s$, $J_b$, and $T_s$) the same as in \cref{fig:buffer_tunnelingtime}. At $T_b=0$ (dark-blue circles), the tunneling time nearly recovers the single-mode result, although drive-induced heating (see \cref{singlemode: Temperature and spectral densities}) begins to limit it slightly in the 4–5 GHz range. Raising the buffer-bath temperature to $T_b=50$ mK (red triangles) introduces small dips that interrupt the staircase-like increase. These dips correspond to level crossings between $\epsilon_{i_s,j_b=1}$ and $\epsilon_{j_s,j_b=0}$ with $i_s \leq 3$ and $j_s \geq 8$, as discussed above.

Indeed, at finite buffer temperature, the bath induces leakage from levels $\nu=(i_s,0_b)$ into $\mu=(i_s,1_b)$, facilitated by matrix elements $X_{\mu,\nu,k=1}^b \sim \mathcal{O}(1)$. This can be seen explicitly by considering $i_s=0,1$ and noting that the leading buffer Floquet matrix elements ($X^b$) enter the master equation as
\begin{align} \begin{split} \label{eq: heating_buffer_matrix_element}
&X_{(0_s,1_b), (0_s, 0_b),k=1}^b\ket{\phi_{(0_s,1_b)}}\bra{\phi_{(0_s, 0_b)}} \\ &+ X_{(1_s,1_b), (1_s, 0_b),k=1}^b\ket{\phi_{(1_s,1_b)}}\bra{\phi_{(1_s, 0_b)}}\\ &\approx X_{(0_s,1_b), (0_s, 0_b),k=1}^b \left(\ket{\beta+\alpha}\bra{\beta+\alpha}\right.\\ &\left. + \ket{\beta-\alpha}\bra{\beta-\alpha}\right)\ket{1_b}\bra{0_b}, \end{split} \end{align}
where the Floquet modes are approximated as 
$\ket{\phi_{i_s,j_b}}\propto(\ket{\beta+\alpha}\pm\ket{\beta-\alpha})\ket{j_b}$ for $i_s=0,1$ and $j_b=0,1$. As a result, even though the dominant buffer Floquet matrix elements do not induce direct tunneling---thanks to destructive interference, as seen in \cref{eq: heating_buffer_matrix_element}---they nonetheless lead the occupation of the buffer mode: $\ket{0_b} \to \ket{1_b}$. Because this process probes the bath at $\Delta_{\mu \nu k } = \omega_b \approx \omega_d$, it is enabled by thermal photons at the drive frequency. The resulting states with finite buffer excitation, shifted by in frequency the ac-Stark effect resulting from the drive, can then become resonant with highly excited modes $\ket{\phi_{i_s,0_b}}$ with $i_s \geq 8$ and hybridize strongly with them, see \cref{fig:buffer_negative_detuning}(a). This facilitates tunneling across the double-well barrier $\ket{\beta \pm \alpha} \to \ket{\beta \mp \alpha}$. This mechanism becomes more evident at larger buffer mode temperature, see the light-blue squares in \cref{fig:buffer_negative_detuning}(b) obtained for $T_b=350$~mK, where the tunneling time shows pronounced dips due to resonances of the one-excitation buffer manifold.

In summary, the presence of the buffer mode can explain the recent experimental observations of the breakdown of the Kerr-cat's tunneling time. When the buffer lies above the drive frequency, resonances involving the low-lying spectrum spectrum are responsible for the qubit’s premature demise. On the other hand, in the negative-detuning case it is buffer heating, together with resonances affecting the modes with finite buffer excitation, that causes the sudden decrease in tunneling time.


\section{SNAIL array modes} 
\label{sec: array modes}

In the previous section, we demonstrated that including the buffer mode in the circuit description can explain the experimentally observed breakdown of the staircase pattern of the cat's tunneling time with increasing drive amplitude. In this section, we explore the role of internal degrees of freedom of the SNAIL circuit that we have dropped in the single-phase approximation of \cref{eq:doublesnail_singlemode}. These are high-frequency collective modes of the double SNAIL which we have so far implicitly assumed to remain in their ground state. In the context of fluxonium qubit, similar array modes have been theoretically shown to give rise to additional multiphoton resonances involving excitations of the array and to lower the onset of measurement-induced state transitions~\cite{singh:2025}. Furthermore, once populated, these modes contribute to dephasing over long timescales~\cite{Viola2015, singh:2025}, something which can significantly degrade the coherence time of the Kerr-cat qubit~\cite{Putterman2022, labaymora:2025}. 

Here, we show that the array modes of the SNAIL can similarly induce multiphoton resonances leading to a breakdown of the staircase pattern of the tunneling time analogous to that caused by the buffer mode. The occurrence of such resonances depends sensitively on the frequencies of the array modes, which are determined by circuit design. Importantly, these resonances arise mainly due to the quadratic coupling between the SNAIL mode and the array modes which is induced by the stray ground capacitance in the array, which can in principle be engineered to remain negligible.

In the double-SNAIL circuit of \cref{fig:circuits}(a) there are six interacting array collective modes. The lowest-energy mode corresponds to the one retained in the single-phase approximation of \cref{eq:doublesnail_singlemode}. Expressed in terms of the fluxes across the six large junctions of the two SNAILs, $\{\theta_j\}_{j=1}^6$, the reduced phase variable of this mode is given by $\varphi = (2\pi/\Phi_0) \sum_{j=1}^6 \theta_j$, and we refer to it as the symmetric mode (see \cref{fig:circuit_array_modes}). The remaining five collective modes are nonlinear oscillators with fundamental frequencies around the plasma frequency of the large junctions, $\sqrt{8 E_{C_J} E_J}$, which is much higher than the frequency $\omega_{01}/2\pi \approx \SI{6.094}{GHz}$ of the symmetric mode~\cite{Ferguson:2013}. Owing to this large detuning and typically weak coupling, they generally interact with the symmetric mode only dispersively~\cite{Ferguson:2013, Viola2015, singh:2025}. Further details can be found in \cref{appendix: array modes}.

However, one of these modes stands out due to its comparatively lower plasma frequency and significantly stronger charge coupling to the symmetric mode~\cite{Ferguson:2013,Viola2015,singh:2025}. To keep the analysis tractable while capturing the dominant effect, we retain only this mode. Its reduced phase coordinate is defined in terms of the fluxes across the large junctions as  $\varphi_- = (2\pi/\Phi_0)[\theta_1 + \theta_2 + \theta_3 - (\theta_4 + \theta_5 + \theta_6)]$ and changes sign under exchange of the two SNAILs; accordingly, we call it the antisymmetric mode. The Hamiltonian describing the coupled symmetric and antisymmetric modes reads
\begin{equation} \label{eq: array mode Hamiltonian}
\begin{split}
\hat H_{sa} =& 4 E_C \hat n^2 + 4 \beta E_C \hat n_-^2  + g \hat n \hat n_-\\
&-6 E_J \cos\left(\frac{\hat \varphi}{6} \right) \cos\left(\frac{\hat \varphi_-}{6} \right) \\
&-2\alpha E_J \cos\left(\frac{\hat \varphi}{2} + \varphi_\textrm{ext} \right) \cos\left(\frac{\hat \varphi_-}{2} \right),
\end{split}
\end{equation}
where $\beta$ is a factor relating the two capacitive energies. The two modes are coupled both through the cosine nonlinearities and via the quadratic term $g \hat n \hat n_-$, which arises from stray ground capacitances in the array; see \cref{appendix: array modes} for details.

In the presence of a drive, whenever the antisymmetric-mode frequency lies near an integer multiple of the drive frequency, the Hamiltonian $\hat H_{sa}$ can lead to similar multiphoton processes than those in the case of the buffer mode that disrupt the spectral kissing and the coherence of the Kerr-cat qubit. 
Indeed, $\hat H_{sa}$ can be rewritten as
\begin{equation}
\begin{split}
\hat H_{sa} =& 4 E_C \hat n^2 + 4 \beta E_C \hat n_-^2 + \frac{E_{J,-}}{2}\hat \varphi_-^2 + g \hat n \hat n_-\\
&-6 E_J \cos\left(\frac{\hat \varphi}{6} \right) -2\alpha E_J \cos\left(\frac{\hat \varphi}{2} + \varphi_\textrm{ext} \right) \\
&+ \hat H_{\text{nl.},-}
+\hat H_{\text{nl. coupl.}}
\end{split}
\end{equation}
where we have introduced the effective Josephson energy of the antisymmetric mode
\[
\begin{split}
    E_{J,-} &= E_J\left[\frac{1}{6} \cos\left(\frac{\varphi_\text{min}}{6}\right) + \frac{\alpha}{2} \cos\left(\frac{\varphi_\text{min}}{2} + \varphi_\text{ext}\right)\right] \\
    &\approx 0.16 E_J.
\end{split}
\]
This Hamiltonian has the same form as in the buffer-mode case except for two extra terms: (i) the nonlinear part of the extra mode's potential, $\hat H_{\text{nl.}, -}$, which does not play an important role, and (ii) a weak nonlinear coupling, $\hat H_\text{nl. coupl.}$, between the extra mode and the symmetric mode.

The spectrum remains free of unwanted resonances if the plasma frequency of the antisymmetric mode $\omega_- \equiv \sqrt{8 \beta E_C E_{J,-}}$ lies below the nearest integer multiple of $\omega_d$. However, if $\omega_-$ exceeds such a multiple, multiphoton resonances emerge just as in the buffer‑mode case. To assess their impact, we first choose parameters such that $\beta\approx16.1$, which sets $\omega_- \approx 2\pi \times \SI{24.43}{GHz}$, exceeding $2\omega_d$ by $2\pi \times \SI{54}{MHz}$. Since for small to moderate $g$ the resonances are weak, we probe a worst‐case scenario setting $g/2\pi = \SI{100}{MHz}$, thereby making the capacitive coupling the dominant interaction.

\begin{figure}[t!]
    \centering
    \includegraphics{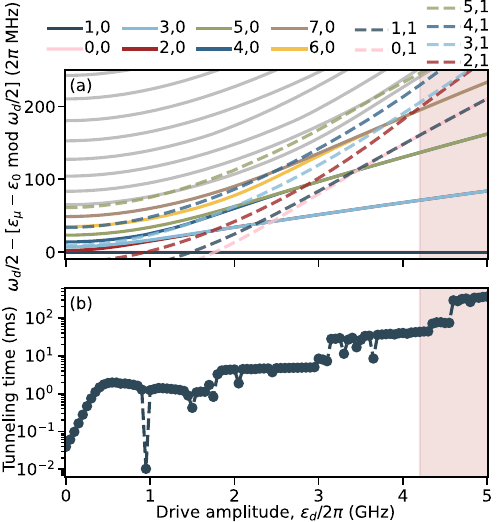}
    \caption{(a) Spectral kissing of the quasienergies in the presence of the antisymmetric Josephson-junction array mode, whose frequency lies above $2\omega_d$ by $2\pi \times \SI{54}{MHz}$. Solid lines indicate the lowest quasienergies corresponding to modes with zero excitation in the array's antisymmetric mode, while dashed lines correspond to modes with one excitation. No crossings are interrupting the spectral kissing of the relevant quasienergies above $\sim \SI{4.2}{GHz}$ indicated by the shaded area.
    (b) The corresponding tunneling time as a function of drive amplitude.
    }
    \label{fig:array_mode}
\end{figure}

\begin{figure}[t!]
    \centering
    \includegraphics{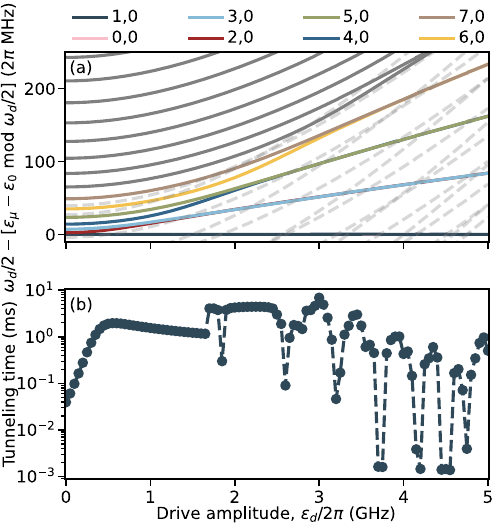}
    \caption{(a) Spectral kissing of the quasienergies when the Josephson-junction array mode is detuned by $2\pi\times\SI{330}{MHz}$ above $2\omega_d$. Solid lines indicate the lowest quasienergies corresponding to modes with zero excitation in the array's antisymmetric mode, while dashed lines correspond to some modes with one excitation and two excitations.
    (b) The corresponding tunneling time as a function of drive amplitude.
    }
    \label{fig:array_mode_330}
\end{figure}

As in the previous section, we first diagonalize the Hamiltonian in \cref{eq: array mode Hamiltonian} to obtain the dressed eigenstates $\{\ket{i_s, j_a}\}$, where $i_s$ and $j_a$ label excitations in the symmetric and antisymmetric modes, respectively. We then add the drive to the symmetric mode and compute the Floquet spectrum as a function of drive amplitude, shown in \cref{fig:array_mode}(a).
As with the case of the buffer mode, the pairwise spectral kissing among the quasienergies $\epsilon_{n_s, 0_a}$ is interrupted by avoided crossings with quasienergies $\epsilon_{m_s, 1_a}$. Here, for clarity, we have chosen to label the Floquet spectrum following the asymptotic modes across each crossing. We also note that, just like with the buffer mode, there are no selection rules: all crossings seen in \cref{fig:array_mode}(a) are actual avoided crossings.

This spectrum reveals two key differences compared to the case of the buffer mode. First, for drive amplitudes above $\sim \SI{4.2}{GHz}$ (shaded area), there are no avoided crossings disrupting the quasidegenerate pairs 0–1, 2–3, 4–5, and 5–6, which implies that the tunneling time should remain long throughout this range.
Second, at lower drive amplitudes, hybridization with the antisymmetric mode involves only low-lying symmetric states. In contrast, in the buffer-mode case, the Kerr-cat is excited to high-energy levels, resulting in significant leakage out of the metapotential double well.

Moving on to the dissipative dynamics, we use the same Floquet–Markov master equation as in \cref{sec:single mode}.  Since the drive port couples much more strongly to the symmetric mode than to the higher-energy modes \cite{Viola2015}, we assume only the symmetric mode to be lossy, coupled to the environment via the charge operator $\hat{n}$. We take the same environment spectral density and frequency-dependent temperature as in \cref{sec:single mode}.
\Cref{fig:array_mode}(b) shows the tunneling time as a function of drive amplitude, computed from the assignment error probability. As before, it exhibits a staircase-like increase interrupted by sharp dips associated with resonances in the Floquet spectrum. A prominent dip occurs at $\varepsilon_d/2\pi \approx \SI{1}{GHz}$ due to the resonance between $(1_s,0_a)$ and $(2_s,1_a)$. Crucially, for this set of parameters, the tunneling time does not collapse in the high-drive regime ($\varepsilon_d/2\pi \gtrsim \SI{4}{GHz}$, shaded area), in contrast to the buffer-mode case.
In this range, where resonances would otherwise be most detrimental, none involve the relevant almost-degenerate quasienergy pairs. As a result, the tunneling time increases with the drive amplitude without disruption.

However, increasing $\beta$ to $16.43$ leads to a detuning $(\omega_- - 2\omega_d)/2\pi \approx \SI{330}{MHz}$, and the array mode begins to play a role similar to the buffer mode analyzed in \cref{subsec: buffer mode spectral kissing,subsec: buffer tunneling time}. \Cref{fig:array_mode_330}(a) shows the spectral kissing, now interrupted by numerous quasienergies corresponding to states with one or two excitations in the array mode (dashed lines). Crucially, these resonances occur predominantly at large drive amplitudes, in contrast to the small-detuning case discussed above. As shown in \cref{fig:array_mode_330}(b), the tunneling time shows a series of dips which become more prominent with increasing drive amplitude, with multiphoton resonances become wider and more frequent.

It is important to note that for very small ground capacitances (small $g$, e.g., $\lesssim 2\pi \times \SI{10}{MHz}$), multiphoton resonances with the array modes are substantially suppressed, and their impact on the tunneling time is strongly attenuated.

Finally, the array mode considered here---with a resonance frequency near $2\omega_d$ and charge-coupled to the Kerr-cat qubit---plays a role analogous to the second harmonic of a $\lambda/2$ resonator buffer or of a 3D cavity box mode~\cite{Frattini2024,ding2025}. Consequently, all conclusions drawn for the array mode apply equally to the buffer mode’s second harmonic.

\section{Conclusion} \label{sec: conclusion}

We show that, within the single-mode approximation, the staircase pattern of the Kerr-cat qubit's tunneling time persists up to high drive amplitudes. In other words, in the absence of additional modes, the Kerr-cat qubit is intrinsically robust. This conclusion is reached by using a Floquet–Markov master equation that accounts for quasidegeneracies in the spectrum and treats nonperturbatively both the Josephson junction nonlinearity and the drive. In this framework, the plateaus in tunneling time as a function of drive amplitude arise from the coherent addition of dissipative transitions between quasidegenerate level pairs: at each spectral kissing point, tunneling between the wells of the metapotential is suppressed due to destructive interference between two relaxation pathways. Furthermore, we find that the tunneling time does not reliably indicate the Kerr-cat qubit's coherence time $T_Z$, which is determined from the initial decay of the logical Pauli operator defined as $\hat{X}_L = \ket{\beta+\alpha}\bra{\beta+\alpha} - \ket{\beta-\alpha}\bra{\beta-\alpha}$.

To reproduce the experimentally observed breakdown of the staircase pattern, we account for additional modes present in the implementation of the Kerr-cat qubit. Specifically, we consider the impact of the buffer mode through which the Kerr-cat is driven, the collective array modes of the double SNAIL used in the Kerr-cat qubit circuit, and the presence of a stray geometric inductance (\cref{appendix: inductance}). For parameters close to those of the experiments in Refs.~\cite{Frattini2024, dealbornoz2024}, we find that these additional modes give rise to multiphoton resonances in the quasienergy spectra. Such resonances are known to underlie ionization of highly-excited hydrogen atoms \cite{Breuer1989_Floquet_Ionization} and drive-induced transitions in transmon \cite{MIST_1, MIST_2,Shillito2022Dynamics,Cohen2023,Dumas2024,connolly:2025,dai:2025,Fechant2025,Wang2025, Mingkang2025} and fluxonium qubits \cite{Nesterov2024,singh:2025}. In the Kerr-cat, the resonances originate from the buffer and Josephson array modes, and are directly responsible for the breakdown of the staircase pattern. 
Importantly, because large tunneling times rely on near-degeneracies among higher excited levels, resonances involving those levels impact Kerr-cat coherence. This is to be contrasted with other qubits, such as the transmons and fluxoniums, where only resonances of the computational states directly affect the qubit. 
These observations come with a silver lining: the robustness of the Kerr-cat qubit in the absence of additional modes suggests that it should be possible to engineer the system in a way that restores this intrinsic protection. A key design element is the sign of the buffer–drive detuning. Placing the buffer above the drive frequency produces numerous resonances that directly impact the cat states and the relevant excited states below the double-well barrier, whereas tuning it below the drive shifts the resonances to levels containing buffer excitations. In latter case, suppressing thermal photons at the drive frequency---for instance, by using directional couplers as attenuators, as in Ref.~\cite{ding2025}---helps mitigate the impact of these unwanted resonances on the tunneling time. This asymmetry between positive and negative detuning arises from the negative anharmonicity of the double-SNAIL element. Beyond the sign of the detuning, another important factor is its absolute magnitude. Our analysis suggests that reducing the detuning from hundreds to tens of megahertz could be beneficial for the Kerr-cat in the large photon number regime, since the unwanted resonances occur primarily at low drive amplitudes, like in the case of the array mode at small detuning (c.f. \cref{fig:array_mode}). A smaller buffer-drive detuning has the additional advantage of increasing the drive delivered to the SNAILs for the same input power and a fixed buffer–SNAIL coupling.

Other neighboring modes in a circuit implementation beyond those considered here can similarly lead to a breakdown of the Kerr-cat qubit if their frequencies lie in proximity to the qubit frequency or its harmonics. For instance, the experiment in Ref.~\cite{Hajr2024} uses a band-stop filter whose internal modes frequencies sit around the Kerr-cat qubit frequency \cite{Ahmed_private_communication}. Our analysis suggests that similar multiphoton resonances---and the resulting saturation of the tunneling time---are expected at large drive amplitudes if one of these filter modes sits about $ \sim 100$~MHz above the qubit. Along the same lines, a spectator
qubit charge‑coupled to the primary Kerr‑cat qubit, as in the experiments of Refs.~\cite{Frattini2024,dealbornoz2024}, could also induce unwanted resonances, though only at much higher drive amplitudes owing to the typically small qubit–qubit coupling. Hence, a detailed characterization of the Kerr‑cat qubit’s microwave environment---such as that performed in Ref.~\cite{dai:2025} in context of high-frequency readout---is essential for identifying parameter regimes that restore the intrinsic robustness of the Kerr-cat qubit.
Moreover, mitigating intrawell leakage—through, for instance, leakage‐reduction techniques \cite{Putterman2022,Venkatraman:2024,Haxell:APS2025,Adinolfi:APS2025}—can further extend both the tunneling time and the coherence time $T_Z$, narrowing the discrepancy between them.

To further improve our predictions of the tunneling time and their comparison to experiment, it would be valuable to include $1/f$ noise by averaging over fluctuations of the external flux around the operating point. In practice, this involves computing the distribution of tunneling times obtained from the Floquet–Markov treatment for different flux realizations. This form of noise has been shown to affect primarily the initial rise of the tunneling time, leaving its behavior at higher drive amplitudes essentially unchanged~\cite{Frattini2024,Hajr2024}. As discussed in \cref{singlemode: Temperature and spectral densities}, having a more refined model---ideally informed by experiments---of the spectral density and effective temperature seen by the system would help more accurately match the height of the plateaus of the staircase. Moreover, although some avoided crossings may not affect the tunneling time when the system is not biased near a resonance induced by the drive, they still pose a problem for the adiabatic preparation of cat states. All avoided crossings affecting the Kerr-cat's quasienergies $\epsilon_{0_s,0_a}$ and $\epsilon_{1_s,0_a}$ must be traversed rapidly enough for the system to cross them diabatically. Landau-Zener transition probabilities can be directly computed from the Floquet spectrum using the methods presented here (see, e.g., Ref.~\cite{Dumas2024, Wang2025}), which could inform future experiments.

Finally, although our analysis focuses on the Kerr‐cat qubit, the methods developed here apply broadly to any strongly driven system with quasidegenerate spectra. Examples include superconducting circuits under subharmonic drives and subject to dissipation, as well as protected qubits operating in double‐well potentials, where quasidegeneracies similarly govern coherence and tunneling dynamics.

\section{Acknowledgments}
The authors are grateful to Manuel Muñoz-Arias, Nicholas E. Frattini, Benjamin L. Brock, Rodrigo G. Cortiñas and Max Schäfer for helpful discussions. This material is based upon work supported by the U.S. Department of Energy, Office of Science, National Quantum Information Science Research Centers, Quantum Systems Accelerator. Additional support is acknowledged from NSERC, the Ministère de l’Économie et de l'Innovation du Québec, the Fonds de recherche du Québec – Nature et technologie and the Canada First Research Excellence Fund, and the National Agency for Research and Development (ANID) through FONDECYT Postdoctoral Grant No.~3250130.

\let\oldaddcontentsline\addcontentsline
\renewcommand{\addcontentsline}[3]{}
\bibliography{articles}
\let\addcontentsline\oldaddcontentsline


\appendix


\section{Numerical details of the Floquet analysis} \label{appendix: single mode details}

\begin{figure}[t]
    \centering
\includegraphics{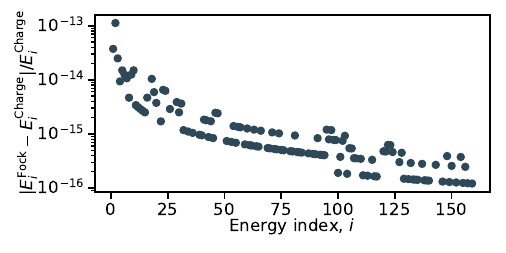}
\caption{
Relative error in the energies obtained from numerical diagonalization of the double-SNAIL Hamiltonian $\hat H_s$ (see \cref{eq:doublesnail_singlemode}), comparing results from the Fock basis and the charge basis. Convergence with respect to the maximum charge ($n_\text{max} = 400$) in $\hat n = \sum_{n=-n_\text{max}}^\text{max} \ket{n}\bra{n}$ is verified independently. The Fock basis is truncated at $n_\text{cut} = 250$. As is typical for periodic-potential Hamiltonians, diagonalization in the Fock basis does not converge with increasing $n_\text{cut}$; therefore, comparison with the charge basis is necessary to assess accuracy.
}
\label{fig:Fock vs charge comparison}
\end{figure}

In this Appendix, we provide additional numerical details supporting the simulations presented in Sec.~\ref{sec:single mode} of the main text. The single-mode Hamiltonian $\hat{H}_s$ introduced in \cref{singlemode_Hamiltonian} exhibits a $12\pi$ phase periodicity and should ideally be diagonalized in either the charge or phase basis. However, for experimentally relevant parameters, the potential wells are sufficiently deep that diagonalization in the Fock basis provides a spectrum nearly identical—up to numerical precision—to that obtained in the charge basis, at least for the lowest $\sim 160$ energy levels, see \cref{fig:Fock vs charge comparison}. Given the convenience of the Fock basis for phase-space representations, we adopt it throughout this work.

Our numerical simulations employ a Hilbert space truncated to 250 Fock states for diagonalization, from which the lowest 160 energy levels are retained for Floquet-state tracking. The relatively large Hilbert space size is necessary due to the small Kerr nonlinearity of the double-SNAIL mode, $K/2\pi = 1.18$ MHz. This weak nonlinearity allows the potential to support approximately 530 bound states, nearly all of which---except those close to the top of the potential well
---experience a positive ac-Stark shift.  Truncating below this number introduces an artificial negative Stark shift in the highest retained levels, causing resonances with lower energy levels
as the drive amplitude increases. Such truncation effects can ultimately impact the low-lying levels critical for Kerr-cat physics. For instance, retaining only 120 levels introduces truncation errors for drive amplitudes exceeding $\varepsilon_d/2\pi \approx 8.5$ GHz. Throughout this work, we consistently retain 160 levels, which we have verified to be sufficient for drive amplitudes up to at least $\varepsilon_d/2\pi = 10$~GHz. A more thorough discussion on truncation effects in strongly driven systems can be found in Ref.~\cite{Breuer1989_Floquet_Ionization}.

\begin{figure}[t!]
    \centering
\includegraphics{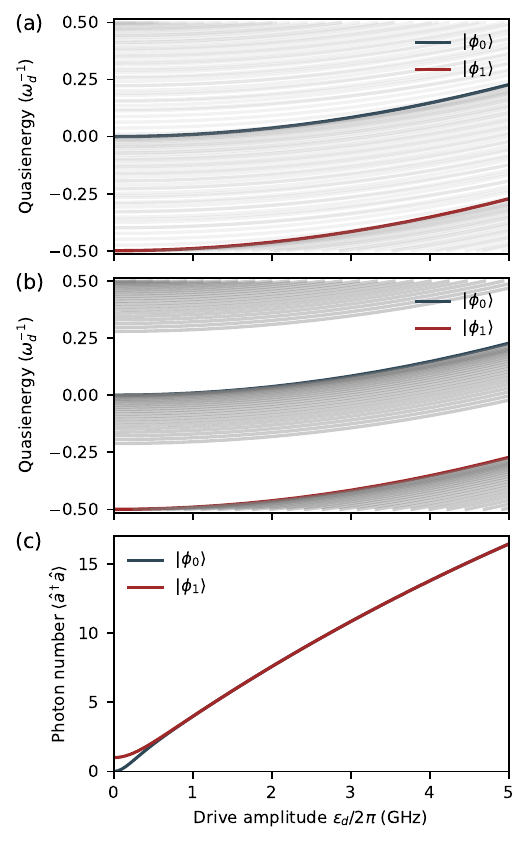}
\caption{ (a) Floquet quasienergies of the lowest 100 modes plotted as a function of drive amplitude. The lowest Floquet modes $\mu=0$ and $\mu=1$ are indicated by dark blue and red lines, respectively. (b) Same plot for the lowest 50 modes, showing more clearly that the spectrum separates into two bands centered at $0$ and $-\omega_d/2$ (mod $\omega_d$), corresponding to even and odd states. (c) Average photon number $\langle\hat{a}^\dagger\hat{a}\rangle$ in the $\mu=0$ (dark blue) and $\mu=1$ (red) Floquet modes, shown as a function of drive amplitude.}
\label{fig:fullspectrum_photonnumber}
\end{figure}

In \cref{fig:fullspectrum_photonnumber}(a) we plot the Floquet spectrum as a function of drive amplitude keeping the first 100 modes and highlighting the first two Floquet modes in dark blue and red, respectively. When restricting to the lowest 50 modes, see \cref{fig:fullspectrum_photonnumber}(b), we see clearly that the spectrum shows two distinct bands: one centered around zero energy modulo $\omega_d$, corresponding to the even Floquet modes, and another centered near $-\omega_d/2$ modulo $\omega_d$, associated with the odd Floquet modes. In \cref{fig:fullspectrum_photonnumber}(c), we present the average photon number $\langle \hat{a}^\dagger \hat{a} \rangle$ for the two lowest Floquet modes, corresponding to the cat states of interest. This allows mapping between drive amplitude and photon number (in the laboratory frame) within the cat-state manifold. Also, as expected, the even (blue line) cat state emerges from the 0-photon Fock state while the off (red line) cat states from the 1-photon state.

Having computed the Floquet spectrum as a function of drive amplitude, we can relate it to quantities measured experimentally in Refs.~\cite{Frattini2024, Venkatraman_deltacat, Hajr2024, ding2025}. In these experiments, an additional weak probe tone described by the Hamiltonian $\delta \hat{H}(t)=\zeta \cos(\omega_s t)\hat{n}\equiv \delta\Omega(t)\hat{n}$ at frequency $\omega_s$ is used alongside the two-photon drive. The oscillator field is monitored via heterodyne detection, from which the tunneling time is extracted \cite{Frattini2024,dealbornoz2024,ding2025}; see \cref{subsec: Results on tunneling single mode} and~\cref{appendix: time dynamics}. Starting from coherent states $\ket{\beta\pm \alpha}$, transitions induced by the probe drive outside the metapotential well manifest as  heterodyne signal amplitude corresponding to the opposite phase-space region ($\beta\mp \alpha$) \cite{Frattini2024, dealbornoz2024, ding2025}. Here, we show that these transitions rates and thus the measured heterodyne signal are directly proportional to the Floquet matrix elements $X_{\mu \nu k}$. Indeed,  writing the total Hamiltonian as
\begin{equation}
\hat{H}_{\text{tot}}(t) = \hat{H}(t) + \delta\hat{H}(t),
\end{equation}
with $\hat{H}(t)$ defined in Eq.\eqref{singlemode_Hamiltonian} and following Ref.~\cite{Bilitewski2015}, the transition amplitude from the coherent state $\ket{\beta+\alpha}\approx(\ket{\phi_0}+\ket{\phi_1})/\sqrt{2}$ to Floquet mode $\ket{\phi_\mu(t)}$ ($\mu\neq 0,1$), at first order of $\delta \hat{H}$, is given by
\begin{align}
\begin{split}
|\langle\phi_\mu(t)|\psi(t)\rangle| &=\frac{1}{\sqrt{2}}\left|\sum_k X_{\mu 0 k}\int_0^t ds\: \delta\Omega(s)e^{i\Delta_{\mu 0 k}s}\right. \\
&\quad+\left.\sum_k X_{\mu 1 k}\int_0^t ds\: \delta\Omega(s)e^{i\Delta_{\mu 1 k}s}\right|,
\end{split}
\end{align}
where $\ket{\psi(t)} = \mathcal{T}e^{-i\int_0^t ds \hat{H}_\text{tot}(s)}|\beta+\alpha \rangle$ is the instantaneous state.
Thus, the probe drives transitions only when $\omega_s \approx \Delta_{\mu 0k}$ or  $\omega_s \approx \Delta_{\mu 1k}$, with rates proportional to $X_{\mu 0k}$ and $X_{\mu 1k}$, respectively.
Since the readout of the Kerr-cat qubit relies on discriminating well occupancy (left or right in phase space), only those transitions that mediate interwell tunneling produce an enhanced signal; states that undergo spectral kissing no longer induce tunneling and thus become “invisible” to this measurement scheme \cite{Frattini2024,ding2025}.

\section{Partial secular Floquet Markov master equation} \label{appendix: master equation}

In this appendix, we derive the master equation used throughout this work by extending the standard Floquet–Markov formalism \cite{Blumel:1991,Kohler:1997,Grifoni:1998} to accommodate quasidegenerate spectra. We begin from the total system–bath Hamiltonian
\begin{equation}
    \hat{H}_{SB}(t) = \hat{H}(t)+\sum_i \omega_i \hat{b}^\dagger_i\hat{b}_i+ \hat{p} \sum_i g_i \hat{p}_i,
\end{equation}
where $\hat H(t)$ is defined in \cref{singlemode_Hamiltonian} of the main text, and $\hat{p}= i(\hat{a}^\dagger-\hat{a})=\hat{n}/n_{\textrm{zpf}}$ and $\hat{p}_i=i(\hat{b}^\dagger_i-\hat{b}_i)$ are the charge operators of the double‐SNAIL and each bosonic bath mode, respectively. This Hamiltonian thus describes charge–charge coupling between the driven SNAIL circuit and its electromagnetic environment; see \cref{fig:circuits}(a) and (c). Following Ref.~\cite{BRE:2002}, and working in the interaction picture under the Born–Markov approximation, the master equation can be expressed as 
\begin{align} \label{Nakajima-Zwanzig}
    \begin{split}
    \frac{d\hat{\rho}^{I}}{dt} =& \int\limits_0^{+\infty} d\tau \sum_i g_i^2 \\
    &\times \left[ \langle  \hat{p}_i^{I}(t)  \hat{p}_i^{I}(t-\tau)\rangle_B \left(  \hat{p}^{I}(t-\tau)\hat{\rho}^{I} \hat{p}^{I}(t)\right.\right.\\
        &\quad \left.\left. -  \hat{p}^{I}(t)  \hat{p}^{I}(t-\tau)\hat{\rho}_I \right)
        \right] + \text{h.c}.
    \end{split}
\end{align}
Here, any operator $\hat{O}$ in the interaction picture is denoted as $\hat{O}^{I} = \mathcal{U}^\dag (t)\hat{O} \mathcal{U}(t)$ with
\begin{equation}
\begin{split}
\mathcal{U}(t) &=\mathcal{T} \exp\left[-i\int _0^t (\hat{H}(s)+\sum_i \omega_i \hat{b}^\dagger_i\hat{b}_i)ds\right] \\
&= \hat U(t) e^{-it \sum_i \omega_i \hat b_i^\dag \hat b_i}.
\end{split}
\end{equation}
Assuming the bath to be in a thermal state, the correlation functions are 
\begin{equation} \label{correlation_func}
\begin{split}
&\langle  \hat{p}_i^{I}(t)  \hat{p}_i^{I}(t-\tau)\rangle_B =\\
&n_{\textrm{th}}(\omega_i)e^{i\omega_i \tau} + (n_{\textrm{th}}(\omega_i)+1)e^{-i\omega_i \tau},
\end{split}
\end{equation}
where $n_{\textrm{th}}(\omega_i)= 1/(e^{\hbar \omega_i/(k_BT(\omega_i))}-1)$ is the average occupation of bath mode $i$.

\Cref{Nakajima-Zwanzig} can be applied to any time-dependent system, as long as the dissipation is weak and exhibits short-time correlations. Given the periodicity of the Hamiltonian, $\hat{H}(t+T) = \hat{H}(t)$, we can now simplify the master equation by decomposing the charge operator into the Floquet-mode basis at time $t$ as
\begin{align}
    \begin{split} \label{eq: charge_op}
        \hat{p} &= \sum_{\mu \nu} \bra{\phi_\mu (t)}\hat{p}\ket{\phi_\nu(t)}\ket{\phi_\mu(t)}\bra{\phi_\nu (t)}\\
        &= \sum_{\mu \nu k}X_{\mu \nu k} e^{+ik\omega_dt} \ket{\phi_\mu(t)}\bra{\phi_\nu (t)},
    \end{split}
\end{align}
where the Fourier coefficients $X_{\mu \nu k}$ are defined in Eq.\eqref{eq:matrix_elements}. Since the Floquet modes at time $t$ are related to $t=0$ by $\ket{\phi_\mu(t)} = e^{i\epsilon_\mu t} \hat U(t) \ket{\phi_\mu}$, in the interaction picture the decomposition becomes
\begin{equation} \label{eq:charge_op_interaction}
    \hat{p}^{I}= \sum_{\mu \nu k}X_{\mu \nu k} e^{+i\Delta_{\mu \nu k} t} \ket{\phi_\mu}\bra{\phi_\nu},
\end{equation}
where $\Delta_{\mu \nu k} =\epsilon_\mu-\epsilon_\nu+k\omega_d$.
Substituting Eqs.\eqref{correlation_func} and \eqref{eq:charge_op_interaction} into Eq.\eqref{Nakajima-Zwanzig}, and neglecting the Lamb–shift terms that simply renormalize the coherent dynamics, we obtain
\begin{align} \label{eq: redfield}
    \begin{split}
        &\frac{d\hat{\rho}^{I}}{dt} = \frac{1}{2}\sum_{\mu \nu k}\sum_{\mu'\nu' k'}X_{\mu \nu k}X_{\mu' \nu' k'}e^{i(\Delta_{\mu \nu k}-\Delta_{\mu'\nu'k'})t}\times\\
    & J(\Delta_{\mu \nu k})n_{\textrm{th}}(\Delta_{\mu \nu k})\left[\ket{\phi_\mu}\bra{\phi_\nu}\hat{\rho}^{I}\ket{\phi_{\nu'}}\bra{\phi_{\mu'}}\right.\\-&\left.\ket{\phi_{\nu'}}\langle \phi_{\mu'}|\phi_{\mu}\rangle \bra{\phi_\nu}\hat{\rho}^{I}\right]+ \frac{1}{2}\sum_{\mu \nu k}\sum_{\mu'\nu' k'}X_{\mu \nu k}X_{\mu' \nu' k'}\times\\
     & e^{i(\Delta_{\mu \nu k}-\Delta_{\mu'\nu'k'})t}J(-\Delta_{\mu \nu k})(n_{\textrm{th}}(-\Delta_{\mu \nu k})+1)\times\\&\left[\ket{\phi_\mu}\bra{\phi_\nu}\hat{\rho}^{I}\ket{\phi_{\nu'}}\bra{\phi_{\mu'}}-\ket{\phi_{\nu'}}\langle \phi_{\mu'}|\phi_{\mu}\rangle\bra{\phi_\nu}\hat{\rho}^{I}\right] + \text{h.c.}\\
    \end{split}
\end{align}
Here, the bath’s spectral density is defined as 
\begin{equation}
    J(\omega)=2\pi \sum_i g_i^2\delta(\omega-\omega_i),
\end{equation}
which vanishes for $\omega<0$. To cast Eq.~\eqref{eq: redfield} into Lindblad form, one typically invokes the full secular approximation \cite{Oelschlagel1993,Grifoni:1998}, retaining only terms with $( \mu',\nu',k')=(\mu,\nu,k) $. This approximation is valid only if different Floquet transition frequencies are well separated, i.e.,
\begin{equation} \label{full secular}
    |\Delta_{\mu' \nu' k'}-\Delta_{\mu \nu k}|\gg |X_{\mu'\nu'k'} X_{\mu \nu k} | \kappa(\Delta_{\mu \nu k})
\end{equation}
for $(\mu', \nu' ,k')\neq (\mu , \nu , k)$ with $\kappa(\Delta_{\mu \nu k})$ defined in \cref{kappa}. However, in the Kerr-cat qubit, the spectrum exhibits multiple near-degeneracies (“spectral kissing”, see \cref{fig:metapotential}(b)), violating \cref{full secular}. Instead, we employ the partial secular approximation: transitions satisfying Eq.~\eqref{full secular} are discarded, but those with 
\begin{equation} \label{partial secular}
    |\Delta_{\mu' \nu' k'}-\Delta_{\mu \nu k}|\lesssim |X_{\mu'\nu'k'} X_{\mu \nu k} | \kappa(\Delta_{\mu \nu k})
\end{equation}
are retained, see for instance Refs.~\cite{Jeske2015,Farina:2019,Cattaneo:2019,Trushechkin2021, Jurcevic2022}. For these quasidegenerate pairs we approximate $e^{i(\Delta_{\mu \nu k}-\Delta_{\mu'\nu'k'})t}\approx 1$, consistent with the Born–Markov approximation (i.e., neglecting terms $\mathcal{O}(g_i^4)=\mathcal{O}(\kappa^2)$ \cite{Jeske2015,Farina:2019, Cattaneo:2019,Trushechkin2021}). 

For compactness, we group transitions into classes indexed $[\mu, \nu, k]$, where $(\mu', \nu', k') \in [\mu, \nu, k]$ if $|\Delta_{\mu\nu k} - \Delta_{\mu'\nu'k'}| \lesssim |X_{\mu'\nu'k'} X_{\mu \nu k} |\kappa(\Delta_{\mu\nu k})$. In terms of these classes, the  master equation in the interaction‐picture becomes
\begin{equation}
    \frac{d\hat{\rho}^{I}}{dt}=\mathcal{L}\hat{\rho}^{I},
\end{equation}
with the Lindbladian
\begin{align}
\mathcal{L} = \sum_{[\mu,\nu,k]} \kappa(\Delta_{\mu\nu k}) \mathcal{D}\left[\sum_{(\mu',\nu',k') \in [\mu,\nu,k]} X_{\mu'\nu'k'} \ket{\phi_{\mu'}}\bra{\phi_{\nu'}}\right]. 
\end{align} 
 In this expression, the first sum runs over the distinct classes of transitions, and the second sum runs over quasidegenerate transitions within a given class. 

By construction, this Lindbladian naturally accounts for strong drive effects through the Floquet quasienergies $\Delta_{\mu\nu k}$ and Floquet modes $\ket{\phi_{\mu}}$, including all perturbative and nonperturbative corrections beyond the rotating‐wave approximation \cite{petrescu2020, Verney2019, Cohen2023, Burgleman2022,Carde2025}. Moreover, the coefficients $X_{\mu\nu k}$ represent transition amplitudes between Floquet modes  $\ket{\phi_\nu}\to\ket{\phi_\mu}$ assisted by $|k|$ photons from the drive. Although the Lindbladian itself connects time-independent Floquet modes, the full temporal dependence of the driven dynamics is encoded in these coefficients through the time-dependent Floquet modes, see \cref{eq:matrix_elements}. Thus, the master equation captures not only the stroboscopic evolution but also the fast micromotion dynamics.

In the absence of quasidegeneracies—i.e., when each class $[\mu,\nu,k]$ contains only a single transition—the master equation reduces to the standard Floquet-Markov secular form. Crucially, when quasidegeneracies occur, the resulting dissipator 
\begin{align} \mathcal{D}\left[\sum_{(\mu',\nu',k') \in [\mu,\nu,k]} X_{\mu'\nu'k'} \ket{\phi_{\mu'}}\bra{\phi_{\nu'}}\right] 
\end{align}
captures interference effects among transitions within the same class. This contrasts sharply with the incoherent sum of dissipators \begin{align} 
\sum_{(\mu',\nu',k')} |X_{\mu'\nu'k'}|^2 \mathcal{D}\left[ \ket{\phi_{\mu'}}\bra{\phi_{\nu'}} \right], 
\end{align} 
which results from applying the usual secular approximation.

It is instructive to compare our partial-secular Floquet–Markov equation \cref{Lindbladian} with other extensions of the Lindblad formalism for quasidegenerate spectra (e.g., Refs.~\cite{Kirifmmode:2018,Kleiherbers:2020,Mozgunov:2020,McCauley:2020,Nathan2020}). In our treatment, the partial-secular approximation necessarily introduces an explicit threshold for grouping transitions, see \cref{partial secular}; in our regime of parameters, we define two transitions $\Delta_{\mu \nu k}$ and $\Delta_{\mu' \nu' k'}$ as quasidegenerate when $|\Delta_{\mu \nu k}-\Delta_{\mu' \nu' k'}|/2\pi<100$ kHz, i.e., less than ten times the qubit decay rate. By contrast, alternative approaches avoid any fixed threshold but at the expense of retaining an explicitly time-dependent Lindbladian even in the interaction picture—making relevant quantities such as the Lindbladian gap not well-defined. Here, the Lindbladian in \cref{Lindbladian} remains time‐independent. Moreover, when expressed in the Floquet‐mode basis $\{\ket{\phi_\mu}\}$, it is highly sparse, making both its diagonalization and the simulation of the driven dynamics efficient in memory usage and computation time.

\section{Calculation of the tunneling time} \label{appendix: time dynamics}

To obtain the tunneling time as is done experimentally via heterodyne detection we use an ideal likelihood discriminator. 
First, let us denote $\hat \rho_\pm(t)$ the evolved state at time $t$ where the initial condition is $\hat \rho_\pm(0) = \ket{\beta\pm\alpha} \bra{\beta\pm \alpha}$. At each time step, we obtain the Husimi-Q distributions of the evolved states denoted by $Q_\pm[\gamma](t)=\bra{\gamma}\hat \rho_\pm(t) \ket{\gamma}/\pi$ with $\ket{\gamma}$ a coherent state. We then construct the log-likelihood function $\lambda(t) = \ln[Q_+(t)/Q_-(t)]$. 

When the system is initialized in $\hat \rho_+(0)$, the probability of correctly assigning the outcome ``$+$'' at time $t$ from a single-shot heterodyne experiment is
\begin{equation}
P(+|+) = \int d^2 \gamma \, \Theta[\lambda(t)] Q_+(t),
\end{equation}
where $\Theta$ is the Heaviside function. Likewise, the probability of correctly assigning the outcome ``$-$'' from a single-shot readout when the state is initialized in $\hat \rho_-(0)$ is
\begin{equation}
P(-|-) = \int d^2 \gamma \, \Theta[-\lambda(t)] Q_-(t).
\end{equation}
The (averaged) assignment error is therefore $\frac{1}{2}[P(-|+) + P(+|-)] = 1 - \frac{1}{2}[P(+|+) - P(-|-)]$. As in the experiments of  Refs.~\cite{Frattini2024,Hajr2024,dealbornoz2024,ding2025}, we extract the tunneling time by fitting a single exponential decay to 
\begin{equation}
X_+(t) = P(+|+)- P(-|+) = 1-2 P(-|+) 
\end{equation}
for the initial condition $\hat \rho_+(0)$, and 
\begin{equation}
X_-(t) = P(+|-)- P(-|-) = 2 P(+|-)-1   
\end{equation}
for $\hat \rho_-(0)$. Numerically, we find that these two tunneling time estimates are almost indistinguishable, consistent with the aforementioned experimental observations \cite{Frattini2024,Hajr2024}. In the main text, we report their average.

\section{Approximate Floquet Hamiltonian} \label{appendix: SW on Shirley}
 
In the main text, we compute the Floquet spectrum and Floquet matrix elements $X_{\mu \nu k}$ by exact numerical diagonalization of the evolution operator over one period.
Here, we derive approximate analytical expressions for the Floquet Hamiltonian and matrix elements. Although not strictly required for our conclusions, this analysis reveals how key quantities scale with drive amplitude and the order of nonlinearity.

This appendix proceeds as follows. First, we move to a displaced and rotating frame, and identify the small parameter for the perturbative treatment. We then map the time-dependent periodic Hamiltonian onto a time-independent Hamiltonian using the Shirley (or Fourier) space~\cite{Shirley1965Solution}. Next, we obtain an analytical Floquet Hamiltonian via a perturbative block-diagonalization using the standard time-independent Schrieffer–Wolff (SW) transformation~\cite{Bravyi2011}. Finally, we return to the laboratory frame, obtain approximate expressions for the Floquet quasienergies and modes, and evaluate the corresponding Floquet matrix elements $X_{\mu \nu k}$, explicitly showing their interference structure. We note that an alternative approach avoids the mapping to Shirley space and instead applies a time-dependent SW transformation directly in the original time domain~\cite{Nakagawa2014,Bukov2015,Theis2017,petrescu2020,Jaya2022,Petrescu2023}. This technique has been recently used to analyze $T_1$ degradation in a dissipatively stabilized cat qubit~\cite{Carde2025}, and to derive effective Lindblad operators for the Kerr-cat~\cite{Venkatraman:2024}.

We begin by expanding $\hat H_s$ from \cref{eq:doublesnail_singlemode} in a power series around the minimum of its potential, yielding~\cite{Frattini2017}
\begin{equation}
\hat H_s = \omega_0 \hat a^\dag \hat a + \sum_{n\geq 3} \frac{g_n}{n}(\hat a + \hat a^\dag)^n,
\end{equation}
where the coefficients are given by $g_n = E_J c_n \varphi_\text{zpf}^n/(n-1)!$. Here, $c_n$ denotes the $n$th derivative of the cosine potential evaluated at the minimum. Using the relation $E_J = \omega_0 / (2 c_2 \varphi_\text{zpf}^2)$, we can rewrite the coefficients as $g_n = \omega_0 c_n \varphi_\text{zpf}^{n-2}/[2c_2 (n-1)!]$. This makes explicit that the coefficients of the nonlinearity scale as $g_3 \sim \mathcal{O}(\omega_0 \varphi_\text{zpf})$, $g_4 \sim \mathcal{O}(\omega_0 \varphi_\text{zpf}^2)$, and so on. These are naturally small parameters governed by the dimensionless zero-point fluctuations of the phase, $\varphi_\text{zpf}$. The perturbative expansion introduced below is organized in powers of $\varphi_\text{zpf}$.

We then add the charge drive to $\hat H_s$, resulting in the time-dependent Hamiltonian
\begin{equation}
\begin{split}
\hat H(t) =& \omega_0 \hat a^\dag \hat a + \sum_{n\geq 3} \frac{g_n}{n}(\hat a + \hat a^\dag)^n \\
&+ \frac{i\varepsilon_d}{2\varphi_\text{zpf}} \cos(\omega_d t + \lambda) (\hat a^\dag - \hat a),
\end{split}
\end{equation}
where we have included a phase delay $\lambda$ for generality. Following Ref.~\cite{Frattini2024}, we eliminate the drive term and fold it into the nonlinear part of the potential by transforming the Hamiltonian with a time-dependent displacement operator $\hat D[\beta(t)]=\exp[\beta(t) \hat a^\dag - \bar \beta(t) \hat a]$, where the displacement amplitude is
\begin{equation}
\begin{split}
\beta(t) =& \frac{\varepsilon_d}{4i \varphi_\text{zpf}} \left(\frac{e^{-i(\omega_d t+ \lambda)}}{\omega_0 - \omega_d} + \frac{e^{i(\omega_d t+\lambda)}}{\omega_0 + \omega_d} \right) \\
=& \frac{\omega_d + \omega_0}{2\omega_d} \Pi e^{-i\omega_d t} + \frac{\omega_d - \omega_0}{2\omega_d} \bar \Pi e^{+i\omega_d t}.
\end{split}
\end{equation}
Here, the coefficient $\Pi$ is given by
\begin{equation} \label{eq: General Pi}
\Pi =\frac{i\varepsilon_d \omega_d e^{-i\lambda}}{2 \varphi_\mathrm{zpf} (\omega_d^2 - \omega_0^2)},
\end{equation}
which becomes useful below. In this section of the appendix, we use an overbar to denote complex conjugation.

We further move to a rotating frame at frequency $\omega_d/2$. The Hamiltonian in the displaced and rotating frame takes the form
\begin{equation} \label{eq: initial Hamiltonian for SW in Shirley space}
\begin{split}
\hat H_d(t) \equiv& \hat R^\dag \left[ \hat D^\dag \hat H \hat D - i \hat D^\dag \dot{\hat D} \right] \hat R -i \hat R^\dag \dot{\hat R}\\
=&(\omega_0 - \omega_d/2)\hat a^\dag \hat a \\
&+ \sum_{n\geq 3} \frac{g_n}{n}(\hat a e^{-i\frac{\omega_d}{2}t} + \Pi e^{-i\omega_d t} + \text{h.c.})^n,
\end{split}
\end{equation}
where $\hat R(t) = \exp[-i(\omega_dt/2) \hat a^\dag \hat a]$. Crucially, in this frame, all energy scales become small compared to the new fundamental drive frequency $\omega_d/2$ in the oscillating terms. Moreover, for all parameter regimes considered in this work, the displacement amplitude satisfies $|\Pi| \lesssim 1$.

A straightforward rotating-wave approximation (RWA) applied to this Hamiltonian yields two second order in $\varphi_\text{zpf}$ a squeezed Kerr oscillator, as in \cref{eq:H squeezed Kerr}, with two-photon drive amplitude $\varepsilon_2 = g_3 \Pi$ and self-Kerr coefficient $K = (3/2)g_4$. The drive frequency must be tuned to account for the ac-Stark shift, satisfying $\omega_d/2 = \omega_0 + 3g_4(1 + 2|\Pi|^2)$. However, while this RWA captures the basic structure, a more accurate description can be obtained via a SW perturbative expansion. Once we obtain the approximate Floquet Hamiltonian in this displaced and rotated frame, we construct the Floquet Hamiltonian in the laboratory frame by undoing the rotation and displacement transformations, as explained below in this section.

The usual time-dependent SW transformation for periodically driven systems consists of finding a hermitian generator $\hat G(t)$ that makes the Hamiltonian time-independent in the transformed frame~\cite{Bukov2015}. Denoting the resulting Hamiltonian as $\hat H_F^d$, this condition is expressed as 
\begin{equation} \label{eq: non-stroboscopic Floquet Hamiltonian}
e^{i\hat G(t)}\hat H_d(t)e^{-i \hat G(t)} - i e^{i\hat G(t)} \partial_te^{-i \hat G(t)} = \hat H_F^d.
\end{equation}
While the original Hamiltonian in \cref{singlemode_Hamiltonian} is periodic with period $T = 2\pi / \omega_d$, the displaced and rotated Hamiltonian $\hat H_d(t)$ is $2T$-periodic, and so the generator $\hat G(t)$ is $2T$-periodic as well. Importantly, $\hat G(t)$ does not vanish at stroboscopic times, $\hat G(2nT) \ne 0$ (where $n$ is an integer), and is thus referred to as the \textit{non-stroboscopic kick operator}. The resulting time-independent Hamiltonian $\hat H_F^d$ is known as the \textit{non-stroboscopic Floquet Hamiltonian}~\cite{Bukov2015}. A feature of this formalism is that $\hat H_F^d$ is independent of a choice of an initial time. This is in contrast to the conventional (stroboscopic) Floquet Hamiltonian, which is obtained from the one-period evolution operator via $(i/T) \ln \hat U(T+t_0, t_0)$, and thus depends explicitly on the choice of initial time $t_0$.

Rather than working directly with the time-dependent expression in \cref{eq: non-stroboscopic Floquet Hamiltonian}, we find it convenient to construct the SW generator in the Shirley space~\cite{Shirley1965Solution}. In this extended Hilbert space, the operator $\hat H_d(t) - i\partial_t$ is mapped to the Shirley Hamiltonian
\begin{equation} \label{eq: Shirley equation}
\begin{split}
\hat H_S =& \frac{\omega_d}{2} \hat m \otimes \hat{\mathbf{1}} + \sum_{n=-\infty}^{\infty} \hat{\mathbf{1}}_{-n} \otimes \hat H_{n},
\end{split}
\end{equation}
where the $\hat H_n$ are the Fourier components of the original time-dependent Hamiltonian in the displaced and rotated frame,
\begin{equation} 
\hat H_d(t) =\sum\limits_{n=-\infty}^\infty \hat H_{n}e^{in\frac{\omega_d}{2}t},
\end{equation}
In this formalism, the full Hamiltonian $\hat H_S$ acts on a tensor product of two spaces: the infinite-dimensional replica space (first factor) and the original system Hilbert space (second factor). In \cref{eq: Shirley equation}, $\hat m = \sum_{m=-\infty}^{+\infty} m \ket{m}\bra{m}$ plays the role of a “photon-number” operator in the replica space, except that its spectrum extends over all integers and is therefore unbounded from below (no vacuum state). The operators $\hat{\mathbf{1}}_n = \sum_{m} \ket{m}\bra{m+n}$ act as ladder operators that shift the replica index by $n$. The 
$n$th Fourier component (harmonic) of the drive frequency couples sectors whose replica indices differ by $n$ via these shifts.

The generator is mapped to Shirley space as
\begin{equation} \label{eq: Sw generator in Shirley}
\hat G(t) = \sum_n \hat G_n e^{i n (\omega_d/2)t} \to \hat G = \sum_n \hat{\mathbf 1}_{-n} \otimes \hat G_n.
\end{equation}
where $\hat G = \hat G^\dagger$ and $\hat G_n^\dagger = \hat G_{-n}$ to ensure hermiticity. As in standard SW perturbation theory, we construct the generator order by order, requiring that the transformed Hamiltonian be diagonal. In this case, however, we only demand diagonalization in the replica index, not in the full Shirley space. Specifically, we enforce the structure
\begin{equation} \label{eq: rotated Shirley Hamiltonian}
\hat H_{S}' = e^{i\hat G} \hat H_S e^{-i\hat G} \stackrel{!}{=} \frac{\omega_d}{2} \hat m\otimes \hat{\mathbf{1}} + \hat{\mathbf{1}} \otimes \hat H_F^d,
\end{equation}
where the right-hand side is the Shirley-space representation of a time-independent Hamiltonian $\hat H_F^d$ in the Hilbert space. This ensures that $\hat H_F^d$ captures the effective dynamics up to the desired order.

Let us now identify the small parameter that controls the perturbative expansion. In the standard SW transformation, the expansion parameter is given by the ratio of the coupling strength between subspaces to their energy detuning. In our case, the smallest detuning in the diagonal part of $\hat H_S$ is set by $\omega_0$, while the off-diagonal couplings have magnitude $g_n$, so the relevant small parameter is $g_n / \omega_0 \propto \varphi_\text{zpf}^{n-2}$. Crucially, since $g_{n+1}/ g_{n} \propto \varphi_\text{zpf}$, we can treat $\varphi_\text{zpf}$ as the single small parameter controlling the entire expansion. This greatly simplifies the analysis, avoiding the need to track each $g_n$ individually as an independent parameter at every order in perturbation theory.

We thus expand each operator $\hat{H}_n$ and $\hat{G}_n$ in powers of the small parameter $\varphi_{\text{zpf}}$ as
\begin{equation}
\begin{split}
\hat H_n =& \sum\limits_{k\geq 0} \hat H_n^{(k)} \\
\hat G_n =& \sum\limits_{k\geq 1} \hat G_n^{(k)},
\end{split}
\end{equation} 
where the superscript $(k)$ denotes terms of order $\varphi_{\text{zpf}}^k$. Note that the zeroth-order term of $\hat{G}_n$ vanishes, $\hat{G}_n^{(0)} = 0$, since the SW transformation needs to be reduced to the identity at zero order.

We expand $\hat H_S'$ in \cref{eq: rotated Shirley Hamiltonian} as
\begin{equation}
\begin{split}
&\hat H_S'=\\
&\left( \hat H_S^{(0)} \right) \\
&+\left(\hat H_S^{(1)} + i [\hat G^{(1)}, \hat H_S^{(0)}]\right) \\
&+ \left( \hat H_S^{(2)} + i[\hat G^{(1)}, \hat H_S^{(1)}]  - \frac{1}{2} [\hat G^{(1)}, [\hat G^{(1)}, \hat H_S^{(0)}]]\right. \\
&\left. + i[\hat G^{(2)}, \hat H_S^{(0}] \right) \\
& +\dots,
\end{split}
\end{equation}
where each parentheses group terms of the same perturbative order. As usual in a SW transformation, the generator at order $k$, $\hat{G}^{(k)} = \sum_n \hat{\mathbf{1}}_{-n} \otimes \hat{G}_n^{(k)}$, is chosen such that the transformed Hamiltonian is diagonal at order $k$---in this case, diagonal in the replica index.

\subsection{Zero order}

At zeroth order, the transformed Shirley Hamiltonian is simply
\begin{equation}
\begin{split}
\hat H_{S}'^{(0)} = \hat H_S^{(0)} = \frac{\omega_d}{2} \hat m \otimes \mathbf{1} + \mathbf 1 \otimes \hat H_0^{(0)},
\end{split}
\end{equation}
where $\hat{H}_0^{(0)} = (\omega_0 - \omega_d/2)\hat{a}^\dagger \hat{a}$.
Thus, the Floquet Hamiltonian at zeroth order is simply $\hat H_F^{d (0)} = \hat H_0^{(0)}$, as expected.

\subsection{First order}

At first order, we require $\hat{G}^{(1)}$ to cancel the off-diagonal terms in the replica index that appear in
\begin{equation} \label{eq: first order terms}
\hat H_S^{(1)} + i [\hat G^{(1)}, \hat H_S^{(0)}].
\end{equation}
Since $\hat{H}_S^{(0)}$ is diagonal and $\hat{G}^{(1)}$ is off-diagonal, the commutator $[\hat{G}^{(1)}, \hat{H}_S^{(0)}]$ is also off-diagonal. Therefore, the diagonal part of $\hat{H}_S'^{(1)}$ becomes
\begin{equation}
\hat H_S'^{(1)} = \text{diag}\left( \hat H_S^{(1)}\right)  = \hat{\mathbf{1}} \otimes \hat H_0^{(1)}
\end{equation}
with
\begin{equation}
\hat H_0^{(1)} = g_3 (\Pi \hat a^{\dag 2} + \bar \Pi \hat a^2).
\end{equation}
Thus, the first-order Floquet Hamiltonian is $\hat{H}_F^{d(1)} = \hat{H}_0^{(1)}$. To first order in $\varphi_\text{zpf}$, this reproduces the two-photon drive term $\varepsilon_2 = g_3 \Pi$, consistent with the RWA.

For $\hat G^{(1)}$ to cancel the off-diagonal terms in \cref{eq: first order terms}, we need that, 
\begin{equation} \label{eq: first order terms seperated by n}
0 \stackrel{!}{=} \hat H_{n}^{(1)} - i n \frac{\omega_d}{2} \hat G_{n}^{(1)} + i(\omega_0 - \frac{\omega_d}{2})[\hat G_{n}^{(1)}, \hat a^\dag \hat a],
\end{equation}
for all $n\neq0$.

We note that the first-order index-off-diagonal terms $\hat H_{n}^{(1)}$ are:
\begin{equation}
\begin{split}
\hat H_{1}^{(1)} =& g_3 [\hat a^{\dag 2} \hat a + (1+2|\Pi|^2) \hat a^\dag]\\
\hat H_{2}^{(1)} =& g_3 \bar \Pi  (2\hat a^\dag \hat a +1 + |\Pi|^2) \\
\hat H_{3}^{(1)} =& \frac{g_3}{3} (\hat a^{\dag 3} + 3 \bar \Pi^2 \hat a) \\
\hat H_{4}^{(1)} =& g_3 \bar \Pi \hat a^{\dag 2} \\
\hat H_{5}^{(1)} =& g_3 \bar \Pi^2 \hat a^\dag\\ 
\hat H_{6}^{(1)} =& \frac{g_3}{3} \bar \Pi^3.
\end{split}
\end{equation}
From these, we obtain
\begin{equation}
\begin{split}
\hat G_{1}^{(1)} =& \frac{-ig_3}{\omega_0}[\hat a^{\dag 2} \hat a + (1+ 2|\Pi|^2) \hat a^\dag] \\
\hat G_{2}^{(1)} =& \frac{-ig_3}{\omega_d}\bar \Pi(2\hat a^\dag \hat a + 1 + |\Pi|^2)\\
\hat G_{3}^{(1)} =& \frac{-ig_3}{9\omega_0}\hat a^{\dag 3} - \frac{ig_3}{2\omega_d - \omega_0}\bar \Pi^2 \hat a \\
\hat G_{4}^{(1)} =& \frac{-ig_3}{2\omega_0 + \omega_d} \bar \Pi \hat a^{\dag 2} \\
\hat G_{5}^{(1)} =& \frac{-i g_3}{\omega_0+ 2\omega_d} \bar \Pi^2 \hat a^\dag \\
\hat G_{6}^{(1)} =& \frac{-ig_3}{9\omega_d} \bar \Pi^3.
\end{split}
\end{equation}

\subsection{Second order}

The second-order terms in the SW expansion are
\begin{equation} \label{eq: second-order SW equation}
\begin{split}
&\hat H_S^{(2)} + i[\hat G^{(1)}, \hat H_S^{(1)}]  - \frac{1}{2} [\hat G^{(1)}, [\hat G^{(1)}, \hat H_S^{(0)}]] \\
&+ i[\hat G^{(2)}, \hat H_S^{(0)}].
\end{split}
\end{equation}
The second-order generator $\hat G^{(2)}$, which we write below, is determined by requiring that the off-diagonal terms in the replica index cancel out. The remaining diagonal terms simplify significantly for our Hamiltonian and reduce to
\begin{equation}
\hat{\mathbf 1}_0 \otimes \left(\hat H_0^{(2)} + \frac{i}{2}\sum_n [\hat G_n^{(1)}, \hat H_{-n}^{(1)}]\right),
\end{equation}
yielding the second-order contribution to the Floquet Hamiltonian:
\begin{equation}
\hat H_F^{d(2)} = \Delta \hat a^\dag \hat a  -K \hat a^{\dag 2} \hat a^2,
\end{equation}
with
\begin{equation}
\begin{split}
\Delta =& 3 g_4 (1+2|\Pi|^2) \\
&- 4 g_3^2 \left[\frac{5+6|\Pi|^2}{3\omega_0} + \frac{|\Pi|^2}{2\omega_0 + \omega_d}\right] \\
K =& -\frac{3g_4}{2} + \frac{10 g_3^2}{3\omega_0}.
\end{split}
\end{equation}

\begin{figure}[t!]
    \centering
\includegraphics{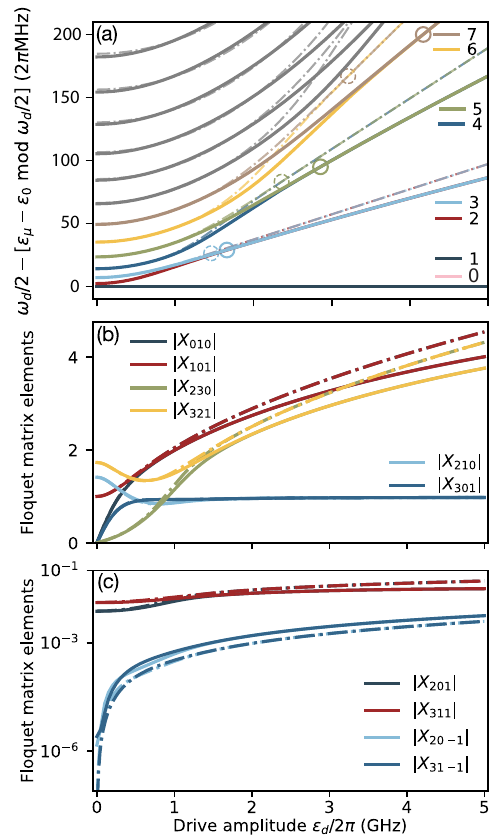}
\caption{
Comparison between exact numerical results (solid) and analytical approximations (dash-dotted) for the low-lying quasienergy spectrum (a) and the relevant Floquet matrix elements (b,c).}
\label{fig:numerics_vs_analytics}
\end{figure}

To second order, the Floquet Hamiltonian in the displaced and rotated frame takes the form
\begin{equation}\label{eq: Floquet Hamiltonian to second order}
\begin{split}
\hat H_F^d =& \left(\omega_0 - \frac{\omega_d}{2} + \Delta \right)\hat a^\dag \hat a - K \hat a^{\dag 2} \hat a^2 \\
&+ \varepsilon_2 \hat a^{\dag 2} + \bar{\varepsilon}_2 \hat a^2 + \mathcal O(\varphi_\text{zpf}^3),
\end{split}
\end{equation}
where $\varepsilon_2 = g_3 \Pi$ is the effective two-photon drive amplitude. Finally, we note that setting $\Pi = 0$ and $\omega_d=0$ reduces the $\hat H_F^d$ to an approximate second-order diagonalization of the undriven SNAIL Hamiltonian in \cref{eq:doublesnail_singlemode}.
\cref{fig:numerics_vs_analytics}(a) compares the exact (full lines) and analytical (dash-dotted lines) low-lying quasienergy spectra. Although the agreement is good at low drive amplitude, discrepancies become noticeable at larger drives. In particular, the “spectral kissing” between pairs (shown by the circles) occurs at lower drive amplitudes in the approximate spectrum. This suggests that the perturbative treatment may underestimate the extent of the tunneling time plateaus shown in \cref{fig:Tunneling_and_leakage}(a).

Since the Floquet matrix elements at second order in $\varphi_{\mathrm{zpf}}$ depend on the generator at the same order, we now write the second-order generator explicitly. It is determined by demanding that
\begin{equation} \label{eq: determines 2nd-order G_n}
\begin{split}
0 \stackrel{!}{=}& \hat H_n^{(2)} + \frac{i}{2}\sum_m [\hat G_{n+m}^{(1)}, \hat H_{-m}^{(1)}] \\
&- \frac{in\omega_d}{2}\hat G_n^{(2)} + i\left(\omega_0 - \frac{\omega_d}{2}\right)[\hat G_n^{(2)}, \hat a^\dag \hat a]
\end{split}
\end{equation}
is satisfied for all $n\neq 0$. Since the operators in $\hat G_n^{(2)}$ need to be proportional to those of $\hat H_n^{(2)}$, simple ans\"atze to the form of the generators readily give a solution to \cref{eq: determines 2nd-order G_n}. They are given by
\begin{equation}
\begin{split}
\hat G_1^{(2)} 
=& -i (c_{1, a^{\dag 3}}\Pi \hat a^{\dag 3} + c_{1, a^\dag a^2} \bar \Pi \hat a^\dag \hat a^2 + c_{1,a} \bar \Pi \hat a) 
\\
\hat G_2^{(2)}
=& -i (c_{2, a^{\dag 3} a} \hat a^{\dag 3} \hat a + c_{2,a^{\dag 2}} \hat a^{\dag 2} + c_{2, a^2} \bar \Pi^2 \hat a^2) 
\\
\hat G_3^{(2)}
=& -i (c_{3,a^{\dag 2} a} \bar \Pi \hat a^{\dag 2}\hat a + c_{3, a^\dag} \bar \Pi \hat a^\dag)
\\
\hat G_4^{(2)}
=&
-i(c_{4, a^{\dag 4}} \hat a^{\dag 4} + c_{4, a^\dag a} \bar \Pi^2 \hat a^\dag \hat a)
\\
\hat G_5^{(2)}
=&
-i(c_{5, a^{\dag 3}} \bar \Pi \hat a^{\dag 3} + c_{5,a} \bar \Pi^3 \hat a)
\\
\hat G_6^{(2)}
=&
-i c_{6, a^{\dag 2}} \bar \Pi^2 \hat a^{\dag 2}
\\
\hat G_7^{(2)}
=& -i c_{7, a^\dag} \bar \Pi^3 \hat a^\dag.
\end{split}
\end{equation}
The coefficients $c_{n, \bullet}$ are real and can found in \cref{sec: Explicit coefficients c_{n, ...}}.

We note that the SW generator and Floquet Hamiltonian obtained in this section are valid not only for the degenerate cat, i.e., when the drive frequency is chosen such that $\omega_d = 2|\epsilon_0-\epsilon_1|$. The expressions are also valid for other drive frequencies, as studied in Ref.~\cite{dealbornoz2024}, provided that $|\omega_d/2 - \omega_0| \ll |\omega_d - \omega_0|$.


\subsection{Floquet Hamiltonian in the laboratory frame}

Since we are interested in quantities evaluated in the laboratory frame-such as quasienergies and Floquet modes-we first relate the state in the displaced rotating frame, $\ket{\psi(t)}_d$, to its counterpart in the laboratory frame $\ket{\psi(t)}$,
\begin{equation}
\ket{\psi(t)}_d = \hat R^\dag(t) \hat D^\dag(t) \ket{\psi(t)}.
\end{equation}
This state evolves under the Hamiltonian $\hat H_d(t)$ via
\begin{equation}
\begin{split}
\ket{\psi(t)}_d =& \mathcal T \exp\left[-i\int_0^t d\tau \, \hat H_d(\tau)\right] \\
&\times \ket{\psi(0)}_d
\end{split}
\end{equation}
where $\hat H_d(t)$ is given in \cref{eq: initial Hamiltonian for SW in Shirley space}. Since $\hat R(0) = \hat{\mathbf{1}}$ but $\hat D(0) \neq \hat{\mathbf{1}}$, the initial state in the displaced and rotating frame is
\begin{equation}
\ket{\psi(0)}_d = \hat D^\dag(0) \ket{\psi(0)}.
\end{equation}

The SW transformation constructed in the previous subsections defines a generator $\hat G(t)$ such that the state
\begin{equation}
\ket{\psi(t)}_{d, G} \equiv e^{i\hat G(t)} \ket{\psi(t)}_d
\end{equation}
evolves under the time-independent Floquet Hamiltonian $\hat H_F^d$ introduced in \cref{eq: non-stroboscopic Floquet Hamiltonian} according to
\begin{equation}
\begin{split}
\ket{\psi(t)}_{d,G} =&  e^{-i t\hat H_F^d}\ket{\psi(0)}_{d,G} \\
=& e^{-i t\hat H_F^d} e^{i\hat G(0)} \hat D^\dag (0) \ket{\psi(0)}.
\end{split}
\end{equation}
We thus recover the state in the lab frame as
\begin{equation} \label{lab_to_displaced_rotating}
\begin{split}
\ket{\psi(t)} =& \hat U(t) \ket{\psi(0)},
\end{split}
\end{equation}
with an explicit form of the lab-frame propagator given by
\begin{equation} \label{eq: Analytical expression for the unitary propagator in the lab frame}
\hat U(t) = \hat D(t) \hat R(t) e^{-i\hat G(t)} e^{-it \hat H_F^d} e^{i \hat G(0)} \hat D^\dag(0).
\end{equation}
Evaluating this operator at $t = T$, we obtain an explicit form for the stroboscopic Floquet propagator in the lab frame, $\hat U(T) \equiv \exp(-i T \hat H_F)$, which is the operator used in our numerical calculations. Since $\hat D(T) = \hat D(0)$, and using the identity $\hat R(T) e^{-i \hat G(T)} = e^{-i\hat G(0)} \hat R(T)$, we can rewrite the propagator as
\begin{equation}
\hat U(T) = \hat D(0) e^{-i \hat G(0)} \hat R(T) e^{-i T \hat H_F^d} e^{i\hat G(0)} \hat D^\dag(0).
\end{equation}
Next, we observe that $\hat R(T) = e^{-i\pi \hat a^\dag\hat a}$ is the parity operator and commutes with $\hat H_F^d$, which is symmetric; see \cref{eq: Floquet Hamiltonian to second order}. Thus, we can write the Floquet-mode decomposition
\begin{equation}
\hat U(T) = \sum_\mu e^{-i T \epsilon_{\mu}} \ket{\phi_{\mu}} \bra{\phi_\mu},
\end{equation}
with
\begin{equation}
\ket{\phi_{\mu}(0)} \equiv \hat D(0) e^{-i \hat G(0)}  \ket{\phi_\mu^d}
\end{equation}
relating the lab-frame Floquet modes to the displaced-and-rotated frame Floquet modes $\ket{\phi_\mu^d}$, and 
\begin{equation}
e^{-i T \epsilon_{\mu}} \equiv e^{-i T[\epsilon_\mu^d + \frac{\omega_d}{2} \text{mod}(\mu, 2)]}
\end{equation}
relating the lab-frame Floquet qusaienergies to the displaced-and-rotated-frame Floquet eigenenergies $\epsilon_\mu^d$. Here, $\epsilon_\mu^d$ and $\ket{\phi_\mu^d}$ are eigenenergies and eigenstates, respectively, of the Hamiltonian $\hat H_F^d$. 
It can be shown that the Floquet modes in the lab frame---which evolve under $e^{it \epsilon_{\mu}} \hat U(t)$---are $T$-periodic functions of time.

Let us clarify the relationship between the eigenenergies $\epsilon_\mu^d$ of the displaced-and-rotated-frame Floquet Hamiltonian defined in \cref{eq: non-stroboscopic Floquet Hamiltonian} and the quasienergies in the lab frame $\epsilon_{\mu}$. The eigenenergies $\epsilon_\mu^d$ are not folded---they can grow unbounded---while the lab-frame quasienergies $\epsilon_{\mu}$ are defined modulo the drive frequency, and therefore lie within a Brillouin zone, which we take to be $[-\omega_d/2, +\omega_d/2)$. See \cref{fig:fullspectrum_photonnumber}(a) for the corresponding numerical spectrum in the lab frame. This distinction arises because the non-stroboscopic Floquet Hamiltonian $\hat H_F^d$ is not defined via the stroboscopic propagator, unlike the standard Floquet Hamiltonian, whose quasienergies are inherently folded into the Brillouin zone. In our case, the appropriate folding to obtain the lab-frame quasienergies from the eigenvalues $\epsilon_\mu^d$ of $\hat H_F^d$ is given by:
\begin{equation} \nonumber
\epsilon_{\mu} = \begin{cases}
\left[(\epsilon_\mu^d + \frac{\omega_d}{2})\mod \omega_d\right] - \frac{\omega_d}{2} & \text{if } \mu \text{ is even}, \\
\left[\epsilon_\mu^d \mod \omega_d \right] - \frac{\omega_d}{2} & \text{if } \mu \text{ is odd}.
\end{cases}
\end{equation}

\subsection{Floquet matrix elements for the master equation}

In the previous subsection, we constructed the Floquet spectrum in the laboratory frame in terms of the spectrum of the squeezed Kerr Hamiltonian, for which approximate analytical expressions are known (see, e.g., Ref.~\cite{Chamberland2022}). Building on this, we now derive approximate expressions for the matrix elements $X_{\mu \nu k}$ appearing in \cref{eq:matrix_elements}, which govern the dynamics in the Floquet-Markov master equation.

We begin by noting that
\begin{equation} \label{floquetmodes lab vs displacedrotated}
\ket{\phi_{\mu}(t)} = e^{i t (\epsilon_{\mu} - \epsilon_\mu^d)} \hat D(t) \hat R(t) e^{-i\hat G(t)}\ket{\phi_\mu^d(0)}.
\end{equation}
Rather than applying these operators to the Floquet modes on the right-hand side, we apply them to the annihilation operator, working to second order in $\varphi_\text{zpf}$. We define
\begin{equation} \label{eq: definition tilde a_n}
\begin{split}
\tilde a(t) \equiv& e^{i\hat G(t)}\hat R^\dag(t) \hat D^\dag (t)\hat a \hat D(t) \hat R(t) e^{-i\hat G(t)} \\
=& \hat a e^{-i\frac{\omega_d}{2}t} + \beta(t) +i [\hat G^{(1)}(t), \hat a]e^{-i\frac{\omega_d}{2} t} \\
&+ i[\hat G^{(2)}(t), \hat a]e^{-i \frac{\omega_d}{2}t} \\
&- \frac{1}{2} [\hat G^{(1)}(t), [\hat G^{(1)}(t), \hat a]] e^{-i \frac{\omega_d}{2}t} \\
=& \sum_n \tilde a_n e^{in \frac{\omega_d}{2} t},
\end{split}
\end{equation}
with the following Fourier components to second order in $\varphi_\text{zpf}$:
\begin{widetext}
\begin{equation}
\begin{split}
\tilde a_{-6} =& (c_{5,a} + d_{-6,1}) \Pi^3 \\
\tilde a_{-5} =& (c_{4, a^\dag a} + d_{-5, a}) \Pi^2 \hat a \\
\tilde a_{-4} =& - \frac{g_3}{\omega_d + \omega_0} \Pi^2
+ (c_{3, a^{\dag 2} a} + d_{-4, a^2}) \Pi \hat a^2
\\
\tilde a_{-3} =& - \frac{2g_3}{\omega_d} \Pi \hat a 
+ (c_{2, a^{\dag 3} a} + d_{-3, a^3}) \hat a^3 + (2 c_{2, a^2} + d_{-3, a^\dag}) \Pi^2 \hat a^\dag
\\
\tilde a_{-2} =& \frac{\omega_d + \omega_0}{2\omega_d} \Pi - \frac{g_3}{\omega_d - \omega_0} \hat a^2 
+ (2 c_{1, a^\dag a^2} + d_{-2, a^\dag a}) \Pi \hat a^\dag \hat a + (c_{1,a} + d_{-2, 1}) \Pi 
\\
\tilde a_{-1} =& \hat a  
+ d_{-1, a^\dag a^2} \hat a^\dag \hat a^2 + d_{-1, a} \hat a
\\
\tilde a_0 =& \frac{g_3}{\omega_d - \omega_0} (2\hat a^\dag \hat a + 1 + 2|\Pi|^2) 
+(-3 c_{1, a^{\dag 3}} + d_{0, a^{\dag 2}}) \Pi \hat a^{\dag 2} + (-c_{1, a^\dag a^2} + d_{0, a^2}) \bar \Pi \hat a^2
\\
\tilde a_1 =& \frac{2g_3}{\omega_d} \bar \Pi \hat a 
+ (-3 c_{2, a^{\dag 3} a} + d_{1, a^{\dag 2} a}) \hat a^{\dag 2} \hat a + (-2c_{2,a^{\dag 2}} + d_{1, a^\dag}) \hat a^\dag
\\
\tilde a_{2} =&  \frac{\omega_d - \omega_0}{2\omega_d}\bar \Pi + \frac{g_3}{3(\omega_d - \omega_0)}\hat a^{\dag 2} 
+ (-2 c_{3, a^{\dag 2}a} + d_{2, a^\dag a}) \bar \Pi \hat a^\dag \hat a + (-c_{3, a^\dag} + d_{2,1}) \bar \Pi
\\
\tilde a_3 =& \frac{2g_3}{3\omega_d - 2\omega_0}\bar \Pi \hat a^\dag 
-4 c_{4, a^{\dag 4}} \hat a^{\dag 3} + (-c_{4, a^\dag a} + d_{3, a})\bar \Pi^2 \hat a
\\
\tilde a_4 =& \frac{g_3}{3\omega_d - \omega_0} \bar \Pi^2
+ (- c_{5, a^{\dag 3}} + d_{4, a^{\dag 2}}) \bar \Pi \hat a^{\dag 2} \\
\tilde a_5 =& (-2 c_{6, a^{\dag 2}} + d_{5, a^\dag}) \bar \Pi^2 \hat a^\dag \\
\tilde a_6 =& d_{6,1} \bar \Pi^3.
\end{split}
\end{equation}
\end{widetext}
The real coefficients $d_{n,\bullet}\propto g_3^2$---whose explicit expressions are given in \cref{sec: Explicit coefficients d_{n, ...}}---originate from the second-order nested commutator of $\hat a$ with $\hat G^{(1)}$ in \cref{eq: definition tilde a_n}. Specifically, this nested commutator contributes to $\tilde a_n$ with $-\frac{1}{2}[\hat G^{(1)}_{n+1-m},[\hat G^{(1)}_{m}, \hat a]]$.

We now define the charge components $\tilde p_n$ via
\begin{equation}
\begin{split} \label{eq: p_k coeffecients}
i[\tilde a^\dag(t) - \tilde a(t)] =& \sum_n i(\tilde a_{-n}^\dag - \tilde a_n) e^{i n \frac{\omega_d}{2}t} 
\\
\equiv& \sum_n \tilde p_n e^{i n \frac{\omega_d}{2}t}.
\end{split}
\end{equation}
By construction, the charge components satisfy $\tilde p_{-n} = \tilde p_n^\dag$. We do not write them out explicitly, as they can be straightforwardly constructed from the $\tilde a_n$ operators. These components enter the Floquet matrix elements of interest as
\begin{equation} \label{eq: time-dependent Floquet matrix elements}
\begin{split}
&i\bra{\phi_{\mu}(t)}(\hat a^\dag - \hat a)\ket{\phi_{\nu}(t)} = \\
&e^{it (\epsilon_{\nu} - \epsilon_{\mu} + \epsilon_\mu^d - \epsilon_\nu^d)} \sum_n \bra{\phi_\mu^d} \tilde p_{n} \ket{\phi_\nu^d} e^{in \frac{\omega_d}{2} t}.
\end{split}
\end{equation}

Using these results, we can now compute some of the matrix elements. Following Ref.~\cite{Chamberland2022}, we use the approximate expression
\begin{equation} \label{eq: approximate floquet modes}
\begin{split}    
\ket{\phi_\mu^d} &\approx \frac{1}{\sqrt{2}}\left[\hat D(\alpha) + (-1)^{\left\lfloor \frac{\mu+1}{2} \right\rfloor} \hat D(-\alpha)\right]\bigg|\left\lfloor \frac{\mu}{2} \right\rfloor\bigg\rangle,
\end{split}
\end{equation}
where the ket on the right-hand-side is a Fock state, to represent the low-lying eigenstates of the squeezed-Kerr Hamiltonian~\cref{eq:H squeezed Kerr}. More accurate results can be obtained by instead using the exact numerical eigenstates of the approximate Floquet Hamiltonian \cref{eq: Floquet Hamiltonian to second order}. 

Let us also introduce the definition
\begin{equation} \label{eq: thetamunu}
\theta_{\mu,\nu} \equiv \epsilon_{\nu} - \epsilon_{\mu} + \epsilon_\mu^d - \epsilon_\nu^d.
\end{equation}
We begin with modes 0 and 1, which always kiss. We have that $\theta_{0,1} = -\omega_d/2$ and $\theta_{1,0} = \omega_d/2$ [see \cref{fig:fullspectrum_photonnumber}(a)], thus
\begin{equation} \label{eq: Floquet matrix elements 01}
\begin{split}
&X_{010} \\
&=\bra{\phi_0^d}\tilde p_1 \ket{\phi_1^d} \\
&\approx i\Bigg[ (1 + 2 c_{2. a^{\dag 2}} + d_{-1,a} - d_{1,a^\dag}) \bar \alpha - \frac{2g_3}{\omega_d} \bar \Pi \alpha \\
&+ (3 c_{2, a^{\dag 3}a} + d_{-1, a^\dag a^2} - d_{1, a^{\dag 2} a}) \bar \alpha |\alpha|^2 \Bigg],
\end{split}
\end{equation}
and the same approximate expression for $X_{101} = \bra{\phi_1^d}\tilde p_1 \ket{\phi_0^d}$. As a result, we find that $X_{010}-X_{101}\approx 0$; see \cref{fig:Floquet_matrix_elements}(a). 

When levels 2 and 3 kiss, we get $\theta_{2,3} = -\omega_d/2$ and $\theta_{3,2} = \omega_d/2$. We find
\begin{equation} \label{eq: Floquet matrix elements 23}
\begin{split}
&X_{230} \\
&= \bra{\phi_2^d}\tilde p_1 \ket{\phi_3^d} \\
&\approx i\Bigg[ (1 + 2 c_{2. a^{\dag 2}} + d_{-1,a} - d_{1,a^\dag}) \bar \alpha - \frac{2g_3}{\omega_d} \bar \Pi \alpha \\
&+ (3 c_{2, a^{\dag 3}a} + d_{-1, a^\dag a^2} - d_{1, a^{\dag 2} a}) \bar \alpha (2 + |\alpha|^2) \Bigg],
\end{split}
\end{equation}
and the same approximate expression for $X_{321} = \bra{\phi_3^d} \tilde p_1 \ket{\phi_2^d}$. Only the third terms in \cref{eq: Floquet matrix elements 23} and \cref{eq: Floquet matrix elements 01} differ.

In general, when levels $\mu$ (even) and $\mu+1$ (odd) kiss, we have
\begin{equation} \label{eq: Floquet matrix elements mu,mu+1}
\begin{split}
&= X_{\mu (\mu+1) 0} \\
&= \bra{\phi_\mu^d}\tilde p_1 \ket{\phi_{\mu+1}^d} \\
&\approx i\Bigg[ (1 + 2 c_{2. a^{\dag 2}} + d_{-1,a} - d_{1,a^\dag}) \bar \alpha - \frac{2g_3}{\omega_d} \bar \Pi \alpha \\
&+ (3 c_{2, a^{\dag 3}a} + d_{-1, a^\dag a^2} - d_{1, a^{\dag 2} a}) \bar \alpha (2 \mu + |\alpha|^2) \Bigg],
\end{split}
\end{equation}
and the same approximate expression for $X_{(\mu+1) \mu 1}$. Therefore, when levels $\mu$ (even) and $\mu+1$ (odd) kiss, we have $|X_{\mu (\mu+1) 0} - X_{(\mu+1) \mu 1}|\approx 0$ in the large $|\alpha|$ regime where the analytical expressions for the Floquet modes \cref{eq: approximate floquet modes} are valid. \Cref{fig:numerics_vs_analytics}(b) shows a comparison between exact numerical results (solid lines) and these analytical approximations (dash-dotted lines) for some of these Floquet matrix elements. The exact results are computed in the lab frame, as in \cref{fig:Floquet_matrix_elements}, and the approximate results are obtained by evaluating the $\tilde p_n$ operators with the numerical eigenstates of the approximate Floquet Hamiltonian \cref{eq: Floquet Hamiltonian to second order} for more accurate estimates.

The matrix elements associated with the main source of leakage are $X_{210}$ and $X_{301}$, which correspond to single-photon absorption from the bath. We have $\theta_{2,1} = -\omega_d/2$ and $\theta_{3,0} = \omega_d/2$, thus
\begin{equation}
\begin{split}
&X_{210} \\
&= \bra{\phi_2^d} \tilde p_1 \ket{\phi_1^d} \\
&\approx i\Bigg[ (1 + 2 c_{2, a^{\dag 2}} + d_{-1,a} - d_{1,a^\dag}) \\
&+ (3 c_{2, a^{\dag 3}a} + d_{-1, a^\dag a^2} - d_{1, a^{\dag 2} a}) 2|\alpha|^2 \Bigg],
\end{split}
\end{equation}
and the same approximate expression for $X_{301} = \bra{\phi_3^d} \tilde p_1 \ket{\phi_0^d}$.  
We also compare these matrix elements with the exact results in \Cref{fig:numerics_vs_analytics}(b).

Other matrix elements of interest are $X_{201}$ and $X_{311}$, which have contributions from two-photon absorption ($\propto \hat a^{\dag 2}$) and drive-induced dephasing ($\propto |\Pi|\hat a^\dag \hat a$) processes in $\tilde p_2$. In both matrix elements we have $\theta_{2,0}=\theta_{3,1}=0$, thus
\begin{equation} \label{eq:X_201}
\begin{split}
&X_{201} \\
&= \bra{\phi_2^d}\tilde p_2 \ket{\phi_0^d} \\
&\approx i \Bigg[ - \frac{4g_3}{3(\omega_d - \omega_0)} 2\bar \alpha \\
&+ (2 c_{1, a^\dag a^2} + 2 c_{3, a^{\dag 2}, a} + d_{-2, a^\dag a} - d_{2, a^\dag a}) \bar \Pi \alpha
\Bigg],
\end{split}
\end{equation}
and the same approximate expression for $X_{311} = \bra{\phi_3^d} \tilde p_2 \ket{\phi_1^d}$. These analytical results are compared with the exact ones in \cref{fig:numerics_vs_analytics}(c).

We note that, at finite temperature, two-photon absorption is present even for $\varepsilon_d=0$ (i.e., $\Pi=0$), due to the third-order nonlinearity ($\propto g_3$) of the double SNAIL. This is a generic feature of circuit implementations whose charge operator has a nonvanishing matrix element between states of the same parity—for example, between $\ket{2_s}$ and $\ket{0_s}$; consequently, this leakage channel is absent in circuits that preserve parity symmetry, as in Refs.~\cite{Puri:2017,bhandari2024}.

Finally, as noted in the main text, certain matrix elements responsible for drive-induced leakage persist even at zero temperature. In particular, $X_{20,-1}$ and $X_{31,-1}$ arise from $\tilde p_{-2}$, which—like $\tilde p_{2}$—contains a drive-induced dephasing term $\propto |\Pi| \hat a^\dag \hat a$. The dephasing term in $\tilde p_2$, which emerges only at finite temperature, describes the conversion of a bath photon at $\omega_d$ into a drive photon, enabled by higher-order wave mixing of the SNAIL. By contrast, the corresponding term in $\tilde p_{-2}$ describes the reverse process—conversion of a drive photon into an environmental excitation at $\omega_d$—and therefore persists even at zero temperature, provided the bath has finite spectral density near $\omega_d$. Crucially, both absorption and emission processes carry which-SNAIL-level information and thus lead to dephasing. In the logical cat-state basis, this term couples $\ket{\phi_0^d}$ to $\ket{\phi_2^d}$ and $\ket{\phi_1^d}$ to $\ket{\phi_3^d}$, thereby inducing leakage~\cite{Venkatraman:2024}, with rates given by
\begin{equation}
\begin{split}
&X_{20,-1} \\
&= \bra{\phi_2^d} \tilde p_{-2} \ket{\phi_0^d} \\
&\approx -i (2 c_{1, a^\dag a^2} + 2 c_{3, a^{\dag 2}, a} + d_{-2, a^\dag a} - d_{2, a^\dag a}) \Pi \alpha,
\end{split}
\end{equation}
with the same approximate expression for $X_{31,-1}$. These analytical results are compared with exact ones in \cref{fig:numerics_vs_analytics}(c). In both \cref{fig:numerics_vs_analytics}(b–c), the analytical results reproduce the interference pattern similar to the exact results, albeit it starts at lower drive amplitudes. However, at large drive amplitudes a slight but noticeable difference in the matrix elements remains between the analytical and exact results, suggesting that the analytical treatment may yield a slightly different estimate of the intrawell leakage rate and tunneling time.

\section{Perturbative expansion accounting for the buffer mode}
\label{appendix: buffer mode}

We revisit the perturbative expansion used to derive the Floquet Hamiltonian in the single-mode approximation, now including the buffer mode in the description. We show that accurately capturing the hybridization between low-lying Floquet modes and those with a single buffer excitation discussed in \cref{subsec: buffer mode spectral kissing} requires carrying the expansion to very high orders in the system's nonlinearities, rendering the perturbative approach impractical for quantitative predictions. As an illustrative example, we show that the leading-order coupling between $\ket{\phi_{7_s,0_b}}$ and $\ket{\phi_{17_s,1_b}}$ appears only at order $\mathcal{O}(\varphi_{\textrm{zpf}}^9)$.

We begin with the SNAIL–buffer Hamiltonian $\hat{H}'$ in \cref{Hamiltonian_buffer} and move to the interaction picture via the unitary transformation $\hat U(t) e^{-i\omega_b t \hat a_b^\dag \hat a_b}$. Using the Floquet modes and the expression for time-dependent matrix elements of the SNAIL charge operator in \cref{eq: time-dependent Floquet matrix elements}, this interaction picture Hamiltonian can be expressed as
\begin{equation}
\begin{split}
&\hat H'_\textrm{int} = ign_\textrm{zpf}(\hat a_b^{\dag} e^{i\omega_b t} - \hat a_b e^{-i\omega_b t}) \\
&\times \sum_{\mu, \nu,n} e^{i(\epsilon_\mu^d - \epsilon_\nu^d + n(\omega_d/2))t} \bra{\phi_\mu^d}\tilde p_n \ket{\phi_\nu^d} \ket{\phi_\mu}\bra{\phi_\nu}.
\end{split}
\end{equation}
The dominant process responsible for the transition $\ket{\phi_{7_s,0_b}} \to \ket{\phi_{17_s,1_b}}$ corresponds to the term with $n = -2$, since the term $\hat a_b^\dag \ket{\phi_{17_s}} \bra{\phi_{7_s}}$ becomes resonant when the detuning $\epsilon_{7_s}^d - \epsilon_{17_s}^d$ approaches $\omega_b - \omega_d=\Delta_{bd}$, a condition reached when the ac-Stark shift becomes substantial, as explained in the main text.

We can now deduce the lowest order in $\varphi_\textrm{zpf}$ at which $\tilde p_{-2}$ couples these two states. When both states lie under the barrier of the double-well metapotential, the relevant coupling arises from $\hat a^{\dag 6}$ since, following \cref{eq: approximate floquet modes}, the states are approximately given by $\ket{\phi_{7_s}^d} \sim [\hat D(\alpha) + \hat D(-\alpha)]\ket{3}$ and $\ket{\phi_{17_s}^d} \sim [\hat D(\alpha) - \hat D(-\alpha)]\ket{8}$, where $\ket{3}$ and $\ket{8}$ are Fock states~\cite{Chamberland2022}. Five creation operators are needed to raise the Fock number and another one for changing the parity. The operator $\tilde p_{-2}$ is proportional to the commutators $[\hat G_{-3}, \hat a^\dag]$ and $[\hat G_{-1}, \hat a]$, see \cref{eq: definition tilde a_n}. Therefore, the lowest-order terms proportional to $\hat a^{\dag 6}$ originate from the generators $\hat G_{-1}^{(11)} \propto \hat a^{\dag 7} \Pi^4$ and $\hat G_{-3}^{(11)} \propto \hat a^{\dag 6} \hat a \Pi^4$. These generators are proportional to $g_{11}$, leading to an overall scaling of $\mathcal{O}(\varphi_\textrm{zpf}^9)$.

This scaling can be confirmed numerically. The operator $\tilde p_{-2}$ includes terms of the form $\sim \varphi_\text{zpf}^{3k} \hat a^{\dagger 2k} \Pi^{k+1}$. To identify the dominant value of $k$, we numerically evaluate the matrix element $\varphi_\text{zpf}^{3k} |\Pi|^{k+1}\big|\! \bra{\phi_{17}^d} \hat a^{\dagger 2k} \ket{\phi_7^d}\!\big|$, using the numerically obtained eigenstates of the squeezed Kerr Hamiltonian~\cref{eq:H squeezed Kerr}. Doing this, we find that the matrix element peaks at $k = 3$ (not shown), confirming that the leading coupling indeed scales as $\sim \varphi_{\text{zpf}}^9$.

In this illustrative case, knowing which states hybridize at some particular drive amplitude allows us to determine the required expansion order. However, in general, such information is unavailable, making the exact treatment of the full nonlinear potential indispensable.

\section{SNAIL array modes}
\label{appendix: array modes}

\begin{figure}[t!]
    \centering
\includegraphics{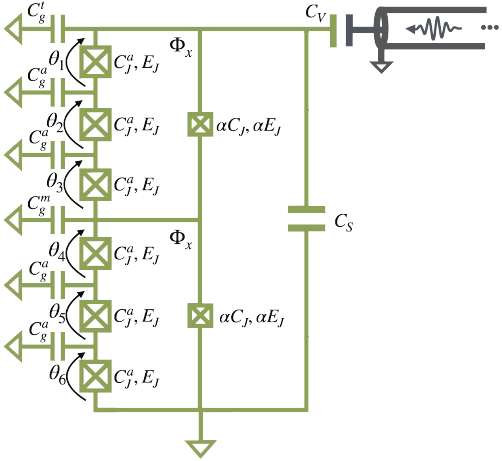}
\caption{Circuit diagram of the shunted double SNAIL capacitively coupled to the drive line. Each island in the array is connected to ground through a parasitic ground capacitance.
}
\label{fig:circuit_array_modes}
\end{figure}

\Cref{fig:circuit_array_modes} shows a more detailed schematic of the double-SNAIL circuit analyzed in \cref{sec: array modes}. We first obtain the Lagrangian in the absence of the ground capacitances $C_g^a$, $C_g^m$, and $C_g^t$; these will be incorporated at a later stage.

Expressed in terms of the fluxes $\{\theta_j \}_{j=1}^6$ across the large junctions, the Lagrangian reads
\begin{equation}
\begin{split}
\mathcal L =& \frac{C_J^a}{2} \sum_{i=1}^6 \dot \theta_i^2  + \frac{C_V}{2} \left(\sum_{i=1}^6 \dot \theta_i - V \right)^2 \\
&+ \frac{C_S}{2} \left( \sum_{i=1}^6 \dot \theta_i \right)^2 \\
&+   \frac{\alpha C_J^a}{2} \left[\left(\sum_{i=1}^3 \dot \theta_i \right)^2 + \left(\sum_{i=4}^6 \dot \theta_i \right)^2 \right] \\
&+ E_J \sum_{i=1}^6 \cos\left(\frac{2\pi}{\Phi_0} \theta_i \right) \\
&+\alpha E_J \cos \left(\frac{2\pi}{\Phi_0}\sum_{i=1}^3 \theta_i + \varphi_\textrm{ext} \right) \\
&+\alpha E_J \cos \left(\frac{2\pi}{\Phi_0}\sum_{i=4}^6 \theta_i + \varphi_\textrm{ext} \right).
\end{split}
\end{equation}
Following Ref.~\cite{Ferguson:2013}, we perform a change of variables based on the symmetries of the circuit that decouples the kinetic energy terms. The circuit is symmetric under permutations within each of the two triplets ${\theta_1, \theta_2, \theta_3}$ and ${\theta_4, \theta_5, \theta_6}$, as well as under an exchange of these two triplets. It is thus convenient to introduce the collective flux variables $\{\phi, \phi_-, \xi_1, \xi_2, \chi_1, \chi_2\}$, which are linear combinations of the original variables defined as
\begin{equation} \label{eq: change basis phase collective modes}
\begin{split}
\theta_i =& \frac{\phi + \phi_-}{6} + \sum_{\mu=1}^2 W_{\mu,i} \xi_\mu, \quad i=1,2,3, \\
\theta_i =& \frac{\phi - \phi_-}{6} + \sum_{\mu=1}^2 W_{\mu,i-3} \chi_\mu, \quad i=4,5,6,
\end{split}
\end{equation}
where $W$ is a $(2\times 3)$ semiorthogonal matrix, with the properties $\sum_{i=1}^3 W_{\mu, i} = 0$ and $\sum_{i=1}^3 W_{\mu, i} W_{\nu, i} = \delta_{\mu \nu}$~\cite{Ferguson:2013}. In particular, we have that the collective flux used in the single-phase approximation corresponds to $\phi = \sum_{i=1}^6 \theta_i$. As discussed below, the other collective mode that plays an important role is $\phi_- = \theta_1 + \theta_2 + \theta_3 - (\theta_4 + \theta_5 + \theta_6)$. While $\phi$ is symmetric under all three symmetries, $\phi_-$ is antisymmetric under the exchange of the triplets. In terms of the new variables, the Lagrangian of the double-SNAIL circuit takes the form
\begin{equation} \label{eq: Lagrangian in collective variables without capacitances}
\begin{split}
\mathcal L =& \frac{C}{2} \dot \phi^2  - C_V \dot \phi V \\
&+ \frac{C_-}{2} \dot \phi_-^2 \\
&+ \frac{C_J^a}{2} \sum_{\mu=1}^2(\dot \xi_\mu^2 + \dot \chi_\mu^2) \\
&+ E_J \sum_{i=1}^3 \cos\left(\frac{\varphi + \varphi_-}{6} + \frac{2\pi}{\Phi_0}\sum_{\mu=1}^2 W_{\mu, i} \xi_\mu \right) \\
&+ E_J \sum_{i=1}^3 \cos\left( \frac{\varphi - \varphi_-}{6} + \frac{2\pi}{\Phi_0} \sum_{\mu=1}^2 W_{\mu, i} \chi_\mu \right) \\
&+2\alpha E_J \cos \left(\frac{\varphi}{2} + \varphi_\text{ext}\right) \cos \left(\frac{\varphi_-}{2} \right).
\end{split}
\end{equation}
Here, we have introduced the reduced fluxes $\varphi = 2\pi \phi/\Phi_0$ and $\varphi_- = 2\pi \phi_-/\Phi_0$, and defined $C = C_J^a (1/6 + \alpha/2) + C_S+C_V$, and $C_- = C_J^a(1/6 + \alpha/2)$.
Note that, as wanted, the kinetic energy terms are now decoupled, and the modes interact only through the nonlinearity. Owing to the properties of the matrix $W$, the potential minimum occurs at $\phi_- = \chi_\mu = \xi_\mu = 0$ and $\varphi = \varphi_\text{min} \neq 0$. Setting $\phi_- = \chi_\mu = \xi_\mu=0$ amounts to ignoring the five high-energy modes and from this we recover the Lagrangian of the single-phase approximation given by
\begin{equation}
\begin{split}
\mathcal L_{\rm SP} =& \frac{C}{2} \dot \phi^2  + \frac{C_V}{2} \left(\dot \phi- V\right)^2\\
&+ 6E_J \cos\left(\ \frac{\varphi}{6} \right) +2\alpha E_J \cos \left( \frac{\varphi}{2} + \varphi_\text{ext} \right).
\end{split}
\end{equation}
This expression immediately leads to the Hamiltonian of \cref{singlemode_Hamiltonian,eq:doublesnail_singlemode}. 

The Hamiltonian without the single mode approximation is also straightforward to obtain from \cref{eq: Lagrangian in collective variables without capacitances}, and from this we easily obtain the plasma frequency of each of the six modes. Before determining  these frequencies, we first identify the location of the minimum of the potential energy. For the parameters used throughout the main text, see the caption of ~\cref{fig:metapotential}---namely, $E_J/2\pi = \SI{272.436}{GHz}$, $E_C/2\pi = \SI{107.8}{MHz}$, $\alpha = 0.046$, and $\varphi_\text{ext} = 2\pi \times 0.33$---the minimum is located at $\varphi_\text{min} = 0.255$. Choosing $(C_S + C_V)/C_J^a = 2.83$ leads to the following plasma frequencies
\begin{equation}
\begin{split}
\omega_{p,+} =& \sqrt{8E_C E_{J,+}} =2\pi \times \SI{6.096}{GHz}, \\
\omega_{p,-} =& \sqrt{8\beta E_C E_{J,-}} =2\pi \times \SI{24.43}{GHz}, \\
\omega_{p,\mu} =& \sqrt{8\beta_\mu E_C E_{J,\mu}} =2\pi \times \SI{26.31}{GHz},
\end{split}
\end{equation}
corresponding to the symmetric mode, the antisymmetric mode, and the remaining four transverse modes, respectively. Here, $\beta = C / C_- \approx 16.1$ and $\beta_\mu = C / C_J^a \approx 3$ are ratios of the total effective capacitance of each mode. The quantity $E_{J,x}$ denotes the curvature of the potential along mode $x$ at the minimum, i.e., $E_{J,x} = E_J \left(\frac{d^2 U}{dx^2}\right)_{\text{min}}$.

Now including the ground capacitances shown in \cref{fig:circuit_array_modes}, the Lagrangian \cref{eq: Lagrangian in collective variables without capacitances} acquires an extra kinetic energy contribution that can be written in matrix form as $(1/2)\dot{\boldsymbol{\theta}}^\intercal \mathbf C_0 \dot{\boldsymbol{\theta}}$, where $\boldsymbol{\theta}^\intercal = (\theta_1, \dots, \theta_6)$ and the (normalized) matrix $\mathbf C_0/C_g^t$ is given by
\begin{equation} \nonumber
\begin{pmatrix}
1 & 1 & 1 & 1 & 1 & 1 \\ 
1 & 1 & 1 & 1 & 1 & 1 \\ 
1 & 1 & 1+2a & 1+2a & 1+2a & 1+2a \\ 
1 & 1 & 1+2a & 1+m+2a & 1+m+2a & 1+m+2a \\ 
1 & 1 & 1+2a & 1+m+2a & 1+m+3a & 1+m+3a \\
1 & 1 & 1+2a & 1+m+2a & 1+m+3a & 1+m+4a
\end{pmatrix},
\end{equation}
with $a=C_g^a/C_g^t$ and $m = C_g^m/C_g^t$; see \cref{fig:circuit_array_modes}. Transforming this matrix to the basis defined by \cref{eq: change basis phase collective modes} and including it in \cref{eq: Lagrangian in collective variables without capacitances} gives the full circuit Lagrangian in terms of the collective modes. After obtaining the full Hamiltonian, we choose the value of the capacitances to obtain the frequencies quoted in \cref{sec: array modes}, for example, $E_C/2\pi = \SI{107.8}{MHz}$, $\beta \approx 16.1$ and $g/2\pi=\SI{100}{MHz}$, which gives a plasma frequency of $\omega_- \approx 2\omega_d + 2\pi \times \SI{54}{MHz}$ for the antisymmetric mode. We do so by considering $C_g^a = C_g^t = C_g^m$ for simplicity, and fixing the ratio $C_g^a/C_J^a = 0.03$ and an updated ratio $(C_S + C_V)/C_J^a = 3.074$. This choice gives us the target value for $\beta$ and fixes the proportionality between $E_C$ and $g$. The only remaining free parameter is $C_J^a$, which fixes their values. The plasma frequencies of the transversal modes become
\begin{equation}
\begin{split}
\omega_{p, \xi_1}/2\pi =& \SI{27.124}{GHz}, \\
\omega_{p, \xi_2}/2\pi =& \SI{27.378}{GHz}, \\
\omega_{p, \chi_1}/2\pi =& \SI{27.113}{GHz}, \\
\omega_{p, \chi_2}/2\pi =& \SI{27.369}{GHz},
\end{split}
\end{equation}
and the charge-charge coupling energies between the symmetric mode and each of these transversal modes are
\begin{equation} \label{eq: coupling array}
\begin{split}
g_{\xi_1}/2\pi =& \SI{21.8}{MHz}, \\
g_{\xi_2}/2\pi =& -\SI{12.6}{MHz}, \\
g_{\chi_1}/2\pi =& \SI{7.7}{MHz}, \\
g_{\chi_2}/2\pi =& \SI{1.5}{MHz}.
\end{split}
\end{equation}
A similar procedure is used to obtain $\beta\approx 16.4$ instead, which gives $\omega_- \approx 2\omega_d + 2\pi \times \SI{330}{MHz}$.

We note that these parameters are obtained from the full Hamiltonian without approximation: the exact inverse of the capacitance matrix is used to construct the Hamiltonian. \Cref{eq: coupling array} confirms that, among the high-energy collective modes, the lowest-energy mode—the antisymmetric mode—couples much more strongly to the symmetric mode than the others. The same conclusion was reached in Refs.~\cite{Ferguson:2013,Viola2015,singh:2025}, where a perturbative expansion in the ground capacitance ratio $C_g^{a}/C_J^{a}$ was used.

\begin{figure*}[t!]
    \centering
\includegraphics{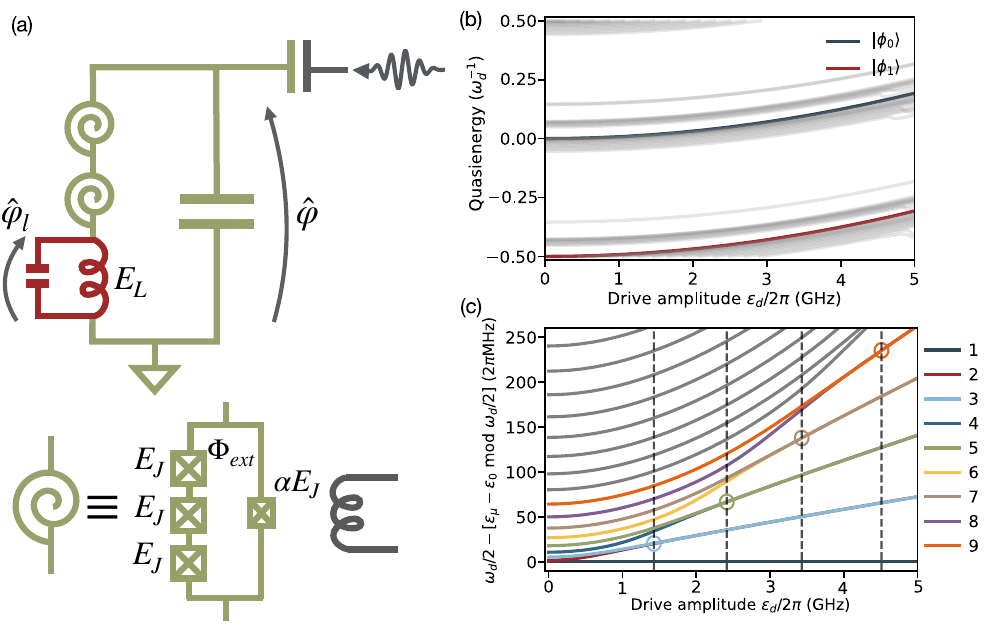}
\caption{(a) Circuit implementation of the Kerr-cat qubit used in Refs.~\cite{Frattini2024,dealbornoz2024,ding2025}, now explicitly including the finite geometric inductance (red). Because the inductive energy $E_L$ is comparable to the SNAIL Josephson energy $E_J$, usual expansions in $E_J/E_L$ cannot be used. Instead, we account for the inductance by retaining the high-frequency inductive mode that arises from the finite self-capacitance (red capacitor).
(b) Floquet spectrum versus drive amplitude for the lowest $50$ levels. The first two Floquet modes---representing the cat states---are highlighted in dark blue and red.
(c) Spectral-kissing pattern in the presence of the serial inductor, matching the behavior observed in the single mode case.} 
\label{fig:inductance}
\end{figure*}

To probe a worst-case scenario, the ground capacitances used here are intentionally chosen to be large, though still within realistic values (a few femtofarads). In practice, these capacitances can be made smaller, see for instance Ref.~\cite{Masluk:2012}, resulting in a weaker charge-charge coupling between the Kerr-cat qubit and the array modes. We therefore hypothesize that, regardless of the array-mode frequencies, it is hybridization with the buffer mode that sets the dominant limitation on Kerr-cat coherence in current experiments.


\section{Stray geometric inductance} \label{appendix: inductance}

In this Appendix, we assess the impact of the serial inductance present in the circuit implementation of the Kerr‐cat qubit in Refs.~\cite{Frattini2024,dealbornoz2024,ding2025} as shown in \cref{fig:inductance}(a). Recent work \cite{Putterman2025} has shown that such inductances can nontrivially renormalize the circuit potential and can degrade the performance of dissipative cat qubits if not properly optimized. Here we show that, although this serial inductance does modify the double‐SNAIL potential \cite{Frattini2021Thesis}, its effect on the Kerr-cat spectrum is minimal and does not affect the Kerr-cat qubit's performance.

To incorporate the serial inductance into the circuit description, one typically eliminates---via a Born–Oppenheimer approximation---the very high-frequency mode formed by the inductor together with its small intrinsic self-capacitance (see the red capacitor in \cref{fig:inductance}(a)). A subsequent perturbative expansion in the usually small parameter $E_J/E_L$, where $E_L$ is the inductive energy, yields an effective single-mode potential that captures the effect of the inductance; see, for example, Ref.~\cite{Kafri2017} for details. That approach leads, for instance, to higher Josephson harmonics in the cosine potential and improves predictions of the onset of measurement-induced transitions in strongly driven circuits~\cite{Fechant2025,Wang2025,Mingkang2025}. However, obtaining an analytical expression for a single-mode potential requires the condition $E_J \ll E_L$. In the Kerr‐cat experiments of Refs.~\cite{Frattini2024,dealbornoz2024,ding2025}, $E_L\sim E_J$, so the $E_J/E_L$ expansion cannot be used. Instead, Ref.~\cite{Frattini2021Thesis} accounted for the inductance by renormalizing the coefficients of the Taylor expansion of the potential around its minimum. While valid for the regime of parameters of interest, this method does not account for the full nonlinear potential and may therefore miss drive-activated multiphoton resonances that could be present if the full circuit nonlinearities are retained~\cite{Dumas2024}.

Here, following Ref.~\cite{Putterman2025}, we forgo any perturbative expansion by explicitly including the high‐frequency inductance mode and retaining the full cosine potential, given by the Hamiltonian
 \begin{equation} \label{eq:inductance}
     \hat{H}_{sl} = 4 E_C \hat{n}^2 + U_s(\hat{\varphi}-\hat{\varphi}_l) +\omega_l\hat{a}_l^\dagger\hat{a}_l,
 \end{equation}
 where $U_s$ is the double‐SNAIL potential from Eq.~\eqref{eq:doublesnail_singlemode}, and 
 \begin{equation}
 \hat{\varphi}_l = \sqrt{\omega_l/2E_L}(\hat{a}_l+\hat{a}_l^\dagger),
  \end{equation}
is the reduced phase of the inductive mode of frequency $\omega_l$ and annihilation operator $\hat{a}_l$. We choose parameters close to those in Refs.~\cite{Frattini2024,dealbornoz2024}, namely $E_L/2\pi= 214.55$ GHz, $E_J/2\pi = 273.28$ GHz and $E_C/2\pi = 129.87$ MHz. In \cref{eq:inductance}, the inductive phase $ \hat{\varphi}_l$ couples to the SNAIL phase $ \hat{\varphi}$ via $U_s$. Diagonalizing $\hat{H}_{sl}$ yields dressed eigenstates that we label $|i_s,j_l\rangle$, where $i_s$ and $j_l$ are the SNAIL and inductance‐mode excitations. For this diagonalization, we use a  Hilbert space of 250 Fock states in the SNAIL mode and 10 Fock states in the inductance mode, then retain the lowest 250 eigenlevels. We then add a charge drive on the SNAIL mode and perform the Floquet analysis as before, yielding Floquet modes $\ket{\phi_{i_s,j_l}}$. In particular, the cat states $\ket{\phi_{0_s,0_l}}$, $\ket{\phi_{1_s,0_l}}$, along with all other relevant Kerr-cat levels, now fully account for the presence of the inductance while at the same time encompass all circuit nonlinearities.

 Because the impedance of a geometric inductance $Z_l = \omega_lL$ cannot exceed the vacuum impedance $Z_{\textrm{vac}}\approx 376.73 \:\Omega$, see for instance Ref.~\cite{Manucharyan2009Fluxonium}, this sets an upper bound on $\omega_l$. Taking $Z_l \approx Z_{\textrm{vac}}$ gives $\omega_l/2\pi \approx 80$ GHz. Although we present results for these values, varying $Z_l$ and $E_L$ and thus $\omega_l$ over other realistic ranges yields the same conclusion.

The Floquet spectrum shown in \cref{fig:inductance}(b) closely resembles that of the single-mode case (Sec.\ref{sec:single mode}), exhibiting no resonances for the drive amplitudes considered here (up to $\approx 20$ photons in the cat manifold). Moreover, because the inductance mode lies at much higher frequency, any levels involving its excitation lie well above the manifold of interest and do not perturb the low-lying spectrum. We therefore conclude that—even though the serial inductance renormalizes the circuit potential (e.g., modifying the self-Kerr and third-order nonlinearities)—the overall structure of the Kerr-cat spectrum---e.g., the spectral-kissing shown in \cref{fig:inductance}(c)---remains unchanged, and we observe the same physics as in the single-mode case.

\section{Explicit coefficients for the second-order Schrieffer–Wolff generator}
\label{sec: Explicit coefficients c_{n, ...}}

\begin{widetext}
\begin{equation}
\begin{split}
c_{1, a^{\dag 3}}
=& \frac{1}{3\omega_0-\omega_d}\left[g_4 - g_3^2 \left( \frac{1}{3\omega_0} + \frac{1}{\omega_d} \right) \right] 
\\
c_{1, a^\dag a^2}
=& \frac{1}{\omega_d - \omega_0}\left[ 3g_4 + g_3^2 \left( \frac{1}{\omega_d} - \frac{1}{\omega_d +2\omega_0} - \frac{5}{6\omega_0} \right) \right] 
\\
c_{1, a}
=& \frac{1}{\omega_d -\omega_0}\Bigg[ 3g_4(1+|\Pi|^2) - g_3^2 \bigg(\frac{1+2|\Pi|^2}{\omega_d} - \frac{|\Pi|^2}{2\omega_d-\omega_0} + \frac{1}{\omega_d + 2\omega_0} + \frac{|\Pi|^2}{2\omega_d + \omega_0} + \frac{1 + 2|\Pi|^2}{\omega_0} - \frac{|\Pi|^2}{\omega_d} 
\\
&+ \frac{1}{3\omega_0} + \frac{|\Pi|^2}{\omega_d + 2\omega_0}
\bigg) \Bigg] 
\\
c_{2, a^{\dag 3} a} 
=& \frac{1}{2\omega_0} \left[g_4 - \frac{4g_3^2}{3\omega_0} \right] 
\\
c_{2, a^{\dag 2}}
=& \frac{1}{4\omega_0} \left\{ 3g_4(1+|\Pi|^2) - 4g_3^2 \left[ \frac{3+2|\Pi|^2}{\omega_0} + |\Pi|^2 \left( \frac{1}{\omega_d +2\omega_0} + \frac{1}{\omega_d} \right) \right]  \right\} 
\\
c_{2, a^{2}}
=& \frac{1}{4(\omega_d-\omega_0)} \left[ 3g_4 + g_3^2 \left( \frac{1}{2\omega_d - \omega_0} - \frac{1}{2\omega_d + \omega_0} - \frac{4}{3\omega_0} \right) \right]
\\
c_{3, a^{\dag 2} a}
=& \frac{1}{\omega_d+ \omega_0} \left[ 3g_4 - g_3^2 \left( \frac{2}{\omega_d + 2\omega_0} + \frac{3}{\omega_0} - \frac{1}{\omega_d} \right)\right]
\\
c_{3, a^\dag}
=& \frac{1}{\omega_d+ \omega_0} \left[ 3g_4(1+|\Pi|^2) -g_3^2 \left( \frac{2(1+|\Pi|^2)}{\omega_d + 2\omega_0} + \frac{|\Pi|^2}{2\omega_d + \omega_0} - \frac{(1+|\Pi|^2)}{\omega_d} + \frac{3(1+2|\Pi|^2)}{\omega_0} \right) \right]
\\
c_{4, a^{\dag 4}}
=& \frac{1}{16\omega_0} \left( g_4 + \frac{4g_3^2}{\omega_0} \right)
\\
c_{4, a^\dag a}
=& \frac{1}{2\omega_d} \left[ 3g_4 + g_3^2  \left( \frac{1}{2\omega_d - \omega_0} - \frac{1}{2\omega_d + \omega_0} - \frac{2}{\omega_0} \right) \right]
\\
c_{5, a^{\dag 3}}
=& \frac{g_4}{4(\omega_d + 3\omega_0)}
\\
c_{5, a}
=& \frac{g_4}{4(3\omega_d - \omega_0)}
\\
c_{6, a^{\dag 2}}
=& \frac{3g_4}{4(\omega_d + \omega_0)}
\\
c_{7, a^{\dag}}
=& \frac{g_4}{3\omega_d + \omega_0}.
\end{split}
\end{equation}
\end{widetext}

\section{Explicit coefficients of operators in the second-order nested commutator $-\frac{1}{2}\sum_m [\hat G_{n+1-m}^{(1)}, [\hat G_{m}^{(1)}, \hat a]]$}
\label{sec: Explicit coefficients d_{n, ...}}

\begin{widetext}
\begin{equation}
\begin{split}
d_{-6,1} =& \frac{g_3^2}{\omega_d (2\omega_d-\omega_0)} \\
d_{-5,a} =& g_3^2 \left( \frac{2}{\omega_d^2} - \frac{1}{\omega_0 (2\omega_d + \omega_0)} - \frac{1}{\omega_0 (2\omega_d - \omega_0)} \right) \\
d_{-4,a^2} =& g_3^2 \left( \frac{3}{\omega_d \omega_0} + \frac{2}{\omega_0(\omega_d + 2\omega_0)} \right) \\
d_{-3, a^3} =& \frac{4 g_3^2}{3\omega_0^2} \\
d_{-3, a^\dag} =& g_3^2 \left( \frac{1}{3\omega_0(2\omega_d + \omega_0)} - \frac{1}{\omega_0 (2\omega_d - \omega_0)}\right) \\
d_{-2, a^\dag a} =& \frac{2g_3^2}{\omega_0}\left( \frac{1}{3(\omega_d + 2\omega_0)} - \frac{1}{\omega_d} \right) \\
d_{-2, 1} =& g_3^2 \left( \frac{1}{3\omega_0 (\omega_d +2\omega_0)} + \frac{|\Pi|^2}{(\omega_d +2\omega_0)(2\omega_d + \omega_0)} - \frac{|\Pi|^2}{\omega_d (2\omega_d - \omega_0)} - \frac{(1+2|\Pi|^2)}{\omega_d \omega_0} \right) \\
d_{-1,a^\dag a^2} =& -\frac{8g_3^2}{9\omega_0^2} \\
d_{-1, a} =& -2g_3^2 \left[ \frac{4}{9\omega_0^2} + |\Pi|^2 \left( \frac{2}{\omega_d^2} - \frac{1}{(\omega_d + 2\omega_0)^2} \right) \right] \\
d_{0, a^{\dag 2}} =& \frac{g_3^2}{3\omega_d \omega_0} \\
d_{0, a^2} =& \frac{g_3^2}{\omega_0} \left( \frac{1}{3(\omega_d + 2\omega_0)} - \frac{3}{\omega_d} \right) \\
d_{1, a^{\dag 2} a} =& \frac{4g_3^2}{\omega_0^2} \\
d_{1 a^\dag} =& \frac{2 g_3^2}{3\omega_0^2} (2+5|\Pi|^2) \\
d_{2,a^\dag a} =& \frac{2g_3^2}{\omega_d \omega_0} \\
d_{2,1} =& g_3^2 \left( \frac{(1+2|\Pi|^2)}{\omega_d \omega_0} + \frac{2|\Pi|^2}{\omega_0 (\omega_d + 2\omega_0)} - \frac{|\Pi|^2}{\omega_d(2\omega_d + \omega_0)} \right) \\
d_{3,a} =& g_3^2 \left( \frac{2}{\omega_d^2} - \frac{1}{\omega_0 (2\omega_d - \omega_0)} - \frac{1}{\omega_0 (2\omega_d + \omega_0)} \right) \\
d_{4, a^{\dag 2}} =& - \frac{g_3^2}{3\omega_d \omega_0} \\
d_{5, a^\dag} =& \frac{g_3^2}{\omega_0}\left( \frac{1}{2\omega_d + \omega_0} - \frac{1}{3(2\omega_d - \omega_0)} \right) \\
d_{6, 1} =& g_3^2 \left( \frac{1}{\omega_d (2\omega_d + \omega_0)} - \frac{1}{(\omega_d + 2\omega_0)(2\omega_d - \omega_0)} \right)
\end{split}
\end{equation}
\end{widetext}

\end{document}